\documentclass[11pt,fleqn]{article}

\usepackage{amsmath,amssymb,amsthm,enumerate,cite}%,backref

\allowdisplaybreaks

\newcommand{\p}{\partial}
\newcommand{\const}{\mathop{\rm const}\nolimits}
\newcommand{\sign}{\mathop{\rm sign}\nolimits}

\newcommand{\thetbn}{\arabic{nomer}}
\newcounter{tbn}

\newcounter{mcasenum}

\newtheorem{theorem}{Theorem}
\newtheorem{lemma}{Lemma}
\newtheorem{corollary}{Corollary}
\newtheorem{proposition}{Proposition}
\newtheorem*{proposition*}{Proposition}
{\theoremstyle{definition}
\newtheorem{definition}{Definition}

\newtheorem{note}{Note}

}

\begin{document}

\par\noindent {\LARGE\bf
Enhanced Group Analysis and Exact Solutions\\ of Variable Coefficient Semilinear
Diffusion Equations with a Power Source
\par}

{\vspace{4mm}\par\noindent \it O.\,O.~Vaneeva$^{\dag 1}$, R.\,O.~Popovych$^{\dag\ddag 2}$ and C. Sophocleous$^{\S 3}$
\par\vspace{2mm}\par}

{\vspace{2mm}\par\noindent\it
${}^\dag$\ Institute of Mathematics of National Academy of Sciences of Ukraine, \\
$\phantom{{}^\dag}$\ 3 Tereshchenkivska Str., Kyiv-4, 01601 Ukraine\\
$^\ddag$\;Fakult\"at f\"ur Mathematik, Universit\"at Wien, Nordbergstra{\ss}e 15, A-1090 Wien, Austria\\
$^\S$\;Department of Mathematics and Statistics, University of Cyprus, Nicosia CY 1678, Cyprus
}
{\noindent\it
$\phantom{{}^\dag{}\;}^1$vaneeva@imath.kiev.ua,
$^2$rop@imath.kiev.ua,
$^3$christod@ucy.ac.cy
\par}

{\vspace{5mm}\par\noindent\hspace*{8mm}\parbox{140mm}{\small
A new approach to group classification problems and more general investigations on transformational
properties of classes of differential equations is proposed.
It is based on mappings between classes of differential equations, generated by families of point transformations.
A~class of variable coefficient (1+1)-dimensional semilinear
reaction--diffusion equations of the general form
$f(x)u_t=(g(x)u_x)_x+h(x)u^m$ ($m\ne0,1$) is studied from the symmetry point of view
in the framework of the approach proposed.
The singular subclass of the equations with $m=2$  is singled out.
The group classifications of the entire class, the singular subclass and their images are performed
with respect to both the corresponding (generalized extended) equivalence groups and all point transformations.
The set of admissible transformations of the imaged class is exhaustively described in the general case $m\ne2$.
The procedure of classification of nonclassical symmetries,
which involves mappings between classes of differential equations, is discussed.
Wide families of new exact solutions are also constructed for equations from the classes under consideration
by the classical method of Lie reductions and by generation of new solutions from known ones for other equations
with point transformations of different kinds
(such as additional equivalence transformations and mappings between classes of equations).
}\par\vspace{5mm}}

\section{Introduction}

Investigation of diffusion equations (with reaction or convection terms or without them) is important
since they are often used as mathematical models of various processes in nature and society.
For example, in biology~\cite{murray2002} one can consider cells, bacteria, chemicals, animals and so on as particles
each of which usually moves around in a random way. Then, a regular motion of their group is assumed as a diffusion
process and often it is not a simple diffusion since there may be an interaction between particles. For simplicity,
biologists often use (1+1)-dimensional continuum model equations for the description of global behaviour
in terms of particle density or concentration. Exact solutions of the model equations allow one to study
particle concentration distribution and character of diffusing. Such solutions can be constructed in a regular
way by the reduction method using Lie symmetry operators. It is one of the reasons for choosing model equations
with nontrivial symmetry properties, i.e., equations admitting Lie symmetry algebra of the maximally possible dimension.
Group classification of a class of reaction--diffusion equations helps to make
appropriate choice of model equations and then to obtain exact solutions.

In the present paper we investigate, from the symmetry point of view, the class of
variable coefficient (1+1)-dimensional semilinear reaction--diffusion equations of the general form
\begin{equation} \label{eqRDfghPower}
f(x)u_t=(g(x)u_x)_x+h(x)u^m,
\end{equation}\looseness=-1
where $f=f(x)$, $g=g(x)$ and $h=h(x)$  are arbitrary smooth functions of the variable~$x$,
$f(x)g(x)h(x)\neq0$, $m$ is an arbitrary constant.
The linear case is excluded from consideration as well-investigated \cite{Ibragimov1994V1;2,Lie1881,Ovsiannikov1982}.
One more reason for this exclusion is that the linear case is singular from the symmetry point of view
and is not connected with the nonlinear case via point transformations.
For this reason we assume through the paper that $m\neq0,1$.
Lie symmetries of equations from class~\eqref{eqRDfghPower}
with constant arbitrary elements $f$, $g$ and $h$
were found in~\cite{Dorodnitsyn1979;1982,Ovsiannikov1959}
within the group classification of the class of equations having the form $u_t=(k(u)u_x)_x+q(u)$.
Class~\eqref{eqRDfghPower} is included in the wider class of quasilinear reaction--diffusion equations
with power nonlinearities
\begin{equation}\label{eqRDfghTwoPower}
f(x)u_t=(g(x)u^nu_x)_x+h(x)u^m,
\end{equation}
investigated for $n\neq0$ in~\cite{VJPS2007}.
A number of special equations from class~\eqref{eqRDfghTwoPower} were successfully used to model phenomenon in physics, chemistry and
biology~\cite{crank1979,kamin&rosenau1982,murray2002,peletier1981,touloukian1970}.
The semilinear subclass~\eqref{eqRDfghPower} is singular in class~\eqref{eqRDfghTwoPower} with respect to symmetry properties 
and cannot be classified with the same method as the subclass with $n\ne0$.
Hence it was excluded from consideration in~\cite{VJPS2007}.

Extended group analysis of class~\eqref{eqRDfghPower} is first carried out in this paper.
Equivalence groups of different kinds are found for class~\eqref{eqRDfghPower}, its subclasses and some related classes.
All the point transformations between cases of Lie symmetry extensions, which are additional to
the transformations from the equivalence groups, are constructed.
As a result, two group classification problems are really solved for each from the above classes---with respect to
the corresponding extended equivalence group and up to general point-transformation equivalence.
The set of admissible transformations in the case $m\ne0,1,2$ is exhaustively described.
Some nonclassical symmetries of equations from class~\eqref{eqRDfghPower} are constructed.
A special procedure of the exhaustive classification of nonclassical symmetries is proposed for this class.
Wide families of exact solutions for the equations under consideration are obtained
by the classical Lie reduction method or generated from known solutions
by additional equivalence transformations or via mappings between classes of equations.

\looseness=-1
There are two main approaches in studying group classification problems in the literature. 
The first one is more algebraic and based on subgroup analysis of the equivalence group associated with 
a class of differential equations under consideration. 
It can be applied if the class is normalized, i.e., if any admissible point transformation in this class 
is generated by a transformation from its equivalence group. 
See \cite{Popovych&Kunzinger&Eshraghi2006} for rigorous definitions of normalized classes and related notions. 
If the class is sufficiently general and possesses a wide equivalence group then the problem is reduced to 
the description of inequivalent realizations of Lie algebras in certain set of vector fields. 
The corresponding theoretical background is presented in detail in~\cite{Basarab-Horwath&Lahno&Zhdanov2001,Zhdanov&Lahno1999;2007}.
It is the approach that was applied by S.~Lie to solve the best known classical group classification problems 
such as the classifications of second order ordinary differential equations~\cite{Lie1891} and
of second order two-dimensional linear partial differential equations~\cite{Ibragimov1994V1;2,Lie1881}.
Recently many classes of $(1+1)$-dimensional evolution equations 
and other important classes of differential equations arising in different sciences
were classified within the framework of the above method
\cite{Abramenko&Lagno&Samojlenko2002,Basarab-Horwath&Lahno&Zhdanov2001,Gagnon&Winternitz1993,
Lagno&Samojlenko2002,Lahno&Spichak&Stognii2002,Lahno&Zhdanov&Magda2006,Popovych&Kunzinger&Eshraghi2006,
Zhdanov&Lahno2005,Zhdanov&Lahno1999;2007}.
The second approach is based on the investigation of compatibility and the direct integration, 
up to the equivalence relation generated by the corresponding equivalence group, 
of determining equations implied by the infinitesimal invariance criterion~\cite{Ovsiannikov1982}. 
This is the most applicable approach but it is efficient only for classes of a simple structure, e.g., 
which have a few arbitrary elements of one or two same arguments or whose equivalence groups are finite-dimensional. 
A number of results on group classification problems investigated within the framework of this approach are 
collected in \cite{BlumanKumei1989,Ibragimov1994V1;2,Ovsiannikov1982} and other books on the subject. 
At the same time, normalized or simple classes do not exhaust all classes of differential equations, 
which are interesting for applications and with the mathematical point of view. 
To solve more group classification problems and to present results in an optimal way, different tools and notions 
(additional equivalence transformations, extended and generalized equivalence groups, conditional equivalence group, 
gauging of arbitrary elements by equivalence transformations, partition of a class to normalized subclasses etc.)
were recently proposed \cite{Ibragimov1994V1;2,Ivanova&Popovych&Sophocleous2006I,Popovych&Kunzinger&Eshraghi2006,Meleshko1994,VJPS2007}.
Their usage had the critical value, in particular, for the complete group classifications of class~\eqref{eqRDfghTwoPower}, where   
$n\neq0$, \cite{VJPS2007} and a class of variable coefficient diffusion--convection equations \cite{Ivanova&Popovych&Sophocleous2006I}. 
Unfortunately, all the methods existing in the literature are not powerful enough  
to exhaustively classify class~\eqref{eqRDfghPower}.

In fact, the ultimate goal of the paper is to present a new approach to group classification problems
and more general investigations on transformational properties of classes of differential equations;
and the group analysis of equations from class~\eqref{eqRDfghPower} is only an illustrative example
on application of different techniques within the framework of the approach proposed.
Mappings between classes of differential equations, generated by families of point transformations,
are involved in the consideration. The approach proposed is based on the simple fact, that
symmetry, transformational and other related properties of differential equations are changed in a predictable way
under such mappings.
A particular case of the mappings between classes of differential equations is given by
mappings of a class into itself, generated by transformations from the corresponding equivalence group
or admissible transformations of the class.

The structure of the paper is as follows:

A necessary theoretical background on mappings between classes of differential equations 
and applications of such mappings to group classification problems
is developed in section~\ref{SectionOnClassMappingsAndGroupClassification}.

In section~\ref{SectionOnEquivTransOfEqRDfghPower_n0} we construct both the usual
and the generalized extended equivalence groups for
class~\eqref{eqRDfghPower} as well as for related classes arising
later under the implementation of the proposed approach to group classification.
The first related class is a subclass of class~\eqref{eqRDfghPower} obtained with the gauge~$g=f$.
It is the gauge that allows us to perform a further gauging via mapping to other classes in the simplest way.
Since all results on symmetries and exact solutions of class~\eqref{eqRDfghPower} can be obtained
from the analogous results for the its subclass associated with the gauge $g=f$,
we restrict ourselves by investigation of this subclass and call it the {\it initial class}.
The second related class is the image of the initial class with respect to a family of nondegenerate point transformations
parameterized by arbitrary elements. So, it is called the {\it imaged class}.
It appears that Lie symmetries of the imaged class are capable to be directly classified only for values of $m\neq2$.
Hence an additional gauge via mapping onto another class is needed for $m=2$.
Such gauge is found and the imaged class with $m=2$ is transformed to the {\it double-imaged one}
which is the third related class arising in our consideration.

Results of section~\ref{SectionOnClassMappingsAndGroupClassification} 
are applied in section~\ref{SectionOnMappingsAndGroupClassificationOfClassUnderConsideration}
to justify the techniques used for the investigation of class~\eqref{eqRDfghPower}.

In section~\ref{SectionOnRDfghLieSymmetries_n0} we carry out, at first, the group classifications of the imaged and
double-imaged classes (subsections~\ref{Subsection_group classification of image class}
and~\ref{Subsection_group classification of double-image class}, respectively)
and then extend the obtained results in subsection~\ref{Subsection_group classification of initial class}
to the classification of Lie symmetries of the initial class.

In section~\ref{SectionOnAdditionalTransformations_n0} additional equivalence transformations
between equations, having extensions of Lie symmetries, are found and used
for performing the group classifications of the imaged classes as well as of the initial class 
with respect to all point transformations.

The form-preserving (admissible) transformations of the imaged class
for the case $m\ne0,1,2$ are studied in section~\ref{SectionClassificationOfAdmTrans_n0}.
As a result, the structure of the sets of admissible transformations of
the imaged class, the initial class and class~\eqref{eqRDfghPower} is described exhaustively
for the nonsingular values of~$m$.

In section~\ref{Section_NonclassicalSymmetries_n0} the approach based on gauging is extended to study nonclassical symmetries.
The properties of equivalence relations on sets of reduction operators and mappings between such sets
are investigated for general classes of differential equations.
A procedure of nonclassical symmetry classification of equations from class~\eqref{eqRDfghPower}
is discussed.
We also single out their subclasses for which reduction operators are derived from known ones for
constant coefficient equations from other classes.
The main idea of the classification procedure is to look for nonclassical symmetries using the same gauges as
for the classification of Lie symmetries, i.e., to find them for the imaged classes and then reconstruct
the corresponding reduction operators for the equations from the initial class.

Both Lie and non-Lie exact solutions for equations from the initial class and related ones are constructed in
section~\ref{Section_OnRDfghExactSolutions_n0} as an illustration of possible applications of the above classification results.
Namely, in subsection~\ref{Subsection_OnRDfghSimilaritySolutions_n0} solutions are obtained by Lie reductions 
involving Lie symmetries found under the group classification.
It is shown in subsection~\ref{Subsection_GenerationOfSolutions_n0} how to use gauging transformations between classes and 
additional equivalence transformations for generating exact solutions for equations from the initial class 
using known solutions of equations with constant coefficients, belonging to the imaged class.

In the conclusion the obtained results are summarized and the problem of group classification 
of a more general class of variable coefficient reaction--diffusion equations is discussed.

\section{Application of class mappings to group classification}\label{SectionOnClassMappingsAndGroupClassification}

Following the procedure of~\cite{Popovych2006c,Popovych&Kunzinger&Eshraghi2006}, we discuss mappings
between classes of systems of differential equations, generated by point transformations.
The behavior of transformational properties of classes under such mappings is studied.
These results form the foundation of our approach to group classification of class~\eqref{eqRDfghPower}.

Let~$\mathcal L_\theta$ be a system $L(x,u_{(p)},\theta(x,u_{(p)}))=0$
of $l$~differential equations %$L^1=0$, \ldots, $L^l=0$
for $m$~unknown functions $u=(u^1,\ldots,u^m)$ and
of $n$~independent variables $x=(x_1,\ldots,x_n)$.
Here $u_{(p)}$ denotes the set of all derivatives of~$u$ with respect to $x$
of order no greater than~$p$, including $u$ as the derivatives of order zero.
$L=(L^1,\ldots,L^l)$ is a tuple of $l$ fixed functions depending on~$x,$ $u_{(p)}$ and~$\theta$,
where $\theta$ denotes the tuple of arbitrary (parametric) functions
$\theta(x,u_{(n)})=(\theta^1(x,u_{(p)}),\ldots,\theta^k(x,u_{(p)}))$ called arbitrary elements
and running the set~$\mathcal{S}$ of solutions of the auxiliary system
\[
S(x,u_{( p)},\theta_{(q)}(x,u_{(p)}))=0, \quad \Sigma(x,u_{(p)},\theta_{(q)}(x,u_{(p)}))\ne0.
\]
This system consists of differential equations and inequalities with respect to $\theta$,
where $x$ and $u_{(p)}$ play the role of independent variables
and $\theta_{(q)}$ stands for the set of all the partial derivatives of $\theta$ of order no greater than $q$
with respect to the variables $x$ and $u_{(p)}$.
(This inequality means that no components of~$\Sigma$ vanish. For simplicity the tuple $\Sigma$ can be
replaced by a single differential function coinciding with the product of its components.)
The tuples $S$ or $\Sigma$ may be empty.
We denote \emph{the class of systems~$\mathcal L_\theta$ with the arbitrary elements $\theta$ running through $\mathcal S$}
as~$\mathcal L|_{\mathcal S}$.

Let $\mathcal L_\theta^i$ denote the set of all algebraically independent differential consequences of $\mathcal L_\theta$,
which have, as differential equations, orders no greater than~$i$. We identify~$\mathcal L_\theta^i$ with
the manifold determined by~$\mathcal L_\theta^i$ in the jet space~$J^{(i)}$.
In particular, $\mathcal L_\theta$ is identified with the manifold determined by~$\mathcal L_\theta^p$ in~$J^{(p)}$.
Then $\mathcal L|_{\mathcal S}$ can be interpreted as a family of manifolds in~$J^{(p)}$, parametrized with
the arbitrary elements $\theta\in\mathcal S$.

\emph{Subclasses} are singled out in the class $\mathcal L|_{\mathcal S}$
with additional auxiliary systems of equations and/or non-vanish conditions
which are attached to the main auxiliary system.

The above definition of a class of differential equations is not really complete.
Gauge equivalence and other nuances have to be taken into account.
See~\cite{Popovych2006c,Popovych&Kunzinger&Eshraghi2006} for details.

For $\theta,\tilde\theta\in\mathcal{S}$ we call the set of point transformations
which map the system~$\mathcal L_\theta$ into the system~$\mathcal L_{\tilde\theta}$
the \emph{set of admissible transformations from~$\mathcal L_\theta$ into~$\mathcal L_{\tilde\theta}$}
and denote it by $\mathrm{T}(\theta,\tilde\theta)$.
The maximal point symmetry group~$G_\theta$ of the system~$\mathcal L_\theta$
coincides with~$\mathrm{T}(\theta,\theta)$.
If the systems~$\mathcal L_\theta$ and $\mathcal L_{\tilde\theta}$ are equivalent with respect to
point transformations then
$\mathrm{T}(\theta,\tilde\theta)=\varphi^0\circ G_\theta=G_{\tilde\theta}\circ\varphi^0$,
where $\varphi^0$ is a fixed transformation from~$\mathrm{T}(\theta,\tilde\theta)$.
Otherwise, $\mathrm{T}(\theta,\tilde\theta)=\varnothing$.
Analogously, the set
$\mathrm{T}(\theta,\mathcal L|_{\mathcal S})=\{\,(\tilde\theta,\varphi)\mid
\tilde\theta\in\mathcal{S},\,%\mathrm{T}(\theta,\tilde\theta)\not=\varnothing,\,
\varphi\in\mathrm{T}(\theta,\tilde\theta)\, \}$
is called the {\em set of admissible transformations of the system~$\mathcal L_\theta$
into the class~$\mathcal L|_{\mathcal S}$}.

\begin{definition}\label{DefOfSetOfAdmTrans}
$\mathrm{T}(\mathcal L|_{\mathcal S})=\{(\theta,\tilde\theta,\varphi)\mid
\theta,\tilde\theta\in\mathcal{S},\,%\mathrm{T}(\theta,\tilde\theta)\not=\varnothing,\,
\varphi\in\mathrm{T}(\theta,\tilde\theta)\}$
is called the {\em set of admissible transformations in~$\mathcal L|_{\mathcal S}$}.
\end{definition}

\begin{note}
The set of admissible transformations was first described by Kingston and Sophocleous
for a class of generalized Burgers equations~\cite{Kingston&Sophocleous1991}.
These authors call transformations of such type {\em form-preserving}
\cite{Kingston&Sophocleous1991,Kingston&Sophocleous1998,Kingston&Sophocleous2001}.
The notion of admissible transformations can be considered as a formalization of their approach.
\end{note}

Now, we introduce notations and notions which are necessary for presentation
of the classical formulation of group classification problems.
$A_\theta$ denotes the maximal Lie invariance (or principal) algebra
of infinitesimal symmetry operators of~$\mathcal L_\theta$ for a fixed $\theta\in\mathcal{S}$.
The common part $A^\cap=A^\cap(\mathcal L|_{\mathcal S})=\bigcap_{\theta\in{\cal S}}A_\theta$ of
all $A_\theta$, $\theta\in\mathcal{S}$, is called
the \emph{kernel of the maximal Lie invariance algebras} of systems from the class $\mathcal L|_{\mathcal S}$.
The  equivalence group of the class~$\mathcal L|_{\mathcal S}$ is denoted by
$G^{\sim}=G^{\sim}(\mathcal L|_{\mathcal S})$.
Roughly speaking, $G^{\sim}$ is the set of admissible transformations
which can be applied to any~$\theta\in\mathcal{S}$.

In the framework of the infinitesimal approach,
the problem of group classification is reformulated as finding all possible
inequivalent cases of extensions for $A_\theta$, i.e.,
as listing all $G^\sim$-inequivalent values of the arbitrary parameters~$\theta$
together with $A_\theta$ satisfying the condition
$A_\theta\ne A^{\cap}$~\cite{Akhatov&Gazizov&Ibragimov1989,Ovsiannikov1982}.
More precisely, the solution of the group classification problem is a list of pairs
$(\mathcal{S}_\gamma,\{A_\theta,\theta\in\mathcal{S}_\gamma\})$, $\gamma\in\Gamma$.
Here $\{\mathcal{S}_\gamma, \gamma\in\Gamma\}$ is a family of subsets of $\mathcal{S}$,
$\bigcup_{\gamma\in\Gamma}\mathcal{S}_\gamma$ contains only $G^{\sim}$-inequivalent values
of $\theta$ with $A_\theta\ne A^\cap$,
and for any $\theta\in\mathcal{S}$ with $A_\theta\ne A^\cap$ there exists $\gamma\in\Gamma$ such that
$\theta\in\mathcal{S}_\gamma\!\!\mod G^{\sim}$.
The structures of the $A_\theta$ are similar for different values of $\theta\in\mathcal{S}_\gamma$ under fixed $\gamma$.
In particular, all $A_\theta$, $\theta\in\mathcal{S}_\gamma$, have the same dimension or
display the same arbitrariness of algebra parameters in the infinite-dimensional case.

The notion of similar differential equations~\cite{Ovsiannikov1982} can be extended to
classes of (systems of) differential equations in a number of ways.
The most direct and evident generalization is given by the following definition.

\begin{definition}\label{DefOnSimilarClasses}
The classes $\mathcal L|_{\mathcal S}$ and $\mathcal L'|_{\mathcal S'}$ are called \emph{similar} if
$n=n'$, $m=m'$, $p=p'$,  $k=k'$ and
there exists a point transformation $\Psi\colon(x,u_{(p)},\theta)\to(x',u'_{(p)},\theta')$
which is projectable on the space of $(x,u_{(q)})$ for any $0\le q\le p$,
and $\Psi|_{(x,u_{(q)})}$ being the $q$-th order prolongation of $\Psi|_{(x,u)}$,
%$\forall\theta\in\mathcal{S}$: $\Psi\theta\in\mathcal{S}'$ and $\forall\theta'\in\mathcal{S}'$: $\Psi^{-1}\theta'\in\mathcal{S}$
$\Psi \mathcal{S}=\mathcal{S}'$
and $\Psi|_{(x,u_{(p)})}\mathcal L_\theta=\mathcal L'_{\Psi\theta}$ for any $\theta\in\mathcal{S}$.
\end{definition}

Here and in what follows the action of such point transformation $\Psi$ in the space of $(x,u_{(p)},\theta)$ on arbitrary elements from $\mathcal{S}$
as $p$th-order differential functions is given by the formula:
\begin{gather*}
\tilde\theta=\Psi\theta \quad\mbox{if}\quad
\tilde\theta(x,u_{(p)})=\Psi^\theta\Bigl(\Theta(x,u_{(p)}),\theta\bigl(\Theta(x,u_{(p)})\bigr)\Bigr),
\end{gather*}
where $\Theta=(\mathrm{pr}_p\Psi|_{(x,u)})^{-1}$ and
$\mathrm{pr}_p$ denotes the operation of standard prolongation of a point transformations
to the derivatives of orders not greater than~$p$.

Roughly speaking, similar classes consist of similar equations with the same similarity transformation.

\begin{proposition}
Similar classes have similar sets of admissible transformations.
Namely, a similarity transformation $\Psi$ from the class $\mathcal L|_{\mathcal S}$ into the class $\mathcal L'|_{\mathcal S'}$
generates a one-to-one mapping $\Psi^\mathrm{T}$
from $\mathrm{T}(\mathcal L|_{\mathcal S})$ into $\mathrm{T}(\mathcal L'|_{\mathcal S'})$
via the rule $(\theta'\!,\tilde\theta'\!,\varphi')=\Psi^\mathrm{T}(\theta,\tilde\theta,\varphi)$
if $\theta'=\Psi\theta$, $\tilde\theta'=\Psi\tilde\theta$ and
$\varphi'=\Psi|_{(x,u)}\circ\varphi\circ\Psi|_{(x,u)}{}^{-1}$.
Here $(\theta,\tilde\theta,\varphi)\in\mathrm{T}(\mathcal L|_{\mathcal S})$,
$(\theta'\!,\tilde\theta'\!,\varphi')\in\mathrm{T}(\mathcal L'|_{\mathcal S'})$.
\end{proposition}

Similar classes of differential equations have also similar equivalence groups.
More precisely, if classes are similar with respect to a transformation of a certain kind
(e.g., a point transformation of the independent variables, the dependent variables,
their derivatives and the arbitrary elements)
then equivalence groups formed by the equivalence transformations of the same kind are similar with respect to this transformation.

It is obvious that lists of Lie symmetries under group classification
with respect to any of the above equivalence relations are similar for similar classes.
Namely, if a transformation $\Psi$ realizes the similarity of the class $\mathcal L'|_{\mathcal S'}$ to
the class $\mathcal L|_{\mathcal S}$ (see definition~\ref{DefOnSimilarClasses}) and
\[\{\{(\theta,A_\theta), \theta\in\mathcal{S}_\gamma\}, \mathcal{S}_\gamma\subset\mathcal{S}, \gamma\in\Gamma\}\]
is a classification list for the class $\mathcal L|_{\mathcal S}$ then
\[
\{\{(\Psi\theta,(\Psi|_{(x,u)})_*A_\theta), \Psi\theta\in\Psi\mathcal{S}_\gamma\}, \Psi\mathcal{S}_\gamma\subset\mathcal{S}', \gamma\in\Gamma\}
\]
is a classification list for the class $\mathcal L'|_{\mathcal S'}$.
Here $(\Psi|_{(x,u)})_*$ is the mapping induced by the transformation~$\Psi$
in the set of vector fields on the space $(x,u)$ (push-forward of vector fields).
$A_\theta$ is the maximal Lie invariance (or principal) algebra of infinitesimal symmetry operators of~$\mathcal L_\theta$.

The set of transformations used in definition~\ref{DefOnSimilarClasses}
can be extended via admitting different kinds of dependence on arbitrary elements
as in the case of equivalence groups.
As a rule, similar classes of systems have similar properties from the group analysis point of view.
In the case of point similarity transformations, the properties really are the same up to similarity.
If~$\Psi$ is a point transformation in the space of $(x,u_{(p)},\theta)$ then these classes practically have the same transformational properties.

If the transformation~$\Psi$ is identical with respect to $x$ and $u$ then
$\mathcal L'_{\Psi\theta}=\mathcal L_\theta$ for any $\theta\in\mathcal{S}$, i.e., in fact
the classes $\mathcal L|_{\mathcal S}$ and $\mathcal L'|_{\mathcal S'}$ coincide as sets of manifolds in a jet space.
We will say that the class $\mathcal L'|_{\mathcal S'}$ is a \emph{re-parametrization} of the class $\mathcal L|_{\mathcal S}$,
associated with the re-parametrizing transformation~$\Psi$.
In the most general approach, $\Psi$ can be assumed an arbitrary one-to-one mapping from~$\mathcal{S}$ to $\mathcal{S}'$,
satisfying the condition $\mathcal L'_{\Psi\theta}=\mathcal L_\theta$ for any $\theta\in\mathcal{S}$.
Note that the number of arbitrary elements in $\mathcal{S}'$ might not coincide with the one in~$\mathcal{S}$.
Transformational properties may be broken under generalized re-parametrizations.

An example of non-point re-parametrization often applied in group analysis is given
by the classes $\{\mathcal I=\theta(\mathcal J)\}$ and $\{\mathcal J=\hat\theta(\mathcal I)\}$,
where $\mathcal I$ and $\mathcal J$ are $k$-tuples of fixed functionally independent expressions of~$x$ and~$u_{(p)}$,
$\theta$ and $\hat\theta$ are arbitrary $k$-ary $k$-vector functions with nonzero Jacobians.
The corresponding mapping between the sets of arbitrary elements is to take the inverse function to each set element.
The advantage of such re-parametrization is that preserves transformational properties of classes.

Similarity of classes implies a one-to-one correspondence between the associated sets of arbitrary elements.
If this feature is neglected, we result to a more general and complicated notion of mappings between classes of differential equations.

\begin{definition}\label{DefOnImageClasses}
The class $\mathcal L'|_{\mathcal S'}$ is called a \emph{point-transformation image} of
the class $\mathcal L|_{\mathcal S}$ if
$n=n'$, $m=m'$, $p=p'$,  $k=k'$ and
there exists a family $\bar\varphi$ of point transformations $\varphi_\theta\colon(x,u)\to(x',u')$ parametrized by $\theta\in\mathcal{S}$
and satisfying the following condition:
For any $\theta\in\mathcal{S}$ there exists $\theta'\in\mathcal{S}'$
and, conversely, for any $\theta'\in\mathcal{S}'$ there exists $\theta\in\mathcal{S}$
such that $\mathrm{pr}_p\varphi_\theta\mathcal L_\theta=\mathcal L'_{\theta'}$.
\end{definition}

We will say that the family $\bar\varphi$ realizes the point-transformation mapping
of the class $\mathcal L|_{\mathcal S}$ onto the class $\mathcal L'|_{\mathcal S'}$, and
will identify the family with the mapping of classes and with the associated mapping of the sets of arbitrary elements.
Thus, the formula  $\mathrm{pr}_p\varphi_\theta\mathcal L_\theta=\mathcal L'_{\theta'}$ can be abbreviated as
$\bar\varphi\theta=\theta'$.

In the case of similar classes the family realizing the corresponding point-transformation mapping in fact consists of a unique element,
i.e., the transformation does not depend on the arbitrary elements.

A point-transformation image inherits certain transformational properties from its class-preimage.
There is also a converse connection.
For example, equations from the class-preimage are equivalent with respect to point transformations if and only if
their images are.

\begin{proposition}
A point-transformation mapping between classes of differential equations induces a mapping between
the corresponding sets of admissible transformations. Namely,
if the class $\mathcal L'|_{\mathcal S'}$ is the point-transformation image of
the class $\mathcal L|_{\mathcal S}$ under the family of point transformations
$\varphi_\theta\colon(x,u)\to(x',u')$, $\theta\in\mathcal{S}$, then
the image of $(\theta,\tilde\theta,\varphi)\in\mathrm{T}(\mathcal L|_{\mathcal S})$ is
$(\theta'\!,\tilde\theta'\!,\varphi')\in\mathrm{T}(\mathcal L'|_{\mathcal S'})$, where
$\mathcal L'_{\theta'}=\mathrm{pr}_p\varphi_\theta\mathcal L_\theta$,
$\mathcal L'_{\tilde\theta'}=\mathrm{pr}_p\varphi_{\tilde\theta}\mathcal L_\theta$ and
$\varphi'=\varphi_{\tilde\theta}\circ\varphi\circ(\varphi_\theta)^{-1}$.
\end{proposition}

Moreover, the similar statement in the opposite direction is also true.

\begin{proposition}
The set of admissible transformations of the initial class
$\mathcal L|_{\mathcal S}$ is reconstructed from
the one of its point-transformation image $\mathcal L'|_{\mathcal S'}$.
\end{proposition}

\begin{proof}
Suppose that the family of point transformations $\varphi_\theta\colon(x,u)\to(x',u')$, $\theta\in\mathcal{S}$,
maps the class $\mathcal L|_{\mathcal S}$ onto the class $\mathcal L'|_{\mathcal S'}$.
Let $(\theta'\!,\tilde\theta'\!,\varphi')\in\mathrm{T}(\mathcal L'|_{\mathcal S'})$ and let
$\mathcal L_\theta$ and $\mathcal L_{\tilde\theta}$ be some equations mapped
to $\mathcal L'_{\theta'}$ and $\smash{\mathcal L'_{\tilde\theta'}}$, respectively.
Then $(\theta,\tilde\theta,\varphi)\in\mathrm{T}(\mathcal L|_{\mathcal S})$, where
$\varphi=(\varphi_{\tilde\theta})^{-1}\circ\varphi\circ\varphi_\theta$.
Each admissible transformation of $\mathcal L|_{\mathcal S}$ is obtainable in the above way.
\end{proof}

If the class $\mathcal L'|_{\mathcal S'}$ is only a point-transformation image of
the class $\mathcal L|_{\mathcal S}$ in the sense of definition~\ref{DefOnImageClasses}
without the similarity in the sense of definition~\ref{DefOnSimilarClasses}
then the similarity of their classifications may be broken.
Only in the case of classifications with respect to the entire sets of admissible transformations there always
exists a one-to-one correspondence to hold between the classification lists for the class-image and the class-preimage.
For such a correspondence to hold in the case of classifications with respect to the equivalence groups,
we need additionally to require
that the image of each orbit of $G^{\sim}(\mathcal L|_{\mathcal S})$ in $\mathcal S$ coincides with
an orbit of $G^{\sim}(\mathcal L'|_{\mathcal S'})$ in $\mathcal S'$.
These facts can be applied for the simplification of solving group classification problems.
If one of the classes is classified in a simpler way, possessing, e.g., a set of arbitrary elements
(resp.\ equivalence group, resp.\ set of admissible transformations, etc.) of a simpler structure then
its group classification can be carried out first and can subsequently be used to derive the
classification of the other class.
It is an approach that is applied in the present paper for the group classification
of class~\eqref{eqRDfghPower}.

A specific kind of mappings between classes of differential equations
is given by mappings of classes to their subclasses.
If the class $\mathcal L|_{\mathcal S}$ is mapped onto its subclass $\mathcal L|_{\mathcal S'}$
by the transformation family $\bar\varphi=\{\varphi_\theta,\theta\in\mathcal{S}\}$ then
the tuple $(\theta,\bar\varphi\theta,\varphi_\theta)$ is an admissible transformation
in the class~$\mathcal L|_{\mathcal S}$.
In a particular case, the mapping~$\bar\varphi$ is associated with a subgroup~$H$
of the equivalence group~$G^\sim(\mathcal L|_{\mathcal S})$.
Namely, the mapping is constructed in the following way.
We choose a subclass $\mathcal L|_{\mathcal S'}$ of $\mathcal L|_{\mathcal S}$ in such a way
that each orbit of the action of~$H$ in $\mathcal S$ intersects $\mathcal S'$ in one element sharp.
Then we put $\bar\varphi\theta=\theta'$, where $\{\theta'\}=\mathcal S'\cap H\theta$, i.e.,
any element of an orbit is mapped to the element of the intersection of the orbit and $\mathcal S'$.
As a realization of the transformation~$\varphi_\theta$ we choose $h|_{(x,u)}^\theta$, where
$h$ is an element of~$H$ mapping~$\theta$ to~$\theta'$.
It is useful to identify $\varphi_\theta$ and~$h$.
The system of additional auxiliary conditions $S'=0$, $\Sigma'\ne0$ singled out
the subclass~$\mathcal L|_{\mathcal S'}$ from the class~$\mathcal L|_{\mathcal S}$
is called a \emph{gauge} of arbitrary elements, generated by the subgroup~$H$.
It is obvious that the preimages of each arbitrary element from $\mathcal{S}'$
with respect to the mapping~$\bar\varphi$ associated with~$H$ are
$G^{\sim}(\mathcal L|_{\mathcal S})$-equivalent.
The mapping~$\bar\varphi$ also establishes
a stronger connection between the generalized extended equivalence groups of the initial and imaged classes
under certain conditions on~$H$.

\begin{proposition}\label{PropositionOnEquivClassificationsUnderGauging}
Suppose that $H$ is a normal subgroup of the generalized extended equivalence group~$G^\sim(\mathcal L|_{\mathcal S})$
of the class~$\mathcal L|_{\mathcal S}$;
and each orbit of~$H$ in $\mathcal S$ intersects $\mathcal S'\subset\mathcal S$ in one element sharp.
Let $\bar\varphi$ be the mapping from~$\mathcal L|_{\mathcal S}$ to~$\mathcal L|_{\mathcal S'}$,
associated with~$H$ and
$G^\sim(\mathcal L|_{\mathcal S'})$ denote the generalized extended equivalence group
of the subclass~$\mathcal L|_{\mathcal S'}$.
Then equations from $\mathcal L|_{\mathcal S}$ are $G^{\sim}(\mathcal L|_{\mathcal S})$-equivalent
if and only if their images under the mapping $\bar\varphi$ are  $G^{\sim}(\mathcal L|_{\mathcal S'})$-equivalent.
\end{proposition}

\begin{proof}
We recall that if the mapping $\bar\varphi$ is associated with~$H$ then
the transformation $\varphi_\theta$ for each $\theta\in\mathcal S$ is interpreted as the element of~$H$
realizing the projection of~$\theta$ to~$\mathcal S'$.

For arbitrary $\theta\in\mathcal S$ and each $\Phi\in G^{\sim}(\mathcal L|_{\mathcal S})$
there exists $\Phi'\in G^{\sim}(\mathcal L|_{\mathcal S'})$ such that
$\Phi'\varphi_\theta\theta=\varphi_{\Phi\theta}\Phi\theta$.
Indeed, we can define $\Phi'$ by the formulas $\Phi'\theta'=\varphi_{\Phi\theta'}\Phi\theta'$ and
$\Phi'|_{(x,u)}^{\theta'}=\varphi_{\Phi\theta'}|_{(x,u)}^{\Phi\theta'}\circ\Phi|_{(x,u)}^{\theta'}$
for each $\theta'\in\mathcal S'$.
The transformation~$\Phi'$ is a point transformation with respect to $(x,u)$
for each $\theta'\in\mathcal S'$ as
a composition of point transformations $\varphi_{\Phi\theta'}|_{(x,u)}^{\Phi\theta'}$ and $\Phi|_{(x,u)}^{\theta'}$.
Since $H$ is a normal subgroup of~$G^\sim(\mathcal L|_{\mathcal S})$,
any transformation from~$G^\sim(\mathcal L|_{\mathcal S})$ maps each orbit of~$H$ onto an orbit of~$H$.
So, if $\theta',\tilde\theta'\in\mathcal S'$ and $\theta'\ne\tilde\theta'$ then
$\Phi\theta'$ and $\Phi\tilde\theta'$ belong to different orbits of~$H$.
$H\Phi\theta'\cap\mathcal S'=\{\Phi'\theta'\}$, $H\Phi\tilde\theta'\cap\mathcal S'=\{\Phi'\tilde\theta'\}$. 
Therefore, $\Phi'\theta'\ne\Phi'\tilde\theta'$, i.e., $\Phi'$ generates a one-to-one mapping on $\mathcal S'$.
This means that $\Phi'\in G^\sim(\mathcal L|_{\mathcal S'})$.
In other words, we have proved that images of $G^{\sim}(\mathcal L|_{\mathcal S})$-equivalent equations
with respect to the mapping $\bar\varphi$ are $G^{\sim}(\mathcal L|_{\mathcal S'})$-equivalent.

If $\theta_0,\tilde\theta_0\in\mathcal S$ and $\Phi'\varphi_{\theta_0}\theta_0=\smash{\varphi_{\tilde\theta_0}}\tilde\theta_0$
for some $\Phi'\in G^\sim(\mathcal L|_{\mathcal S'})$ then
there exists $\Phi\in G^\sim(\mathcal L|_{\mathcal S})$ such that $\Phi\theta_0=\tilde\theta_0$.
Indeed, we can consider the transformation $\hat\Phi$ defined by the formula
$\hat\Phi\theta=\varphi_{\theta}^{-1}\Phi'\varphi_{\theta}\theta$ for each $\theta\in\mathcal S$.
It generates a one-to-one mapping on $\mathcal S$, is a point transformation with respect to $(x,u)$
as a composition of point transformations and, therefore, belongs to $G^\sim(\mathcal L|_{\mathcal S})$.
Now $\hat\Phi\theta_0=\varphi_{\theta_0}^{-1}\smash{\varphi_{\tilde\theta_0}}\tilde\theta_0$, i.e.,
$\Phi\theta_0=\tilde\theta_0$ if we put $\Phi=(\smash{\varphi_{\tilde\theta_0}})^{-1}\varphi_{\theta_0}\hat\Phi$.
Therefore, we have proved that preimages of $G^{\sim}(\mathcal L|_{\mathcal S'})$-equivalent equations
with respect to the mapping $\bar\varphi$ are $G^{\sim}(\mathcal L|_{\mathcal S})$-equivalent.
This completes the proof of the proposition.
\end{proof}

This means that the group classification in class~$\mathcal L|_{\mathcal S}$
up to $G^{\sim}(\mathcal L|_{\mathcal S})$-equivalence is reduced
to the group classification in the subclass~$\mathcal L|_{\mathcal S'}$
with respect to its equivalence group~$G^\sim(\mathcal L|_{\mathcal S'})$.
Proposition~\ref{PropositionOnEquivClassificationsUnderGauging} extends to the case of
a gauge of arbitrary elements, generated by a subgroup which are not normal.

\begin{proposition}\label{PropositionOnEquivClassificationsUnderGaugingWithoutNormality}
Suppose that $\{H_\gamma,\gamma\in\Gamma\}$ is a family of subgroups
of the generalized extended equivalence group~$G^\sim(\mathcal L|_{\mathcal S})$ of the class~$\mathcal L|_{\mathcal S}$;
each transformation from~$G^\sim(\mathcal L|_{\mathcal S})$ induces a similarity relation on this family;
and for any $\gamma\in\Gamma$
each orbit of~$H_\gamma$ in $\mathcal S$ intersects $\mathcal S'\subset\mathcal S$ in one element sharp.
Let $\bar\varphi$ be the mapping from~$\mathcal L|_{\mathcal S}$ to~$\mathcal L|_{\mathcal S'}$,
associated with~$H_{\gamma_0}$ for a fixed value $\gamma_0\in\Gamma$ and
$G^\sim(\mathcal L|_{\mathcal S'})$ denotes the generalized extended equivalence group
of the subclass~$\mathcal L|_{\mathcal S'}$.
Then equations from $\mathcal L|_{\mathcal S}$ are $G^{\sim}(\mathcal L|_{\mathcal S})$-equivalent
if and only if their images under the mapping $\bar\varphi$ are  $G^{\sim}(\mathcal L|_{\mathcal S'})$-equivalent.
\end{proposition}

\begin{proof}
It is enough to appropriately modify only first part of the proof of proposition~\ref{PropositionOnEquivClassificationsUnderGauging}
on that images of $G^{\sim}(\mathcal L|_{\mathcal S})$-equivalent equations
with respect to the mapping $\bar\varphi$ are $G^{\sim}(\mathcal L|_{\mathcal S'})$-equivalent.
The difference is in the demonstration of that
the transformation $\Phi'$ defined in the same way generates a one-to-one mapping on $\mathcal S'$.
Here we use the rule of contraries.
Let $\theta',\tilde\theta'\in\mathcal S'$, $\theta'\ne\tilde\theta'$ and $\Phi'\theta'=\Phi'\tilde\theta'$.
The last equality means that $\Phi\tilde\theta'=h\Phi\theta'$ for some $h\in H_{\gamma_0}$.
Then $\tilde\theta'=\Phi^{-1}h\Phi\theta'=h_\Phi\theta'$,
where $h_\Phi=\Phi^{-1}h\Phi$ belongs to a subgroup $H_{\gamma_1}$ from the family $\{H_\gamma,\gamma\in\Gamma\}$,
i.e., $\theta'$ and $\tilde\theta'$ lie on the same orbit of $H_{\gamma_1}$ in $\mathcal S$.
In view of a proposition's condition, the orbit intersects $\mathcal S'$ in one element sharp.
Therefore, $\theta'=\tilde\theta'$ and we have a contradiction.
\end{proof}

\section{Equivalence groups and choice of classes for investigation}
\label{SectionOnEquivTransOfEqRDfghPower_n0}

The first step to solve a group classification problem is to derive the point transformations which
preserve the general form of equations from the class investigated and transform only arbitrary elements.
Such transformations are called {\it equivalence} transformations and form a group~\cite{Akhatov&Gazizov&Ibragimov1989,Ovsiannikov1982}.

The usual equivalence group~$G^{\sim}$ of
class~\eqref{eqRDfghPower} consists of the nondegenerate point
transformations in the space of~$(t,x,u,f,g,h,m)$, which possess the following additional properties.
Firstly, they are projectible on the space of~$(t,x,u)$, i.e., they have the form
\begin{gather*}
(\tilde t,\tilde x,\tilde u)=(T^t,T^x,T^u)(t,x,u), \\[0.5ex]
(\tilde f,\tilde g,\tilde h,\tilde m)=(T^f,T^g,T^h,T^m)(t,x,u,f,g,h,m).
\end{gather*}
Secondly, they transform any equation from class~\eqref{eqRDfghPower} for the
function $u=u(t,x)$ with the arbitrary elements $(f,g,h,m)$ to
an equation from the same class for the function $\tilde u=\tilde
u(\tilde t,\tilde x)$ with the new arbitrary elements~$(\tilde
f,\tilde g,\tilde h,\tilde m)$.

\begin{theorem}\label{equivfgh_usual}
$G^{\sim}$ consists of the transformations
\[\hspace{-.5\arraycolsep}
\begin{array}{l}
\tilde t=\delta_1 t+\delta_2,\quad \tilde x=\varphi(x), \quad
\tilde u=\delta_3 u, \\[1ex]
\tilde f=\dfrac{\delta_0\delta_1}{\delta_3\varphi_x} f, \quad
\tilde g=\dfrac{\delta_0\varphi_x}{\delta_3}\, g, \quad
\tilde h=\dfrac{\delta_0}{\delta_3^m\varphi_x} h, \quad \tilde m=m,
\end{array}
\]
where $\delta_j$, $j=0,\dots,3$,  are arbitrary constants, $\delta_0\delta_1\delta_3\not=0$,
$\varphi$ is an arbitrary smooth function of~$x$ with $\varphi_x\not=0$.
\end{theorem}

It appears that class~\eqref{eqRDfghPower} admits other
equivalence transformations which do not belong to~$G^{\sim}$ and
form, together with the usual equivalence transformations,
a {\it generalized extended equivalence group}. Restrictions on
transformations can be weakened in two directions. We admit that
transformations of the variables $t$, $x$ and $u$ can depend on
arbitrary elements (the prefix ``generalized''~\cite{Meleshko1994}), and this
dependence are not necessarily point and have to become point with
respect to $(t,x,u)$ after fixing values of arbitrary elements.
The explicit form of the new arbitrary elements~$(\tilde f,\tilde
g,\tilde h,\tilde m)$ is determined via
$(t,x,u,f,g,h,m)$ in some non-fixed (possibly, nonlocal) way
(the prefix ``extended''). We construct the complete (in this
sense) generalized extended equivalence group~$\hat G^{\sim}$ of
class~\eqref{eqRDfghPower}, using the direct method
\cite{Kingston&Sophocleous1998,Popovych&Ivanova2004NVCDCEs}.

\begin{theorem}\label{equivfgh}
The generalized extended equivalence group~$\hat G^{\sim}$ of
class~\eqref{eqRDfghPower} is formed by the transformations
\[\hspace{-.5\arraycolsep}
\begin{array}{l}
\tilde t=\delta_1 t+\delta_2,\quad \tilde x=\varphi(x), \quad
\tilde u=\psi(x) u, \\[1ex]
\tilde f=\dfrac{\delta_0\delta_1}{\varphi_x\psi^{2}} f, \quad
\tilde g=\dfrac{\delta_0\varphi_x}{\psi^{2}}\, g, \quad
\tilde h=\dfrac{\delta_0}{\varphi_x\psi^{m+1}} h, \quad
\tilde m=m,
\end{array}
\]
where $\varphi$ is an arbitrary smooth function of~$x$ with $\varphi_x\not=0$,
$\delta_j$, $j=0,1,2$, are arbitrary constants, $\delta_0\delta_1\not=0$, and
$\psi=\psi(x)$ is a (nonvanishing) solution of the second-order nonlinear ODE
\begin{equation}\label{Eq2ndOrderODEforPsi}
\left( \frac{g\psi_x}{\psi^2}\right)_x=0.
\end{equation}
\end{theorem}

The usual equivalence group~$G^{\sim}$ of class~\eqref{eqRDfghPower} is
the subgroup of the generalized extended equivalence group~$\hat G^{\sim}$,
which is singled out with the condition $\psi=\const$.

\begin{note}
Each solution of equation~\eqref{Eq2ndOrderODEforPsi} has the form
$\psi(x)=\bigl(\delta_3\int{dx}/{g(x)}+\delta_4\bigr)^{-1}$, where
$\delta_3$ and $\delta_4$ are constants, $(\delta_3,\delta_4)\ne(0,0)$,
and the integral denotes a fixed antiderivative of $1/g(x)$.
Generally speaking, the constants and the way of taking antiderivative in $\psi$ can depend on arbitrary elements.
Hence it seems better to work with the representation of $\psi$ as a solution of equation~\eqref{Eq2ndOrderODEforPsi}.
\end{note}

Theorems~\ref{equivfgh_usual} and~\ref{equivfgh} imply that the arbitrary element $m$
is invariant under the transformations from $\hat G^{\sim}$.
This allows us to partition class~\eqref{eqRDfghPower} in the subclasses
each of which corresponds to a fixed value of $m$.
The equivalence groups of the subclasses are {\it conditional equivalence groups}
for the whole class~\eqref{eqRDfghPower}.
Only the conditional equivalence group for the value $m=2$ is nontrivial.
For all the other values of~$m$ the corresponding conditional equivalence groups are
trivial since they coincide with the restrictions of $\hat G^{\sim}$ to such fixed values of~$m$.

\begin{theorem}\label{equivfgh_m2}
The class of equations
\begin{gather}\label{eqRDfghPower_m2}
f(x)u_t=(g(x)u_x)_x+h(x)u^2
\end{gather}
admits the generalized extended equivalence group~$\hat G^{\sim}_{m=2}$ consisting of the transformations
\begin{gather*}
\tilde t=\delta_1 t+\delta_2,\quad \tilde x=\varphi(x),\quad
\tilde u=\psi(x) u+\chi(x), \\
\tilde f=\dfrac{\delta_0\delta_1}{\varphi_x\psi^{2}}\,f, \quad 
\tilde g=\dfrac{\delta_0\varphi_x}{\psi^{2}}\,g, \quad 
\tilde h=\dfrac{\delta_0}{\varphi_x\psi^3}\,h,
\end{gather*}
where $\delta_j$, $j=0,1,2$, are arbitrary constants,
$\delta_0\delta_1\not=0$, 
$\varphi$ is an arbitrary smooth function of~$x$ with $\varphi_x\not=0$, %the function
$\psi=\psi(x)$ is a (nonvanishing) solution of the fourth-order nonlinear ODE
\begin{equation}\label{fourth-orderODE}
{\left[\frac g{\psi^2}{\left(\frac{\psi^2}{2h}{\left(
 \frac{g\psi_x}{\psi^2}\right)}_x\right)}_x\right]}_x=
 \frac {\psi}{4h}\left[\left( \frac{g\psi_x}{\psi^2}\right)_x\right]^2
 \end{equation}
 and $\chi=-\dfrac{\psi^2}{2h}\left(\dfrac{g\psi_x}{\psi^2}\right)_x.$
\end{theorem}

The group~$\hat G^{\sim}_{m=2}$ is the generalized extended conditional equivalence group of class~\eqref{eqRDfghPower}
under the condition~$m=2$.
It is really a nontrivial conditional equivalence group since
the equivalence group $\hat G^{\sim}$ of the whole class~\eqref{eqRDfghPower} restricted to the value~$m=2$
is obviously narrower than~$\hat G^{\sim}_{m=2}$,
$\hat G^{\sim}|^{}_{m=2}\varsubsetneq\hat G^{\sim}_{m=2}$.
(Every solution of equation~\eqref{Eq2ndOrderODEforPsi} is a particular solution of equation~\eqref{fourth-orderODE}.)
Note that $G^{\sim}|^{}_{m=2}=G^{\sim}_{m=2}$, i.e.,
the condition $m=2$ gives no extension for the usual equivalence group.

The presence of the arbitrary function $\varphi(x)$ in the equivalence
transformations from $G^{\sim}$ and $\hat G^{\sim}$ allows us to simplify the problem of group classification of
class~\eqref{eqRDfghPower} via reducing the number of arbitrary functions.
For example, the transformation from the group $G^{\sim}$
\[
\tilde t=t,\quad \tilde x=\int_{x_0}^x\frac{dy}{g(y)}+x_0,\quad\tilde u=u
\]
maps each equation from class~\eqref{eqRDfghPower} to the equation
$
\tilde f(\tilde x)\tilde u_{\tilde t}= \tilde u_{\tilde x\tilde x} + \tilde h(\tilde x)\tilde u^m
$
of the same form with the new arbitrary elements
$\tilde f(\tilde x)=f(x)g(x)$, $\tilde g(\tilde x)=1$ and $\tilde h(\tilde x)=g(x)h(x)$.
In other words, we put the {\it gauge} $g=1$ on the arbitrary elements of class~\eqref{eqRDfghPower}.

Generally speaking, every arbitrary functional element of class~\eqref{eqRDfghPower}
(i.e., $f$, $g$ or $h$) can be gauged to the unity or other chosen function by transformations from~$G^{\sim}$.
Despite the fact that the gauge $g=1$ seems most successful,
the problem of group classification under this gauge remains complicated.
In fact, the appropriate way of dealing with class~\eqref{eqRDfghPower} consists of two main steps.
Namely, they are the gauge of the arbitrary elements by equivalence transformations with another auxiliary condition
(which is quite nonobvious) and the subsequent mapping of the gauged subclass to a different class by a family
of point transformations parameterized with arbitrary elements.
The problem of group classification for the resulting class is solved much easier than
the similar problem for class~\eqref{eqRDfghPower}.

In the first step we put the gauge $f=g$ on the arbitrary elements.
In view of theorem 1, this gauge can be provided for each equation from class~\eqref{eqRDfghPower}
by the transformation
\begin{equation}\label{gauge_f=g}
t'=\sign(f(x)g(x))t,\quad
x'=\int_{x_0}^x\sqrt{\left|\frac{f(y)}{g(y)}\right|}\,dy+x_0, \quad
u'=u
\end{equation}
for some fixed value~$x_0$.
The arbitrary elements of corresponding gauged equation in the primed variables take the values
\[
f'(x')=g'(x')=\sign(g(x))\left|f(x)g(x)\right|^{\frac12}, \quad
h'(x')=\sqrt{\left|\dfrac{g(x)}{f(x)}\right|}\,h(x).
\]
For this reason we can restrict ourselves, without loss of generality, to
investigation of the class
\begin{equation}\label{class_f=g}
f(x)u_t= (f(x)u_{x})_{x}+h(x)u^m,
\end{equation}
since all results on Lie symmetries and solutions for this class can be extended to
class~\eqref{eqRDfghPower} with the use of transformation~\eqref{gauge_f=g}.
(See section~\ref{SectionOnMappingsAndGroupClassificationOfClassUnderConsideration} for detailed explanations.)

It is easy to deduce the generalized extended equivalence group for class~\eqref{class_f=g}
from theorem~\ref{equivfgh} and conditional one for the value $m=2$ from theorem~\ref{equivfgh_m2} by setting
$\tilde f=\tilde g$ and $f=g$.
The results are summarized in the following theorems.

\begin{theorem}\label{equivfh}
The generalized extended equivalence group~${\hat G_{f=g}}^{\sim}$ of
class~\eqref{class_f=g} consists of the transformations
\begin{gather*}
\tilde t=\delta_1{}^{\!2} t+\delta_2,\quad \tilde x=\delta_1x+\delta_3,\quad \tilde u=\psi(x) u, 
\quad
\tilde f=\dfrac{\delta_0\delta_1}{\psi^{2}} f, \quad \tilde h=\dfrac{\delta_0}{\delta_1\psi^{m+1}} h, \quad \tilde m=m,
\end{gather*}
where $\delta_j$, $j=0,\dots,3$, are arbitrary constants, $\delta_0\delta_1\ne0$,
and $\psi=\psi(x)$ is a (nonvanishing) solution of the second-order nonlinear ODE
\begin{equation}\label{Eq2ndOrderODEforPsiAfterGauging}
\left( \frac{f\psi_x}{\psi^2}\right)_x=0.
\end{equation}\end{theorem}

\begin{theorem}\label{equivfh_m2}
The class of equations
\begin{gather}\label{eqRDfhPower_m2}
f(x)u_t=(f(x)u_x)_x+h(x)u^2
\end{gather}
admits the generalized extended equivalence group~$\hat
G^{\sim}_{f=g,m=2}$ consisting of the transformations
\begin{gather*}
\tilde t={\delta_1}^2 t+\delta_2,\quad \tilde x=\delta_1x+\delta_3,\quad \tilde u=\psi(x) u+\chi(x), 
\quad
\tilde f=\dfrac{\delta_0\delta_1}{\psi^{2}} f, \quad \tilde h=\dfrac{\delta_0}{\delta_1\psi^3} h,
\end{gather*}
where $\delta_j$, $j=0,\dots,3$, are arbitrary constants,
$\delta_0\delta_1\not=0$. The function $\psi=\psi(x)$ is a (nonvanishing) solution of the
fourth-order nonlinear ODE
\begin{equation}\label{forth-orderODE2}
{\left[\frac f{\psi^2}{\left(\frac{\psi^2}{2h}{\left(
 \frac{f\psi_x}{\psi^2}\right)}_x\right)}_x\right]}_x=
 \frac {\psi}{4h}\left[\left( \frac{f\psi_x}{\psi^2}\right)_x\right]^2
 \end{equation}
 and $\chi=-\dfrac{\psi^2}{2h}\left(\dfrac{f\psi_x}{\psi^2}\right)_x.$
\end{theorem}

Similarly to the situation with class~\eqref{eqRDfghPower}, the conditional equivalence group
$\smash{\hat G^{\sim}_{f=g,m=2}}$ of class~\eqref{class_f=g} under the condition $m=2$
is wider than the restriction $\hat G^{\sim}_{f=g}|_{m=2}^{}$ of the equivalence group $\hat G^{\sim}_{f=g}$ to the value $m=2$.
The usage of equivalence transformations from theorems~\ref{equivfh} and~\ref{equivfh_m2} 
instead of usual equivalence transformations
allows us to essentially simplify the solution of the group classification problem for class~\eqref{class_f=g}.
Note that the usual equivalence groups of classes~\eqref{class_f=g} and~\eqref{eqRDfhPower_m2}
are obtained by means of putting $\psi=\const$ in $\hat G^{\sim}_{f=g}$ and $\smash{\hat G^{\sim}_{f=g,m=2}}$,
respectively. It is obvious that $G^{\sim}_{f=g}|_{m=2}^{}$ and $\smash{G^{\sim}_{f=g,m=2}}$ coincide.

The next step is to make the following change of the dependent variable in class~\eqref{class_f=g},
\begin{gather}\label{gauge}
v(t,x)=\sqrt{|f(x)|}u(t,x).
\end{gather}
As a result, the class of related equations of the form
\begin{equation}\label{class_vFH}
v_t=v_{xx}+H(x)v^m+F(x)v
\end{equation}
is obtained, where the new arbitrary elements $F$ and $H$ are expressed via the formulas
\begin{gather}\label{FH_formulas}
F(x)=-\dfrac {(\sqrt{|f(x)|})_{xx}}{\sqrt{|f(x)|}},\quad H(x)=\dfrac {h(x)\sign f(x)}{(\sqrt{|f(x)|})^{m+1}}.
\end{gather}
Since class~\eqref{class_vFH} is an image of class~\eqref{class_f=g} with respect to
the family of the transformations~\eqref{gauge} parameterized by the arbitrary element~$f$,
we will call them the {\it imaged class} and the {\it initial class}, respectively.

The family of transformations~\eqref{gauge} which is parameterized by the arbitrary element~$f$
generates a peculiar gauge of the arbitrary elements of class~\eqref{class_f=g}.
Namely, every fixed pair $(F,H)$ is an image of a multitude of pairs $(f,h)$.
Moreover, all results on Lie symmetries and exact solutions of class~\eqref{class_vFH}
can be extended to class~\eqref{class_f=g} by the inversion of transformation~\eqref{gauge}.

Using the direct method, we found the generalized extended equivalence groups
of the whole class~\eqref{class_vFH} and its subclass associated with the constraint~$m=2$.
An interpretation of these results is given in the next section.

\begin{theorem}\label{TheoremOnGsimFH}
The generalized extended equivalence group~$\hat G^{\sim}_{FH}$ of class~\eqref{class_vFH}
coincides with the usual equivalence group~$G^{\sim}_{FH}$ of the same class
and is formed by the transformations
\begin{gather*}
\tilde t={\delta_1}^2 t+\delta_2,\quad \tilde x=\delta_1 x+\delta_3, \quad \tilde v=\delta_4 v, 
\quad
\tilde F=\dfrac F{{\delta_1}^2},\quad \tilde H=\dfrac H{{\delta_1}^2{\delta_4}^{m-1}}, \quad \tilde m=m,
\end{gather*}
where $\delta_j$, $j=1,\dots,4$, are arbitrary constants, $\delta_1\delta_4\not=0$.
\end{theorem}

\begin{theorem}
The class of equations
\begin{gather}\label{class_vFH_m2}
v_t=v_{xx}+H(x)v^2+F(x)v
\end{gather}
admits the generalized extended equivalence group
$\hat G^{\sim}_{FH,m=2}$ consisting of the transformations
\begin{gather*}
\tilde t={\delta_1}^2 t+\delta_2,\quad \tilde x=\delta_1 x+\delta_3, \quad \tilde v=\delta_4 v+\chi(x), 
\quad
\tilde F=\dfrac F{{\delta_1}^2}-\dfrac{2H}{{\delta_1}^2{\delta_4}}\chi,\quad \tilde H=\dfrac H{{\delta_1}^2{\delta_4}},
\end{gather*}
where $\delta_j$, $j=1,\dots,4$, are arbitrary constants,
$\delta_1\delta_4\not=0$, $\chi=\chi(x)$ is a solution of the second-order nonlinear ODE
$\chi_{xx}=\delta_4^{-1}H\chi^2-F\chi$.
\end{theorem}

Analogously to classes~\eqref{eqRDfghPower} and~\eqref{class_f=g}, we have that
\[
G^{\sim}_{FH}\bigr|_{m=2}=\hat G^{\sim}_{FH}\bigr|_{m=2}=G^{\sim}_{FH,m=2}\varsubsetneq \hat G^{\sim}_{FH,m=2}.
\]

The sequential gauges finally reduce the group classification problem for class~\eqref{eqRDfghPower} with $m\neq2$
to the simpler group classification problem for class~\eqref{class_vFH}.

In the case $m=2$ an additional gauge via mapping is needed.
This gauge has to be chosen so that the residuary complicated equivalence transformations
containing the function $\psi$ will be mapped to the identical transformation
on the set of the new variables and arbitrary elements.
The gauge can be made with the family of the transformations
\begin{gather}\label{gauge_m=2}
w(t,x)=v(t,x)+\dfrac{F(x)}{2H(x)},
\end{gather}
which maps class~\eqref{class_vFH_m2} onto the class of equations having the form
\begin{equation}\label{class_GH}
w_t=w_{xx}+H(x)w^2+G(x),
\end{equation}
where \begin{gather}\label{FGH_equation}
G(x)=-\left(\dfrac {F(x)}{2H(x)}\right)_{xx}-\dfrac {F(x)^2}{4H(x)}.\end{gather}
We will call class~\eqref{class_GH} the {\it double-imaged class}.
Note that transformations~\eqref{gauge_m=2} are parameterized by
the two arbitrary elements~$F$ and~$H$.

\begin{theorem}
The generalized extended equivalence group of class~\eqref{class_GH}
coincides with the usual equivalence group~$G^{\sim}_{HG}$ of the same class
and is formed by the transformations
\begin{gather*}
\tilde t={\delta_1}^2 t+\delta_2,\quad \tilde x=\delta_1 x+\delta_3, \quad \tilde w=\delta_4 w, 
\quad
\tilde G=\dfrac {\delta_4G}{{\delta_1}^2},\quad \tilde H=\dfrac H{{\delta_1}^2\delta_4},
\end{gather*}
where $\delta_j$, $j=1,\dots,4$, are arbitrary constants,
$\delta_1\delta_4\not=0$.
\end{theorem}

The exhaustive group classifications for classes~\eqref{class_vFH} and~\eqref{class_GH} and then for
class~\eqref{class_f=g} are  carried out in section~\ref{SectionOnRDfghLieSymmetries_n0}.

\begin{note}
Due to physical sense of equation~\eqref{eqRDfghPower}, the function~$u$ should satisfy the condition~$u\ge0$.
In this case we have to demand for the multipliers of~$u$ to be positive in all transformations.
If we avoid positiveness of~$u$ then we have to use the modular of~$u$ as base of powers which are not determined for negative values of base.
The same statement is true for similar expressions in transformations and other places.
The necessary changes in formulas are obvious.
\end{note}

\section{Mappings between classes and group classification of class~\bf\eqref{eqRDfghPower}}
\label{SectionOnMappingsAndGroupClassificationOfClassUnderConsideration}

We interpret the results of section~\ref{SectionOnEquivTransOfEqRDfghPower_n0}
on the general class~\eqref{eqRDfghPower} with $m\ne0,1$ in the framework of mappings
between classes of differential equations, presented in section~\ref{SectionOnClassMappingsAndGroupClassification}.
The interpretation for the subclass~\eqref{eqRDfghPower_m2} corresponding to the singular value $m=2$
is analogous but more complicated.

At first we take the discrete subgroup of the generalized extended equivalence group~$\hat G^{\sim}$,
formed by the equivalence transformations with $\delta_1=\pm1$, $\delta_0=1$, $\delta_2=0$, $\psi=1$ and $\varphi=x$.
Using the alternating of the sign of~$t$, we can always reduce each equation from class~\eqref{eqRDfghPower}
to an equation from the same class, in which the arbitrary elements~$f$ and~$g$ (locally) have equal signs.
The corresponding mapping is given by the formula
\[
\tilde t=\sign(f(x)g(x))t,\quad \tilde x=x,\quad \tilde u=u,\quad
\tilde f=\sign(f(x)g(x))f,\quad \tilde g=g,\quad \tilde h=h,\quad \tilde m=m.
\]
The subclass of equations satisfying the gauge $\sign f=\sign g$ will be marked by~(\ref{eqRDfghPower}$'$).
Its generalized extended equivalence group~$\check G^{\sim}$ consists of the transformations from~$\hat G^{\sim}$ with $\delta_1>0$.
It is obvious that equations from class~\eqref{eqRDfghPower} are $\hat G^{\sim}$-equivalent if and only if
their images in class~(\ref{eqRDfghPower}$'$) are $\check G^{\sim}$-equivalent.

Consider the subgroup~$H_{x_0}$ of the group~$\check G^{\sim}$,
which is formed by the equivalence transformations, where $\delta_0=\delta_1=1$, $\delta_2=0$, $\psi=1$
and $\varphi$ runs through the set of smooth functions having positive derivatives and the same fixed point $x_0$.
Each transformation from~$\check G^{\sim}$ induces a similarity relation on the subgroup family $\{H_{x}\}$,
where $x$ runs through the set of its values.
This fact is not trivial in view of that transformations become point only after we fixed the arbitrary elements.
To prove it, we take an arbitrary equivalence transformation~$\Phi$ of the form adduced in theorem~\ref{equivfgh}
and an arbitrary transformation $\Omega$:
\begin{gather*}
\tilde t=t,\quad \tilde x=\omega(x), \quad \tilde u=u, \quad
\tilde f=\frac1{\omega_x} f, \quad
\tilde g=\omega_x g, \quad
\tilde h=\frac1{\omega_x} h, \quad
\tilde m=m
\end{gather*}
from~$H_{x_0}$.
Note that the transformation~$\bar\Phi=\Phi^{-1}$ which is the inverse of a transformation~$\Phi$ from $\check G^{\sim}$
has the same form with the value
\[
\bar\delta_0=\frac1{\delta_0},\quad
\bar\delta_1=\frac1{\delta_1},\quad
\bar\delta_2=-\frac{\delta_2}{\delta_1},\quad
\bar\varphi=\varphi^{-1},\quad
\bar\psi^{\tilde g}=\frac1{\psi^g\circ\varphi^{-1}}.
\]
Here $\varphi^{-1}$ is the inverse function of $\varphi$.
The superscripts of $\psi$'s denote the values of the arbitrary element~$g$ appearing
in the corresponding equations of form~\eqref{Eq2ndOrderODEforPsi}. Indeed,
\[
\biggl(\frac{\tilde g\bar\psi^{\tilde g}_{\bar x}}{(\bar\psi^{\tilde g})^2}\biggr)_{\bar x}
=-\frac{\delta_0}{\varphi_x}\left(\frac{\varphi_x}{(\psi^g)^2}g\frac{\psi^g_x}{\varphi_x}\right)_x
=-\frac{\delta_0}{\varphi_x}\left(\frac{g\psi^g_x}{(\psi^g)^2}\right)_x
=0.
\]

We show that $\Phi^{-1}\Omega\Phi\in H_{\bar\varphi(x_0)}$.
We mark the values transformed by $\Phi$, $\Omega\Phi$ and $\Phi^{-1}\Omega\Phi$
with the signs of tilde, hat and check, respectively.
\begin{gather*}
\check t=\bar\delta_1\hat t+\bar\delta_2=\frac{\tilde t-\delta_2}{\delta_1}=t,\quad
\check x=\bar\varphi(\hat x)=(\bar\varphi\circ\omega)(\tilde x)=(\bar\varphi\circ\omega\circ\varphi)(x),\\
\check u=\frac{\hat u}{\hat\zeta}=\frac{\tilde u}{\hat\zeta}=\frac\psi{\hat\zeta}u=u,\\
\check f
=\frac{\bar\delta_1\bar\delta_0}{\bar\varphi_{\hat x}}\hat\zeta^2\hat f
=\frac{\bar\delta_1\bar\delta_0}{\bar\varphi_{\tilde x}}\hat\zeta^2\tilde f
=\frac{\bar\delta_1\bar\delta_0}{\bar\varphi_{\tilde x}}\frac{\delta_1\delta_0}{\varphi_x}\frac{\hat\zeta^2}{\psi^2}f
=f,\\
\check g
=\bar\delta_0\bar\varphi_{\hat x}\hat\zeta^2\hat g
=\bar\delta_0\bar\varphi_{\tilde x}\hat\zeta^2\tilde g
=\bar\delta_0\delta_0\bar\varphi_{\tilde x}\varphi_x\frac{\hat\zeta^2}{\psi^2}g
=g,\\
\check h
=\frac{\bar\delta_0}{\bar\varphi_{\hat x}}\hat\zeta^{m+1}\hat h
=\frac{\bar\delta_0}{\bar\varphi_{\tilde x}}\hat\zeta^{m+1}\tilde h
=\frac{\bar\delta_0}{\bar\varphi_{\tilde x}}\frac{\delta_0}{\varphi_x}\frac{\hat\zeta^{m+1}}{\psi^{m+1}}h
=h,
\end{gather*}
where
\[
\hat\zeta=\frac1{\bar\psi^{\hat g}}=\frac1{\bar\psi^{\tilde g}\circ\omega^{-1}}=\psi^g\circ\bar\varphi\circ\omega^{-1},
\]
and, therefore, $\hat\zeta(\hat x)=\psi^g(x)$.
Note that $\bar\psi^{\hat g}$ is a solution of the equation~\eqref{Eq2ndOrderODEforPsi}
associated with the value $\hat g$ of the arbitrary element since
\[
\biggl(\frac{\hat g\bar\psi^{\hat g}_{\hat x}}{(\bar\psi^{\hat g})^2}\biggr)_{\hat x}
=\frac1{\omega_{\tilde x}}\biggl(
\omega_{\tilde x}\frac{\tilde g\bar\psi^{\tilde g}_{\tilde x}}{(\bar\psi^{\tilde g})^2}
\frac1{\omega_{\tilde x}}\biggr)_{\tilde x}
=0.
\]
The function $\bar\varphi\circ\omega\circ\varphi$ has $\bar\varphi(x_0)$ as a fixed point.
Hence, the above formulas mean that $\Phi^{-1}\Omega\Phi\in H_{\bar\varphi(x_0)}$.

For each equation from class~(\ref{eqRDfghPower}$'$)
there exists a unique transformation from~$H_{x_0}$,
which transforms this equation to an equation from class~\eqref{class_f=g},
i.e., with arbitrary elements constrained by the gauge~$\tilde f=\tilde g$.
The transformation corresponds to $\omega(x)=\int_{x_0}^x\sqrt{f(y)/g(y)}\,dy+x_0$.
Therefore, each  orbit of~$H_{x_0}$ in the sets of arbitrary elements of class~(\ref{eqRDfghPower}$'$)
intersects the sets of arbitrary elements of class~\eqref{class_f=g} in one element sharp.
This implies in view of proposition~\ref{PropositionOnEquivClassificationsUnderGaugingWithoutNormality}
that equations from class~(\ref{eqRDfghPower}$'$) are $\check G^{\sim}$-equivalent if and only if
their images in class~\eqref{class_f=g} are ${\hat G_{f=g}}^{\sim}$-equivalent.
The last statement also can be checked directly.

The family of equivalence transformations~\eqref{gauge_f=g} induces a nontrivial mapping
from $\check G^{\sim}$ onto
the generalized extended equivalence group~${\hat G_{f=g}}^{\sim}$ of class~\eqref{class_f=g}.
Indeed, the transformation presented in theorem~\ref{equivfgh} implies
the following transformation for the variables and arbitrary elements with prime,
depending on values of~$f$ and~$g$:
\begin{gather*}
\tilde t'=|\delta_1|t'+\sign(\delta_1fg)\delta_2,
\\
\tilde x'=\sign(\varphi_x)|\delta_1|^{\frac12}x'
+\sign(\varphi_x)|\delta_1|^{\frac12}\int_{\varphi^{-1}(x_0)}^{x_0}\sqrt{\left|\frac{f(y)}{g(y)}\right|}\,dy
+x_0-\sign(\varphi_x)|\delta_1|^{\frac12}x_0,
\\
\tilde u'=\psi^{f'} u', \qquad
\psi^{f'}(x')=\psi^g(x),
\\
\tilde f'=\sign(\varphi_x)\frac{|\delta_1|^{\frac12}}{\psi^2}f',
\quad
\tilde h'=\frac{\delta_0\sign(\varphi_x)}{|\delta_1|^{\frac12}\psi^{m+1}}h'.
\end{gather*}
The function~$\psi^{f'}$ satisfies equation~\eqref{Eq2ndOrderODEforPsiAfterGauging} associated with the value~$f'$
of the arbitrary element~$f$ since
\[
\left(\frac{f'\psi^{f'}_{x'}}{(\psi^{f'}_{x'})^2}\right)_{x'}
=\sign(g)\Bigl|\frac gf\Bigr|^{\frac12}\left(|fg|^{\frac12}\Bigl|\frac gf\Bigr|^{\frac12}\frac{\psi^g_x}{(\psi^g)^2}\right)_x
=\Bigl|\frac gf\Bigr|^{\frac12}\left(\frac{g\psi^g_x}{(\psi^g)^2}\right)_x
=0.
\]

In section~\ref{SectionOnEquivTransOfEqRDfghPower_n0}
class~\eqref{class_f=g} is further mapped onto class~\eqref{class_vFH} by means of the transformation
defined by formulas~\eqref{gauge} and~\eqref{FH_formulas}.
The set of preimages of each equation from class~\eqref{class_vFH} coincides with an orbit
of the subgroup~$H'$ of the group~${\hat G_{f=g}}^{\sim}$ in class~\eqref{class_f=g},
where the subgroup~$H'$ is formed by the transformation from~${\hat G_{f=g}}^{\sim}$
with $\delta_0=\delta_1=1$ and $\delta_2=\delta_3=0$ (see theorem~\ref{equivfh}).
Indeed, the conditions
\[
-\dfrac {(|f(x)|^{\frac12})_{xx}}{|f(x)|^{\frac12}}=-\dfrac {(|\tilde f(x)|^{\frac12})_{xx}}{|\tilde f(x)|^{\frac12}},\quad
\dfrac {h(x)}{|f(x)|^{\frac{m+1}2}}=\dfrac {\tilde h(x)}{|\tilde f(x)|^{\frac{m+1}2}}
\]
imply that $\tilde f=\zeta^2f$ and $\tilde h=\zeta^{m+1}h$, where $\zeta=\delta_4\int\!\frac{dx}{f(x)}+\delta_5$
for some constants $\delta_4$ and $\delta_5$.
Moreover, two equations from class~\eqref{class_f=g} are connected by a transformation from~${\hat G_{f=g}}^{\sim}$ if and only if
their images in class~\eqref{class_vFH} are connected by a transformation from~$G^{\sim}_{FH}$.
Let us recall that the generalized extended equivalence group of class~\eqref{class_vFH}
coincides with its usual equivalence group~$G^{\sim}_{FH}$.

The family of transformations~\eqref{gauge} induces a homomorphism of the equivalence group of class~\eqref{class_f=g}
onto the equivalence group of class~\eqref{class_vFH}.
Namely, the transformation presented in theorem~\ref{equivfh} maps in the transformation from theorem~\ref{TheoremOnGsimFH},
where the new $\delta_4$ equals $\sqrt{|\delta_0\delta_1|}\sign(\psi)$.
The kernel of the homomorphism coincides with the subgroup~$H'$
including, in some sense, the most complicated transformations from~$\hat G^{\sim}_{f=g}$,
which makes this equivalence group generalized and extended.
Hence the image of~$\hat G^{\sim}_{f=g}$ is the usual equivalence group~$G^{\sim}_{FH}$.
(It is not the case for the mapping of $\hat G^{\sim}_{f=g,m=2}$.)

As a result, we construct the chain of mappings
\[
\mbox{class~\eqref{eqRDfghPower}${}\to{}$class~(\ref{eqRDfghPower}$'$)${}\to{}$class~\eqref{class_f=g}${}\to{}$class~\eqref{class_vFH}}.
\]
Each element of the chain is a surjection and possesses the property that
equations from the corresponding initial class are equivalent
with respect to its generalized extended equivalence group if and only if
their images are equivalent with respect to the generalized
extended equivalence group of the corresponding imaged class.
Then the resulting mapping from class~\eqref{eqRDfghPower} onto class~\eqref{class_vFH}
possesses the same property as a composition of mappings which have it.

Summarizing the above interpretation of the results of section~\ref{SectionOnEquivTransOfEqRDfghPower_n0},
we formulate the following statement.

\begin{proposition}\label{PropositionOnEquivClassificationsUnderGaugingForClassEqRDfghPower}
The group classification in class~\eqref{eqRDfghPower} with respect to its
generalized extended equivalence group~$\hat G^{\sim}$ is equivalent to
the group classification in class~\eqref{class_vFH} with respect to
the usual equivalence group~$G^{\sim}_{FH}$ of this class.
A classification list for class~\eqref{eqRDfghPower} can be obtained from
a classification list for class~\eqref{class_vFH} by means of taking a single preimage for each
element of the latter list with respect to the resulting mapping
from class~\eqref{eqRDfghPower} onto class~\eqref{class_vFH}.
\end{proposition}

In the case $m=2$ the similar chain of mappings is longer:
\[
\mbox{class~\eqref{eqRDfghPower_m2}${}\to{}$class~(\ref{eqRDfghPower_m2}$'$)${}\to{}$class~\eqref{eqRDfhPower_m2}${}\to{}$class~\eqref{class_vFH_m2}%
${}\to{}$class~\eqref{class_GH}}.
\]
The proof of necessary properties of the mappings is also more difficult than in the general case
since, for example, equation~\eqref{fourth-orderODE} is more complicated than equation~\eqref{Eq2ndOrderODEforPsi}
and additional summands appear in expressions for equivalence transformations of dependent variables.
The final statement on a connection between the group classifications in classes~\eqref{eqRDfghPower_m2} and~\eqref{class_GH}
are formulated analogously to Proposition~\ref{PropositionOnEquivClassificationsUnderGaugingForClassEqRDfghPower}.

\begin{proposition}\label{PropositionOnEquivClassificationsUnderGaugingForClassEqRDfghPower_m2}
The group classification of class~\eqref{eqRDfghPower_m2} with respect to its
generalized extended equivalence group~$\hat G^{\sim}_{m=2}$ is equivalent to
the group classification of class~\eqref{class_GH} with respect to
the usual equivalence group~$G^{\sim}_{HG}$ of this class.
A classification list for class~\eqref{eqRDfghPower_m2} is constructed from
a classification list for class~\eqref{class_GH} via taking a single preimage for each
element of the latter list with respect to the resulting mapping
from class~\eqref{eqRDfghPower_m2} onto class~\eqref{class_GH}.
\end{proposition}

\section{Lie symmetries}\label{SectionOnRDfghLieSymmetries_n0}

It is shown in sections~\ref{SectionOnEquivTransOfEqRDfghPower_n0}
and~\ref{SectionOnMappingsAndGroupClassificationOfClassUnderConsideration} that the problem of group
classification for class~\eqref{class_f=g} has been reduced to
the similar but simpler problems for class~\eqref{class_vFH} if $m\neq2$ and for class~\eqref{class_GH} if $m=2$.
In the next two subsections we carry out the group classifications of classes~\eqref{class_vFH} and~\eqref{class_GH}.
In subsection~\ref{Subsection_group classification of initial class} the obtained results are used
to derive the group classification of class~\eqref{class_f=g}.

\subsection{Group classification of the imaged class}\label{Subsection_group classification of image class}

The group classification of class~\eqref{class_vFH}, where $m\ne0,1,2$, will be carried out within the framework of
the classical Lie approach~\cite{Olver1986,Ovsiannikov1982} up to the equivalence generated
by the equivalence group~$G^{\sim}_{FH}$ of this class.
We search for operators of the form $\Gamma=\tau(t,x,v)\partial_t+\xi(t,x,v)\partial_x+\eta(t,x,v)\partial_v$
which generate one-parameter groups of point symmetry transformations of equations from class~\eqref{class_vFH}.
It follows from the infinitesimal invariance criterion that
\[
\tau=\tau(t),\quad
\xi=\frac 12 \tau_t x+\sigma(t), \quad
\eta=\left(-\frac 18 \tau_{tt} x^2-\frac 12\sigma_t x+\zeta(t)\right)v,
\]
where $\sigma$ and $\zeta$ are arbitrary smooth functions of~$t$ and
\begin{gather*}
\Bigl(\frac 12\tau_tx+\sigma\Bigr) H_x=\Bigl(\frac {m-1}8\tau_{tt}x^2 +\frac {m-1}2\sigma_tx+(1-m)\zeta -\tau_t\Bigr)H,\\[.5ex]
\Bigl(\frac 12\tau_t x+\sigma\Bigr) F_x=-\tau_tF-\frac 18\tau_{ttt}x^2-\frac 12\sigma_{tt}x+\frac 14\tau_{tt}+\zeta_t.
\end{gather*}
The two last equations include both the residuary uncertainties in coefficients of the operator
and the arbitrary elements of the class under consideration.
We will call them the \emph{classifying equations} since
they enable us to derive the forms of $\tau$, $\sigma$ and $\zeta$
depending on values of $F$ and~$H$.
The split of the classifying equations with respect to the arbitrary elements $F$ and $H$
gives the conditions $\tau_t=0$, $\sigma=\zeta=0$.
Hence, the kernel of Lie invariance groups of class~\eqref{class_f=g} is associated with
the Lie algebra $A^{\rm ker}=\langle\partial_t\rangle$.
All possible $G^{\sim}_{FH}$-inequivalent values of the parameter-functions~$F$ and $H$
admitting extension of Lie symmetry are listed in table~\ref{TableLieSymHF} together with  
bases of the corresponding maximal Lie invariance algebras.

\begin{center}
\footnotesize\setcounter{tbn}{-1}\renewcommand{\arraystretch}{1.6}
\refstepcounter{table}\label{TableLieSymHF}
\textbf{Table~\thetable.}
The group classification of the class $v_t=v_{xx}+H(x)v^m+F(x)v$. $m\neq0,1$; $H(x)\neq0$.
\\[2ex]
\begin{tabular}{|c|c|c|l|}
\hline
N&$H(x)$&$F(x)$&\hfil Basis of $A^{\max}$ \\
\hline
\refstepcounter{tbn}\label{TableLieSymHF_ker}\thetbn&$\forall$&$\forall$&$\partial_t$\\
\hline
\refstepcounter{tbn}\label{TableLieSymHF_exp_qx_2op}\thetbn&
$\delta e^{qx}$&$a_1$&$\partial_t,\,\partial_x+\alpha v\partial_v$\\
\hline
\refstepcounter{tbn}\label{TableLieSymHF_exp_qx_3op}\thetbn&$\delta e^{qx}$&
$-\alpha^2$&$\partial_t,\,\partial_x+\alpha v\partial_v,$\\ &&&
$2t\partial_t+(x-2\alpha t)\partial_x+\bigl(\alpha(x-2\alpha t)+\frac2{1-m}\bigr)_{_{}}v\partial_v$\\
\hline
\refstepcounter{tbn}\label{TableLieSymHF_power_k}\thetbn&$\delta x^k$&$a_2 x^{-2}$&
$\partial_t,\,2t\partial_t+x\partial_x+\frac{k+2}{1-m}\,v\partial_v$\\
\hline
\refstepcounter{tbn}\label{TableLieSymHF_power_exp}\thetbn&$\delta x^ke^{px^2}$&
$-{\beta}^2x^2+\beta\frac {2k+5-m}{1-m}+a_2 x^{-2}$&$\partial_t,\,e^{4\beta t}\bigl[\partial_t+2\beta x \partial_x -
2\beta\bigl(\beta x^2-\frac{k+2}{1-m}\bigr)v \partial_v\bigr]_{_{}}$\\
\hline
\refstepcounter{tbn}\label{TableLieSymHF_exp_px2_2op}\thetbn&$\delta e^{px^2}$&$-{\beta}^2x^2+\beta a_3$&
$\partial_t,\,e^{2\beta t}[\partial_x-\beta xv\partial_v]$\\
\hline
\refstepcounter{tbn}\label{TableLieSymHF_exp_px2_3op}\thetbn&$\delta e^{px^2}$&
$-{\beta}^2x^2+\beta\frac {5-m}{1-m}$&
$\partial_t,\,e^{2\beta t}\bigl[\partial_x-\beta xv\partial_v\bigr],$\\&&&
$e^{4\beta t}\bigl[\partial_t+2\beta x \partial_x -
2\beta\bigl(\beta x^2-\frac{2}{1-m}\bigr)v \partial_v\bigr]_{_{}}$\\
\hline
\end{tabular}
\\[2ex]
\parbox{150mm}{Here $\alpha, \beta, \delta, k, p, q, a_1, a_2, a_3$ are constants satisfying the conditions:
$\alpha=\tfrac q{1-m}$, $\beta=\tfrac {2p}{m-1},$
$\delta=\pm1\bmod G^{\sim}_{FH}$,
$p\neq0,\quad a_1\neq-\alpha^2$, $k^2+a_2^2\neq0$, $q^2+a_1^2\neq0$;
$a_3\neq\frac{5-m}{1-m}$ and additionally $a_3\neq5$ if $m=2$.}
\end{center}

\begin{note}
In table~\ref{TableLieSymHF} the parameter~$m$ is assumed
constrained by the classification supposition $m\neq2$ for a while.
Further we show that a list of $\hat G^{\sim}_{FH,m=2}$-inequivalent cases of Lie symmetry extension
in class~\eqref{class_vFH_m2} ($m=2$) are also exhausted
by the values of $F$ and $G$ presented in table~\ref{TableLieSymHF}.
This is why a special restriction on the constant $a_3$ in the case $m=2$ appears under table~\ref{TableLieSymHF}.
\end{note}

\begin{note}\label{note_on_gauges}
Some constants from table~\ref{TableLieSymHF} can be additionally gauged
by transformations from the equivalence group~$G^{\sim}_{FH}$.
Thus, in cases~\ref{TableLieSymHF_exp_qx_2op} and~\ref{TableLieSymHF_exp_qx_3op} nonzero values of~$q$ 
is gauged to $1$ by the transformation $\tilde t=q^2t,\,\tilde x=qx,\,\tilde v=q^\frac2{1-m}v$.
A similar scale transformation in cases~\ref{TableLieSymHF_power_exp}--\ref{TableLieSymHG_exp_px2_3op}
makes $p=\pm1$.
Other possibilities also exist.
For example, any nonzero value of $a_1$ is mapped to $1$ or $-1$ depending on its sign by
the transformation $\tilde t=|a_1|t$, $\tilde x=|a_1|^{\frac12}x$, $\tilde v=|a_1|^\frac{1}{1-m}v$.
\end{note}

\subsection{Group classification of the double-imaged class}\label{Subsection_group classification of double-image class}

In the case $m=2$ the group classification is performed for the double-imaged class~\eqref{class_GH}
with respect to its equivalence group~$G^{\sim}_{HG}$
and then ported for classes~\eqref{class_vFH} and~\eqref{eqRDfghPower}. 
Suppose that $\Gamma=\tau(t,x,w)\partial_t+\xi(t,x,w)\partial_x+\eta(t,x,w)\partial_w$ 
be a Lie symmetry operator of an equation from class~\eqref{class_GH}.
Analogously to the previous subsection, the infinitesimal invariance condition implies
the expressions for the coefficients of $\Gamma$
\[
\tau=\tau(t),\quad
\xi=\frac 12 \tau_t x+\sigma(t),\quad
\eta=\left(-\frac 18 \tau_{tt} x^2-\frac 12\sigma_t x+\zeta(t)\right)w+\eta^0(t,x)
\]
and the classifying equations
\begin{gather*}
\left(\frac 12\tau_tx+\sigma\right) H_x=\left(\frac 18\tau_{tt}x^2+\frac 12\sigma_tx-\zeta -\tau_t\right)H,\\
2\eta^0H=-\frac 18\tau_{ttt}x^2-\frac 12\sigma_{tt}x+\zeta_t+\frac 14 \tau_{tt},\\
\left(\frac 12\tau_tx +\sigma\right)G_x=-\left(\frac 18 \tau_{tt}x^2+\frac 12
\sigma_tx-\zeta+\tau_t\right)G+\eta^0_t-\eta^0_{xx}.
\end{gather*}
The kernel of the maximal Lie invariance algebras of equations from class~\eqref{class_GH}
coincides with the one-dimensional algebra $\langle\partial_t\rangle$.
All possible $G^\sim_{HG}$-inequivalent cases of extension of the maximal Lie invariance algebras are exhausted by ones
adduced in table~\ref{TableLieSymHG}.

%\begin{table}
{\setcounter{tbn}{-1}\footnotesize
\renewcommand{\arraystretch}{1.6}
\begin{center}
\refstepcounter{table}\label{TableLieSymHG}\textbf{Table~\thetable.}
The group classification of the class $w_t=w_{xx}+H(x)w^2+G(x)$. $H(x)\neq0$.
\\[2ex]
\begin{tabular}{|c|c|c|l|}
\hline
N&$H(x)$&$G(x)$&\hfil Basis of $A^{\max}$ \\
\hline
\refstepcounter{tbn}\label{TableLieSymHG_ker}\thetbn&$\forall$&$\forall$&$\partial_t$\\
\hline
\refstepcounter{tbn}\label{TableLieSymHG_exp_qx_2op}\thetbn&$\delta e^{qx}$&$b_1e^{-qx}$&$\partial_t,\,
\partial_x-qw\partial_w$\\
\hline
\refstepcounter{tbn}\label{TableLieSymHG_exp_qx_3op}\thetbn&$\delta e^{qx}$&$\frac{q^4}{4\delta}
e^{-qx}$&$\partial_t,\,
\partial_x-qw\partial_w,\,$\\
&&&$2t\partial_t+(x+2qt)\partial_x-
\bigl((qx+2q^2t+2)w+\frac{q^2}{\delta}e^{-qx}\bigr)\partial_w$\\
\hline
\refstepcounter{tbn}\label{TableLieSymHG_power_k}\thetbn&$\delta x^k$&$\frac{b_2}{\delta }x^{-k-4}$&
$\partial_t,\,2t\partial_t+x\partial_x-(k+2)w\partial_w$\\
\hline
\refstepcounter{tbn}\label{TableLieSymHG_power_exp}\thetbn&$\delta x^k e^{px^2}$&
$\frac1\delta x^{-k-4}e^{-px^2}P(x)$
&$\partial_t,\,e^{8pt}\bigl[\partial_t+4px \partial_x -$\\
&&&$ 4p\bigl((2px^2+k+2)w+2\frac p\delta(4px^2+2k+3)x^{-k}e^{-px^2}\bigr)\partial_w\bigr]$\\
\hline
\refstepcounter{tbn}\label{TableLieSymHG_exp_px2_2op}\thetbn&$\delta e^{px^2}$&
$\frac{p^2}\delta(4p^2x^4-20px^2+b_3)e^{-px^2}$&
$\partial_t,\,e^{4pt}\bigl[\partial_x-2px(w+2\frac p\delta e^{-px^2})\partial_w\bigr],$\\
\hline
\refstepcounter{tbn}\label{TableLieSymHG_exp_px2_3op}\thetbn&$\delta e^{px^2}$&
$\frac{p^2}\delta(4p^2x^4-20px^2-11)e^{-px^2}$&
$\partial_t,\,e^{4pt}\bigl[\partial_x-2px(w+2\frac p\delta e^{-px^2})\partial_w\bigr],$\\
&&&$e^{8pt}\bigl[\partial_t+4px\partial_x-8p\bigl((px^2+1)w+\frac p\delta(4px^2+3)e^{-px^2}\bigr)\partial_w\bigr]$\\
\hline
\end{tabular}
\\[2ex]
\parbox{150mm}{$\delta=\pm1\bmod G^\sim_{HG}$, $b_1\neq\frac{q^4}{4\delta}$, $(k,b_2)\neq(0,0)$, $p\neq0$, $b_3\neq-11$, 
$q$ is arbitrary constant.\\[1ex]
$P(x)=p^2(2px^2+1)(2px^2-11)x^4+8kp^3x^6+2k(3k-5)p^2x^4+k(k+1)(2k+3)px^2+b_2.$ }
\end{center}

}
%\end{table}

In order to derive the group classification of class~\eqref{class_vFH_m2}, we have to find preimages,
with respect to transformation~\eqref{gauge_m=2}, of each equation from class~\eqref{class_GH}
with arbitrary elements $(H,G)$ adduced in table~\ref{TableLieSymHG} and the
corresponding basis operators of the maximal Lie invariance algebras.
Every value of the arbitrary element $G$ is the image of values of the arbitrary element $F$ running through
the two-parametric general solution of equation~\eqref{FGH_equation}.
Each two equations from class~\eqref{class_vFH_m2}, associated with the pairs $(F^1,H)$ and $(F^2,H)$,
where $F^1$ and $F^2$ are particular solutions of~\eqref{FGH_equation} for the same value of~$G$,
are equivalent with respect to a transformation from the group~$\hat G^{\sim}_{FH,m=2}$.

Constructing particular solutions of ODE~\eqref{FGH_equation} for each pair $(G,H)$ from table~\ref{TableLieSymHG},
we obtain the pairs $(F,H)$ of table~\ref{TableLieSymHF}, where $m=2$. 
The corresponding pairs $(G,H)$ and $(F,H)$ have the same numbers. 
The parameters appearing in these tables are connected by the formulas
\begin{gather*}
a_1=-q^2\pm\sqrt{q^4-4\delta b_1}, \\ 
a_2=-(k+2)(k+3)\pm\sqrt{(k+2)^2(k+3)^2-4b_2}, \\ 
a_3=1\pm\sqrt{5-b_3}.
\end{gather*}
Therefore, the group classification of class~\eqref{class_vFH_m2} with respect to its equivalence group
$\hat G^{\sim}_{FH,m=2}$ is presented by the cases of table~\ref{TableLieSymHF}
after the substitution $m=2$ where it is necessary.

\subsection{Group classification of the initial class}\label{Subsection_group classification of initial class}

Due to family \eqref{gauge} of nondegenerate point transformations mapping class~\eqref{class_f=g} onto class~\eqref{class_vFH},
basis elements of Lie invariance algebras of equations from~\eqref{class_f=g}
can be found from basis elements of Lie invariance algebras of corresponding equations from~\eqref{class_vFH}
by the formula
\begin{gather}\label{tr_op}
\tilde Q=\tau\partial_t+\xi\partial_x+\biggl(\frac{\eta}{\sqrt{|f|}}-\frac{\xi f_x}{2f}u\biggr)\partial_u.
\end{gather}
Here $\tau$, $\xi$ and $\eta$ are coefficients of $\partial_t$, $\partial_x$ and $\partial_v$, respectively,
in the operators from table~\ref{TableLieSymHF}.
The substitution $v=\sqrt{|f|}\,u$ is assumed.

Transformation~\eqref{gauge} is not one-to-one since the preimage set of each equation from class~\eqref{class_vFH}
is a two-parametric family of $\smash{{\hat G_{f=g}}^{\sim}}$-equivalent equations from class~\eqref{class_f=g}.
To solve the problem of group classification for class~\eqref{class_f=g} with respect to transformations
from group~$\hat G^{\sim}_{f=g}$, it is enough to find a single preimage for each equation from class~\eqref{class_vFH}
with values of arbitrary elements listed in table~\ref{TableLieSymHF}.
In other words, for each pair $(F,H)$ from table~\ref{TableLieSymHF}
one should find a pair of functions $(f,h)$ satisfying conditions~\eqref{FH_formulas}.
This problem is solved in two steps. At first, the second-order nonlinear ODE
\begin{equation}\label{eqfF}
\bigl(\sqrt{|f|}\,\bigr)_{xx}+F\sqrt{|f|}=0
\end{equation}
should be integrated with respect to~$f$ for each value of~$F$ from table~\ref{TableLieSymHF}.
Then the corresponding value of~$h$ is easily found from the second condition of~\eqref{FH_formulas}: $h=(\sqrt{|f|})^{m+1}H$.

The general solutions of equation~\eqref{eqfF} can be constructed for all the cases of table~\ref{TableLieSymHF} 
in terms of elementary (cases 1--3) or Whittaker functions (cases 4--6) that gives two-parametric families of $(f,h)$.
As mentioned above, for the complete group classification of class~\eqref{class_f=g}
with respect to its generalized extended equivalence group~$\hat G^{\sim}_{f=g}$
it is enough to take single (simplest) representatives from these families.
The integration for cases 1--3 depends on values of parameters. 
Thus, case~1 is split into cases 1.1 ($a_1=0$), 1.2 ($a_1>0$) and 1.3 ($a_1<0$) of table~\ref{TableLieSym_initial_class}.
Case~2 is split into cases 2.1 ($q=0$) and 1.2 ($q\ne0$). 
Under solving equation~\eqref{eqfF} in case~3, the set of values of the parameter~$a_2$ is partitioned into 
four different subsets ($a_2=0$; $a_2<\frac 14$ and $a_2\ne0$; $a_2=\frac 14$; $a_2>\frac 14$). 
At the same time, the integration results for the three first subsets can be united, up to transformations 
from~$\smash{\hat G^{\sim}_{f=g}}$, to a single case associated with all values $a_2\leqslant\frac 14$. 
Therefore, we finally obtain two cases (3.1 and 3.2) of table~\ref{TableLieSym_initial_class}. 

To complete the group classification of class~\eqref{class_f=g} with respect to its generalized extended equivalence
group $\hat G^{\sim}_{f=g}$ (resp.~$\hat G^{\sim}_{f=g,m=2}$ for $m=2$), for each derived families of pairs $(f,h)$
we have to choose the simplest pair and construct the maximal Lie invariance algebra of the associated equation from class~\eqref{class_f=g}
using formula~\eqref{tr_op}. 
The final results are collected in table~\ref{TableLieSym_initial_class},
where the first number of each case indicates the corresponding case of table~\ref{TableLieSymHF}.

\smallskip
%\begin{table}
{\footnotesize
\renewcommand{\arraystretch}{1.7}
\begin{center}
\refstepcounter{table}\label{TableLieSym_initial_class}\textbf{Table~\thetable.}
The group classification of the class $f(x)u_t=(f(x)u_x)_x+h(x)u^m$, $f(x)h(x)\neq0$.
\\[2ex]
\begin{tabular}{|c|c|c|l|}
\hline
N&$f(x)$&$h(x)$&\hfil Basis of $A^{\rm max}$ \\
\hline
0&$\forall$&$\forall$&$\partial_t$\\
\hline
1.1&$1$&$\delta e^x$&$\partial_t,\,(1-m)\partial_x+ u\partial_u$\\
\hline
1.2&$(\cos x)^2$&$\delta e^{qx}|\cos x|^{m+1}$&$\partial_t,\,\partial_x+\left(\alpha+\tan x\right)u\partial_u$\\
\hline
1.3&$e^x$&$\delta e^{rx}$&$\partial_t,\,\partial_x+\frac{r-1}{1-m}u\partial_u$\\
\hline
2.1&$1$&$\delta$&$\partial_t,\,\partial_x,\, 2t\partial_t+x\partial_x+\frac2{1-m}u\,\partial_u$\\
\hline
2.2&$e^x$&$\delta e^x$&$\partial_t,\,\partial_x,\,
2t\partial_t+(x-t)\partial_x+\frac2{1-m}u\partial_u$\\
\hline
3.1&$x^\lambda$&$\delta x^\gamma$&
 $\partial_t,\,2t\partial_t+x\partial_x+\frac{2-\lambda+\gamma}{1-m}u\partial_u$\\
\hline
3.2&$x(\cos\ln|x|^\rho)^2$&$\delta x^l|\cos\ln|x|^\rho|^{m+1}$&
$\partial_t,\,2t\partial_t+x\partial_x+\left(\rho\tan\ln|x|^\rho+\frac{l+1}{1-m}\right)_{_{}}u\partial_u$\\
\hline
4&$x{}^{-1} f_1(x)^2$&$\delta x^s e^{px^2}|f_1(x)|^{m+1}$&$\partial_t,\,
e^{4\beta t}\Bigl[\partial_t+2\beta x \partial_x-2\beta\bigl(\beta x^2-2\kappa_1+xg_1(x)\bigr)u\partial_u\Bigr]_{_{}}$\\
%4&$x{}^{-1} f_1(x)^2$&$\delta x^s e^{px^2}f_1(x)^{m+1}$&$\partial_t,\,
%e^{4\beta t}\Bigl[\partial_t+2\beta x \partial_x-$\\&&&
%$2\beta\Bigl(2\beta x^2-4\kappa_1+\left(2\mu_1+\frac{s-m+4}{1-m}\right)g_1(x)\Bigr)u\partial_u\Bigr]_{_{}}$\\
\hline
5&$x{}^{-1} f_2(x)^2$&$\delta{x}^{-\frac{m+1}2} e^{px^2}|f_2(x)|^{m+1}$&
$\partial_t,\,
e^{2\beta t}\Bigl[\partial_x-\bigl(4\beta x^2-1-a_3+(a_3+3)g_2(x)\bigr)\dfrac u{2x}\partial_u\Bigr]_{_{}}$\\
\hline
6&$x{}^{-1} f_3(x)^2$&$\delta x^{\,-\frac{m+1}2} e^{px^2}|f_3(x)|^{m+1}$
&$\partial_t,\,
e^{2\beta t}\Bigl[\partial_x-\left(2\beta x^2+\frac{m-3}{1-m}+2\frac{2-m}{1-m}g_3(x)\right)\dfrac ux\partial_u\Bigr],$\\
&&&$e^{4\beta t}\Bigl[\partial_t+2\beta x \partial_x-2\beta\left(2\beta x^2-4\kappa_3+2\frac{2-m}{1-m}g_3(x)\right)u\partial_u\Bigr]_{_{}}$\\
\hline
\end{tabular}
\\[2ex]
\parbox{150mm}{
$\alpha=\frac q{1-m},\,\beta=\frac {2p}{m-1},\,\delta=\pm1,\,p\neq0, \rho\neq0$; $q,\,s$ are arbitrary constants.
$r\neq1,m$. %In case 1.1 $q\neq0$.

\smallskip
\noindent
In case 3.1 $(\lambda,\gamma)\neq\{(0,0),\,(2,m+1)\}$ if $m\not=2$ and
$(\lambda,\gamma)\neq\{(-6,-9),\,(0,0),\,(2,3),\,(8,12)\}$ if $m=2$.

\smallskip

\noindent
$f_1(x)=W(\beta x^2)$, 
where $W$ is a real solution of the Whittaker equation \smallskip $4y^2W''(y)=(y^2\!-4\kappa_1y\!+4\mu_1^2\!-\!1)W(y)$
with $\kappa_1=\frac{s+3}{2(1-m)}$ and $\mu_1=\frac{\sqrt{1-4a_2}}4$. $g_1=f_{1,x}/f_1$. \smallskip 
$f_i(x)=M_{\kappa_i,\mu_i}(\beta x^2)$, $g_i(x)=M_{\kappa_i+1,\mu_i}(\beta x^2)/f_i(x)$, $i=2,3$, 
where $M_{\kappa,\mu}$ is the Whittaker function~\cite{WhittakerWatson}.
$\kappa_2=\frac{a_3}{4}$, $\kappa_3=\frac{5-m}{4(1-m)}$, $\mu_2=\mu_3=\frac14$.

\smallskip
\noindent
In case 4 $s\neq-\frac{m+1}2$ if $a_2=0$ (i.e., if $\mu_1=\frac14$).
In case 5 $a_3\neq\frac{5-m}{1-m}$ and additionally $a_3\neq5$ if $m=2$.}
\end{center}
%\end{table}
}

\begin{note}
The equation from class~\eqref{class_f=g}
with the power coefficients $(f,h)=(x^{\lambda_1},\delta x^{\gamma_1})$
is equivalent to the one with $(\tilde f,\tilde h)=(x^{\lambda_2}, \delta x^{\gamma_2})$
with respect to the transformation
$\tilde t=t$, $\tilde x=x$, $\tilde u =x^{\frac{\lambda_1-\lambda_2}2}u$
which belongs to the group $\hat G^{\sim}_{f=g}$
if and only if $\lambda_2=2-\lambda_1$, $\gamma_2=\gamma_1+(m+1)(1-\lambda_1)$,
$(\lambda_1,\lambda_2)\neq(1,1)$. 
In the case $m=2$ the equations with power coefficients are equivalent if and only if $\lambda_2=2-\lambda_1$ or
$\lambda_2=1\pm\sqrt{49+17\lambda_1{}^{\!2}-24\gamma_1\lambda_1-58\lambda_1+40\gamma_1+8\gamma_1{}^{\!2}}$
and $\gamma_2=\gamma_1+\frac32(\lambda_2-\lambda_1)$.
Using these formulas, we can easily check that equations with $(f,h)=(x^{\lambda},\delta x^{\gamma})$ (case 3.1), 
where $(\lambda,\gamma)=(2,m+1)$ for $m\not=2$ and
$(\lambda,\gamma)=\{(-6,-9),\,(2,3),\,(8,12)\}$ for $m=2$, 
are equivalent to the equation of the same form with $(\lambda,\gamma)=(0,0)$ (case 2.1).
This fact explains the presence of restrictions on values of $\lambda$ and $\gamma$, 
listed below the table~\ref{TableLieSym_initial_class}.
\end{note}

\begin{note}
Whittaker functions are expressed for some values of their parameters via elementary functions.
See, e.g.,~\cite{WhittakerWatson}.
In particular, using the formula $M_{\kappa,-\kappa-\frac12}(z)=e^\frac z 2z^{-\kappa}$,
we can derive conditions on parameters for which
values of arbitrary elements and coefficients of Lie symmetry operators 
in cases 4--6 of table~\ref{TableLieSym_initial_class} are expressed in terms of elementary functions.
Below we list conditions  with corresponding arbitrary elements and Lie symmetry algebras.

In the case 4 with $s=m-4-2(1-m)\mu_1$, where $\mu_1=\frac14\sqrt{|1-4a_2|}$
and, therefore, $\kappa_1=-\mu_1-\frac12$,
we have in terms of $\kappa_1$ that
$(f,h)\sim(x^{-1-4\kappa_1}e^{\beta x^2},\,\delta x^{-3-4m\kappa_1}e^{\beta m x^2})$ and 
\[
A^{\max}=\bigl\langle\partial_t,\,
e^{4\beta t}[\partial_t+2\beta x\partial_x-2\beta(2\beta x^2-4\kappa_1)u\partial_u]\bigr\rangle.
\]
In the case 5 with $a_3=-3$ and, therefore, $m\neq2$ we obtain
$(f,h)\sim(x^2e^{\beta x^2},\,\delta x^{m+1}e^{\beta m x^2})$ and 
\[
A^{\max}=\bigl\langle\partial_t,\,e^{2\beta t}[\partial_x-(2\beta x^2+1)x^{-1}u\partial_u]\bigr\rangle.
\]
In the case 6 with $m=2$ we have 
$(f,h)\sim(x^2e^{2p x^2},\,\delta x^{3}e^{4p x^2})$ and 
\[
A^{\max}=\bigl\langle\partial_t,\,e^{4pt}[\partial_x-(4px^2+1)x^{-1}u\partial_u],\,
e^{8pt}[\partial_t+4px\partial_x -4p(4px^2+3)u\partial_u]\bigr\rangle.
\]
\end{note}

\section{Additional equivalence transformations}\label{SectionOnAdditionalTransformations_n0}

Taking advantage of generalized extended equivalence groups, we have carried out the group classification
of equations from classes~\eqref{class_f=g},~\eqref{class_vFH} and, therefore, from class~\eqref{eqRDfghPower}.
It happens that there exist point transformations between inequivalent, with respect to the corresponding equivalence
groups, cases of symmetry extension. 
Such transformations are called \emph{additional equivalence transformations} (see~\cite{VJPS2007} for details).
They simplify further application of the group classification results.

The independent pairs of point-equivalent cases from table~\ref{TableLieSymHF} and the corresponding transformations
are exhausted by the following:

\medskip
\noindent
$1\mapsto{\tilde1}|_{\smash{\tilde q=0,\,\tilde a_1=a_1+\alpha^2}},\quad 2\mapsto\tilde2|_{\tilde q=0}\colon$
\begin{equation}\label{addit_tr_1}
\tilde t=t,\quad\tilde x=x+2\alpha t,\quad \tilde v=e^{-\alpha x}v;
\end{equation}

\noindent
$4\mapsto\tilde3\colon$\vspace{-1ex}
\begin{equation}\label{addit_tr_2}
\tilde t=-\frac 1{4\beta}e^{-4\beta t},\quad
\tilde x=e^{-2\beta t}x ,\quad
\tilde v=\exp{\left(\frac{\beta}{2}\,x^2+2\beta\frac{k+2}{m-1}\,t\right)}v;
\end{equation}

\noindent
$6\mapsto\tilde2|_{\tilde q=0}\colon$
the transformation~\eqref{addit_tr_2} with $k=0$.

\medskip

In order to find additional equivalence transformations in the initial class, it will be convenient to use 
also another additional equivalence transformation in the imaged class for case~1:

\medskip
\noindent
$1|_{\smash{q^2+a_1(m-1)^2\geqslant0}}\mapsto{\tilde1}|_{\smash{\tilde q=\sqrt{q^2+a_1(m-1)^2},\,\tilde a_1=0}}\colon\quad
\tilde t=t,\quad\tilde x=x+2\sigma t,\quad \tilde v=e^{-\sigma x-(a_1+\sigma^2)t}v.$

\medskip

Hereafter the notations of section~\ref{SectionOnRDfghLieSymmetries_n0} are used:
$\alpha=\frac q{1-m}$, $\beta=\frac{2p}{m-1}$ etc. 
Additionally $\sigma=\frac{q-\tilde q}{1-m}$.

Note that transformation~\eqref{addit_tr_2} with $k=0$ maps the equation~\eqref{class_vFH}
with $F=-{\beta}^2x^2+\beta a_3$ and $H=\delta e^{px^2}$
(case~\ref{TableLieSymHF_exp_px2_2op} of table~\ref{TableLieSymHF})
to the equation
$\tilde v_{\tilde t}=\tilde v_{\tilde x\tilde x}+\delta{\tilde v}^m
-\frac1{4\tilde t}\left(\frac{5-m}{m-1}+a_3\right)\tilde v$. % having a single time-dependent coefficient.

\medskip
The additional equivalence transformations are derived also for the double-imaged class~\eqref{class_GH}:

%\medskip
\noindent
$1\mapsto{\tilde1}|_{\smash{\tilde q=0,\,\tilde b_1=b_1-\frac{q^4}{4\delta}}},\quad 
2\mapsto\tilde2|_{\tilde q=0}\colon\quad
\tilde t=t,\quad\tilde x=x-2qt,\quad\tilde w=e^{q x}w+\dfrac{q^2}{2\delta}$;

%\medskip
\noindent
$4\mapsto\tilde3\colon\quad
\tilde t=-\dfrac 1{8p}e^{-8p t},\quad
\tilde x=e^{-4p t}x ,\quad
\tilde w=e^{4p(k+2)t}\left(e^{p x^2}w+p\dfrac{2px^2+2k+3}{\delta x^k}\right);$

%\medskip
\noindent
$6\mapsto\tilde2|_{\tilde q=0}\colon$\quad the previous transformation with $k=0$.

\medskip

As a result, we obtain the following theorem.

\begin{theorem}\label{TheoremOnClassification_upto_PointTrans_GH_and_VFH}
Up to point transformations, a complete list of extensions of the maximal Lie invariance algebras
of equations from class~\eqref{class_vFH} (resp. class~\eqref{class_GH})
is exhausted by cases $0$, $1|_{q=0}$, $2|_{q=0}$, $3$, and $5$ of table~\ref{TableLieSymHF}
(resp. table~\ref{TableLieSymHG}).
\end{theorem}

The problem of finding additional equivalence transformations for class~\eqref{class_f=g}
and, moreover, the problem of description of all admissible transformations
in this class are much more complicated than the similar problems for the imaged classes.
The optimal way for constructing additional equivalence transformations in the initial class~\eqref{class_f=g} is
to take ``preimages'' of additional equivalence transformations of the imaged classes.

\medskip
\noindent
$1.2\mapsto\tilde{1.1}\colon\quad
\tilde t=\tilde q^2t,\quad\tilde x=\tilde q(x+2\sigma t),\quad
\tilde u=\tilde q^\frac2{1-m}e^{-\sigma x-(1+\sigma^2)t}(\cos x)u,\quad a_1:=1;$

\medskip
\noindent
$1.3|_{4\alpha^2\geqslant1}\mapsto\tilde{1.1}\colon\quad
\tilde t=\tilde q^2t,\quad\tilde x=\tilde q(x+2\sigma t),\quad
\tilde u=\tilde q^\frac2{1-m}e^{(\frac 12-\sigma)x+(\frac 14-\sigma^2)t}u,\quad a_1:=-\frac14;$

\medskip
\noindent
$1.3|_{4\alpha^2<1}\mapsto\tilde{1.3}|_{2\tilde r=m+1}\colon\quad
\tilde t=\nu^2t,\quad\tilde x=\nu(x+2\alpha t),\quad
\tilde u=\nu^\frac2{1-m}e^{(\frac 12-\alpha-\frac\nu2)x-\alpha\nu t}u,$

\medskip
\noindent
where $\tilde q=\sqrt{q^2+a_1(m-1)^2}$, $\nu=\sqrt{1-4\alpha^2}$. 
In the two last cases $q:=r-\frac{m+1}2$.

\medskip
\noindent
$2.2\mapsto{\tilde {2.1}}\colon$\quad
\begin{equation}\label{addit_tr_1fh}
\tilde t=t,~~\tilde x=x+t,~~\tilde u=u;
\end{equation}

%\medskip
\noindent
$4|_{a_2\leqslant\frac14}\mapsto\tilde{3.1}|_{\lambda=1+4\mu_1,\,\gamma=s+(m+1)(1+2\mu_1)}\colon
\quad
\tilde u=\exp\left(\frac{\beta}{2}\,x^2+2\beta(1+2\mu_1-2\kappa_1)t\right)
\dfrac{f_1(x)}{x^{1+2\mu_1}}u,$

\medskip
\noindent
$4|_{a_2>\frac14}\mapsto\tilde{3.2}|_{l=s+m+1,\,\rho=\frac12\sqrt{4a_2-1}}\colon\quad
\tilde u=\exp{\left(\frac{\beta}{2}\,x^2+2\beta(1-2\kappa_1)t\right)}
\dfrac{f_1(x)}{x\cos(\rho\ln|e^{-2\beta t}x|)}u,
$

\medskip
\noindent
$6\mapsto\tilde{2.1}\colon$\quad
\begin{equation}\label{addit_tr_3fh}
\tilde u=\exp{\left(\frac{\beta}{2}\,x^2+\frac{4\beta }{m-1}\,t\right)}\frac{M_{\kappa_3,\frac14}(\beta x^2)}{\sqrt{|x|}}u,
\end{equation}

\noindent
where $\kappa_1=\frac{s+3}{2(1-m)}$, $\kappa_3=\frac{5-m}{4(1-m)}$, $\mu_1=\frac{\sqrt{1-4a_2}}4$.
The function~$f_1$ is defined in table~\ref{TableLieSym_initial_class}.
In the latter three transformations the independent variables are transformed as in~\eqref{addit_tr_2}.

\medskip

We can reduce the classification list adduced in table~\ref{TableLieSym_initial_class}
with the above additional equivalence transformations.
As a result, we obtain the classification of Lie symmetry extensions of class~\eqref{eqRDfghPower}
up to point transformations.
An easier way to derive this classification is
to take a single simplest preimage for each case from the similar classifications for
class~\eqref{class_vFH} and~\eqref{class_GH}.
It does not mean that the additional equivalence transformations are unnecessary at all to be calculated
since they form an important component of classifications up to point transformations.

\begin{theorem}
Up to point transformations, a complete list of extensions of
the maximal Lie invariance algebras of equations from class~\eqref{eqRDfghPower}
is exhausted by cases adduced in table~\ref{TableLieSymhf1}.
\end{theorem}

{\footnotesize
\renewcommand{\arraystretch}{1.7}
\begin{center}
\refstepcounter{table}\label{TableLieSymhf1}\textbf{Table~\thetable.}
The group classification of the initial class up to the point-transformation equivalence.
\\[2ex]
\begin{tabular}{|c|c|l|}
\hline
$f(x)$&$h(x)$&\hfil Basis of $A^{\rm max}$ \\
\hline
$\forall$&$\forall$&$\partial_t$\\
\hline
$1$&$\delta e^x$&$\partial_t,\,(1-m)\partial_x+ u\partial_u$\\
\hline
$e^x$&$\delta e^{\frac{m+1}2x}$&$\partial_t,\,2\,\partial_x-u\partial_u$\\
\hline
$1$&$\delta$&$\partial_t,\,\partial_x,\, 2t\partial_t+x\partial_x+\frac2{1-m}u\,\partial_u$\\
\hline
$x^\lambda$&$\delta x^\gamma$&
 $\partial_t,\,2t\partial_t+x\partial_x+\frac{2-\lambda+\gamma}{1-m}u\partial_u$\\
\hline
$x(\cos\ln|x|^\rho)^2$&$\delta x^l|\cos\ln|x|^\rho|^{m+1}$&
$\partial_t,\,2t\partial_t+x\partial_x+\left(\rho\tan\ln|x|^\rho+\frac{l+1}{1-m}\right)_{}u\partial_u$\\
\hline
$x{}^{-1} f_2(x)^2$&$\delta{x}^{-\frac{m+1}2} e^{px^2}|f_2(x)|^{m+1}$&$\partial_t,\,
e^{2\beta t}\Bigl[\partial_x-\bigl(
4\beta x^2-1-a_3+(a_3+3)g_2(x)\bigr)\dfrac u{2x}\partial_u\Bigr]_{_{}}$\\
\hline
\end{tabular}
\\[2ex]
\parbox{154mm}{\medskip
$\delta=\pm1$, $\rho\neq0$.
$(\lambda,\gamma)\neq\{(0,0),\,(2,m+1)\}$ if $m\not=2$ and
$(\lambda,\gamma)\neq\{(-6,-9),\,(0,0),\,(2,3),\,(8,12)\}$ if $m=2$.
$f_2(x)=M_{\kappa,\frac14}(\beta x^2),\,g_2(x)=M_{\kappa+1,\frac14}(\beta x^2)/f_2(x)$,
$\beta=\frac{2p}{m-1}$, $\kappa=\frac{a_3}{4}$, 
$a_3\neq\frac{5-m}{1-m}$ and additionally $a_3\neq5$ if $m=2$.}
\end{center}
}

\begin{corollary}
Any equation from class~\eqref{eqRDfghPower}, possessing three-dimensional Lie invariance algebra,
is reduced by a point transformation to a constant coefficient equation from the same class.
\end{corollary}

\section{Classification of form--preserving (admissible)\\  transformations}\label{SectionClassificationOfAdmTrans_n0}

Due to special form of equations from class~\eqref{eqRDfghPower}
and the possibility of mapping of class~\eqref{eqRDfghPower} to a class of a simpler structure,
the following problem can be solved completely.
To describe all point transformations each of which
connects a pair of equations from class~\eqref{eqRDfghPower}.
Such transformations are called
\emph{form-preserving}~\cite{Kingston&Sophocleous1998} or
\emph{admissible}~\cite{Popovych&Kunzinger&Eshraghi2006}
\emph{transformations}.
See~\cite{Popovych2006c,Popovych&Kunzinger&Eshraghi2006}
and section~\ref{SectionOnClassMappingsAndGroupClassification} for more rigorous definitions.

The description of admissible transformations is much more complicated problem
than the classification of Lie symmetries and, in certain sense, covers this classification.
The consideration in this section is one of the first implementations
of such description in the case of a non-normalized class.
This is why we present the solution of the problem in detail.
Note that a class of differential equations is called \emph{normalized} if any
admissible transformation in this class belongs to its equivalence group \cite{Popovych2006c,Popovych&Kunzinger&Eshraghi2006}.
The set of admissible transformations of a \emph{semi-normalized class} is generated by
the transformations from the point symmetry groups of initial equations in pairs
and the transformations from the equivalence group of the whole class.
Any normalized class is semi-normalized.
Two systems from a semi-normalized class are transformed into one another by a point transformation
if and only if they are equivalent with respect to~the equivalence group of this class.

Since class~\eqref{eqRDfghPower} can be gauged with transformations
from its usual equivalence group~$G^{\sim}$ to class~\eqref{class_f=g} and then
mapped to class~\eqref{class_vFH},
it is enough for us to solve the similar problem for class~\eqref{class_vFH}.
To do this, we consider a pair of equations from the class under consideration,
i.e., equations~\eqref{class_vFH} and
\begin{equation}\label{class_vFHtilde}
\tilde v_{\tilde t}= \tilde v_{\tilde x\tilde x}
+\tilde H(\tilde x)\tilde v^{\tilde m}+\tilde F(\tilde x)\tilde v, \quad \tilde
m\neq0,1, \quad \tilde H\ne0,
\end{equation}
and assume that these equations are connected via a point
transformation $\mathcal T$ of the general form
\[
\tilde t=T(t,x,v), \quad \tilde x=X(t,x,v), \quad \tilde v=V(t,x,v),
\]
where $|\p(T,X,V)/\p(t,x,v)|\ne0$.
We have to derive the determining equations for the functions~$T$, $X$ and
$V$ and to solve them depending on values of arbitrary elements
in~\eqref{class_vFH} and~\eqref{class_vFHtilde}.

After substitution of expressions for the tilde-variables
into~\eqref{class_vFHtilde}, we obtain an equation in the tildeless
variables. It should be an identity on the manifold determined
by~\eqref{class_vFH} in the second-order jet space over the
space $(t,x\,|\,v)$, where $(t,x)$ and $v$ are assumed independent
and dependent variables, respectively.
The splitting of this identity with respect to
derivatives $v_x,\,v_{tt},\,v_{tx},\,v_{xx}$ implies at first the
equations
$
T_x=T_v=X_v=V_{vv}=0
$
that agrees with results on more general classes of evolution
equations~\cite{Kingston&Sophocleous1998,Popovych&Ivanova2004NVCDCEs,Prokhorova2005}.
Taking into account the above equations, we deduce the following relations:
\begin{gather}T=T(t),\quad X=X(t,x),\quad
V=V^1(t,x)v+V^0(t,x),\nonumber
\\
 T_t={X_x}^2,\quad 2\frac{V^1_x}{V^1}=-\frac{X_t
 X_x}{T_t}+\frac{X_{xx}}{X_{x}},\label{1DetEqForClassificationOfAdmTrans_m_not2}
\\[1ex]
\tilde F V+\tilde H V^{\tilde
m}=\frac{V_v}{T_t}(Fv+Hv^m)+\frac{V_tX_x-V_xX_t-V_{xx}X_x}{T_tX_x},
\label{2DetEqForClassificationOfAdmTrans_m_not2}
\end{gather}
and $T_tX_xV^1\neq0$.
Solving equations~\eqref{1DetEqForClassificationOfAdmTrans_m_not2}, we find that
\[%\label{EqRDfghN0AdmTransGenForm2}
T_t>0, \quad X=\varepsilon\sqrt{T_t} x+\sigma(t),\quad
V^1=\zeta(t)\exp\left(-\frac18\frac{T_{tt}}{T_t}x^2-\frac\varepsilon2\frac{\sigma_t}{\sqrt{T_t}}x\right),
\]
where $\zeta\not=0$, $\varepsilon=\pm1$.
Equation~\eqref{2DetEqForClassificationOfAdmTrans_m_not2} and the linearity of~$V$ with respect to~$v$ imply that
\[
\tilde m=m,\quad \tilde H=\frac{(V^1)^{1-m}}{T_t}H.
\]
Below we restrict ourself with the values $m\ne2$. This is possible since $m$
is invariant under admissible transformations in the class~\eqref{class_vFH} and can be assumed fixed.
The case~$m=2$ is singular and demands separate study with usage of one more mapping like
the investigation of Lie symmetries. It will be a subject of a forthcoming paper.

It follows from~\eqref{2DetEqForClassificationOfAdmTrans_m_not2} under the condition $m\neq2$ that $V^0(t,x)=0$.
After subsequent split of~\eqref{2DetEqForClassificationOfAdmTrans_m_not2} with respect to~$v$,
we obtain the expression for $\tilde F$:
\begin{gather*}
\tilde F =\frac{F}{T_t}+\frac{V^1_tX_x-V^1_xX_t-V^1_{xx}X_x}{T_tX_xV^1}.
\end{gather*}
The equations presenting the expressions for $\tilde H$ and~$\tilde F$ form, in fact, the system of
\emph{classifying equations} which has to be simultaneously solved with respect to $T$, $\sigma$, $\zeta$, $H$ and $F$.

\begin{lemma}\label{LemmaOnRDfghN0FormOfFHforAdmTrans}
Admissible transformations which are not generated by the associated equivalence group~$G^{\sim}_{FH}$ exist
only between equations with the arbitrary elements of the general form
\begin{gather}\label{EqRDfghN0FormOfFHforAdmTrans}
H=\delta|x+\nu|^k e^{px^2+qx}, \quad F=s_2x^2+s_1x+s_0+\frac\kappa{(x+\nu)^2},
\end{gather}
where $k$, $\kappa$, $\delta$, $\nu$, $p$, $q$, $s_2$, $s_1$ and $s_0$ are constants.
The subclass~$\mathcal E$ of such equations is closed under admissible transformations
in the whole class~\eqref{class_vFH}.
\end{lemma}

\begin{proof}
We substitute the expressions for $X$ and~$V^1$ into the classifying equations:
\begin{gather*}
\tilde H T_t\zeta^{m-1}
\exp\left(-\frac{m-1}8\frac{T_{tt}}{T_t}x^2-\frac{m-1}{2\varepsilon}\frac{\sigma_t}{\sqrt{T_t}}x\right)=H,
\\[1ex]
T_t\tilde F =F+\frac{3T_{tt}{}^2-2T_{ttt}T_t}{16T_t{}^2}x^2
-\frac{\sqrt{T_t}}{2\varepsilon}\left(\frac{\sigma_t}{T_t}\right)_tx
+\frac{{\sigma_t}^2+T_{tt}}{4T_t}+\frac{\zeta_t}{\zeta}.
\end{gather*}
The differentiation of the latter equations with respect to~$t$ gives the following system:
\begin{gather*}
\left(\frac\varepsilon2T_{tt}\sqrt{T_t}\,x+T_t\sigma_t\right){\tilde H}_{\tilde x}=\\ \qquad
\left(\frac{m-1}{8}\,T_t\!\left(\frac{T_{tt}}{T_t}\right)_tx^2
+\frac{m-1}{2\varepsilon}\,T_t\!\left(\frac{\sigma_t}{\sqrt{T_t}}\right)_tx
-(m-1)T_t\frac{\zeta_t}{\zeta}-T_{tt}\right){\tilde H},
\\[.5ex]
\left(\frac\varepsilon2T_{tt}\sqrt{T_t}x+T_t\sigma_t\right){\tilde F}_{\tilde x}+T_{tt}\tilde F=\\ \qquad
\left(\frac{3T_{tt}{}^2-2T_{ttt}T_t}{16T_t{}^2}\right)_tx^2
-\frac\varepsilon2\left({\sqrt{T_t}}\left(\frac{\sigma_t}{T_t}\right)_t\right)_t x
+\left(\frac{{\sigma_t}^2+T_{tt}}{4T_t}+\frac{\zeta_t}{\zeta}\right)_t.
\end{gather*}
This system does not contain $F$ and~$H$.
The variable~$x$ is excluded from it by the substitution $x=(\tilde x-\sigma)/\sqrt{T_t}$.
This results in an uncoupled system of two ordinary differential equations
for the functions~$\tilde F$ and~$\tilde H$ depending only on the variable~$\tilde x$.
The variable $t$ can be assumed as a parameter for splitting.
Note that the coefficients of~${\tilde H}_{\tilde x}$ and~${\tilde F}_{\tilde x}$
in the system coincide.

If an equation of the system is an identity with respect to the corresponding unknown function then
the common value of the coefficients equals 0 that implies $T_{tt}=0$, $\sigma_t=0$ and hence
$\zeta_t=0$ since $\tilde H\ne0$.
After integration of the derived equations and necessary substitutions, we obtain exactly formulas for
transformations from the associated equivalence group~$G^{\sim}_{FH}$ (see theorem~\ref{TheoremOnGsimFH}).

Therefore, nontrivial admissible transformations
(i.e., transformations which are not generated by transformations from~$G^{\sim}_{FH}$)
exist only in the case when the classifying equations imply a system of nonidentical equations
with respect to both~$\tilde F$ and~$\tilde H$.
This system necessarily has the form
\begin{gather*}
(a\tilde x+b){\tilde H}_{\tilde x}=(c_2\tilde x{}^2+c_1\tilde x+c_0){\tilde H},\\
(a\tilde x+b){\tilde F}_{\tilde x}=-2a\tilde F+d_2 {\tilde x}^2+d_1{\tilde x}+d_0,
\end{gather*}
where the coefficients $a$, $b$, $c_i$ and $d_i$, $i=0,1,2$, are constants, and $(a,b)\ne(0,0)$.
Moreover, $c_2=d_2=0$ if $a=0$.
After integrating the system, we obtain that
the functions~$\tilde F$ and~$\tilde H$ have form~\eqref{EqRDfghN0FormOfFHforAdmTrans}.
Any transformation satisfying the conditions already derived
does not change the general form~\eqref{EqRDfghN0FormOfFHforAdmTrans} of~$F$ and~$H$,
influencing only the values of constant parameters.
Hence $F$ and $H$ have the same form and, therefore,
the subclass~$\mathcal E$ of equations with $F$ and $H$ of such form
is closed under admissible transformations in the whole class~\eqref{class_vFH}.
\end{proof}

In view of lemma~\ref{LemmaOnRDfghN0FormOfFHforAdmTrans}, the admissible transformations of
class~\eqref{class_vFH} with $m\ne2$ are exhausted by
the admissible transformations generated by transformations from~$G^{\sim}_{FH}$
and the admissible transformations in the subclass~$\mathcal E$.
To complete the investigation, we have to describe
the set $\mathrm{T}(\mathcal E)$ of admissible transformations in the subclass~$\mathcal E$.
Since any admissible transformation with $T_{tt}=0$ and $\sigma_t=0$ is trivial,
it is sufficient to find all elements of $\mathrm{T}(\mathcal E)$ with $(T_{tt},\sigma_t)\ne(0,0)$.

The constants $k$, $\kappa$, $\nu$, $p$, $q$, $\delta$, $s_2$, $s_1$ and $s_0$
can be assumed as the arbitrary elements of the class~$\mathcal E$.
Splitting the classifying equations (i.e., the transformation formulas for $F$ and $H$)
with respect to~$x$, we obtain the classifying equations (or the transformation formulas)
in terms of the new arbitrary elements:
\begin{gather*}
\tilde k=k, \quad \tilde\kappa=\kappa,\\
\tilde\nu=\varepsilon\sqrt{T_t}\,\nu-\sigma \quad\mbox{if}\quad (k,\kappa)\ne(0,0),\\
\tilde p=\frac p{T_t}+\frac{m-1}8\frac{T_{tt}}{T_t{}^2},\quad
\tilde q+2\sigma\tilde p=\frac {\varepsilon q}{\sqrt{T_t}}+\frac{m-1}2\frac{\sigma_t}{T_t},\quad
\tilde\delta\zeta^{m-1}T_t{}^{k/2+1}e^{\tilde p\sigma^2+\tilde q\sigma}=\delta,\\
T_t{}^2\tilde K_2=K_2,\quad
T_t{}^{3/2}(\tilde K_1+2\sigma\tilde K_2)=K_1,\quad
T_t(\tilde K_0+\sigma\tilde K_1+\sigma^2\tilde K_2)=K_0,\quad
\end{gather*}
where
\[
K_2=s_2+\frac{4p^2}{(m-1)^2},\quad
K_1=s_1+\frac{4pq}{(m-1)^2},\quad
K_0=s_0+\frac{q^2+4p(k+2)}{(m-1)^2}-\frac{2p}{m-1}.
\]
$\tilde K_2$, $\tilde K_1$ and $\tilde K_0$ are expressed via the corresponding tilde constants in the same way.
Below each above classifying equation will be denoted by the value transformations of which are described by
this equation, i.e.,
($k$), ($\kappa$), ($\nu$), ($p$), ($q$), ($\delta$), ($K_2$), ($K_1$) and ($K_0$), respectively.

Since $k$ and $\kappa$ are preserved by all admissible transformations, we can partition the class~$\mathcal E$ into
the family $\{\mathcal E_{k\kappa}\}$ of subclasses parameterized by $k$ and $\kappa$
and consider each subclass separately.
Although the values $K_2$, $K_1$ and $K_0$ are changed under admissible transformations, the systems
$K_2=0$, $K_2=K_1=0$ and $K_2=K_1=K_0=0$ as well as their negations are invariant with respect to all of them.
These conditions are convenient to single out different cases and subclasses with nontrivial admissible transformations.

If $K_2\ne0$ (resp. $K_2=0$ and $K_1\ne0$) then equation ($K_2$) (resp. ($K_1$)) implies that
$\tilde K_2\ne0$ (resp. $\tilde K_1\ne0$) and $T_t=\const$.
Then $\sigma=\const$ in view of equation ($K_1$) (resp. ($K_0$)).
Therefore, the subclass singled out in class~$\mathcal E$ by the condition $(K_1,K_2)\ne(0,0)$ is normalized and closed
with respect to point transformations in the whole class~\eqref{class_vFH}.
Its equivalence group is induced by~$G^{\sim}_{FH}$.

In what follows we assume that $K_2=K_1=0$. We consider all possible cases.

\looseness=-1
Under the supposition $K_0\ne0$, equation~($K_0$) implies $\tilde K_0\ne0$ and $T_t=K_0/\tilde K_0=\const$.
If additionally $(k,\kappa)\ne(0,0)$ then $\sigma=\const$ in view of equation~($\nu$), i.e.,
the corresponding subclass is also normalized, possessing an equivalence group induced by~$G^{\sim}_{FH}$,
and closed with respect to point transformations in the whole class~\eqref{class_vFH}.
Therefore, the condition $k=\kappa=0$ is necessary for existence of nontrivial admissible transformations in this case.
It is necessary to consider the two different cases $p=\tilde p=0$ and $p\tilde p\ne0$ for integration of equation~($q$).
Note that the conditions $p=0$ and $\tilde p=0$ (or $p\ne0$ and $\tilde p\ne0$) have to be satisfied simultaneously
because of constraint~($p$).

In the first case we have $\sigma_t=\const$, i.e., finally
\[
T=\delta_1{}^{\!2}t+\delta_2,\quad
\sigma=\delta_5\delta_1t+\delta_3,\quad
\varepsilon=\sign\delta_1,\quad
\zeta=\delta_4\exp\left[\left(-\frac{q\delta_5}{m-1}-\frac{\delta_5{}^2}2\right)t\right],
\]
where $\delta_i$, $i=1,\dots,5$, are arbitrary constants, $\delta_1\delta_4\ne0$.
After all necessary substitutions, we obtain transformations for pairs of equations
from the subclass~$\mathcal E_1$ of the class~$\mathcal E$ with arbitrary elements constrained by the
conditions
\[
k=\kappa=p=s_2=s_1=0, \quad s_0+\frac{q^2}{(m-1)^2}\ne0.
\]
Since the transformations are applicable to any equation from the subclass~$\mathcal E_1$ and
the expression for~$\tilde v$ in them depends on arbitrary element~$q$ then
these transformations form the generalized equivalence group~$G^{\sim}(\mathcal E_1)$ of the subclass~$\mathcal E_1$.
This also implies that the subclass~$\mathcal E_1$ is normalized in the generalized sense and closed
with respect to point transformations in the class~\eqref{class_vFH}.
The transformations from~$G^{\sim}(\mathcal E_1)$ with $\delta_5\ne0$ are not induced
by transformations from~$G^{\sim}_{FH}$.
Therefore, $G^{\sim}(\mathcal E_1)$ is a nontrivial conditional generalized equivalence group
of the whole class~\eqref{class_vFH}.

The case $p\tilde p\ne0$ is studied in similar way.
The complete set of the constraints imposed on the arbitrary elements is
\[
k=\kappa=s_2=s_1=0, \quad p\ne0,\quad s_0+\frac{q^2+4p(k+2)}{(m-1)^2}-\frac{2p}{m-1}\ne0.
\]
The subclass singled out by these constraints will be denoted by~$\mathcal E_2$.
We obtain
\begin{gather*}
T=\delta_1{}^{\!2}t+\delta_2,\quad
\sigma=\delta_5\delta_1\exp\frac {4pt}{m-1}+\delta_3\delta_1,\quad
\varepsilon=\sign\delta_1,\\
\zeta=\delta_4\exp\left[\frac{-1}{m-1}\left(p\delta_5{}^2\exp\frac {8pt}{m-1}-p\delta_3{}^2
+q\delta_5\exp\frac {4pt}{m-1}+q\delta_3\right)t\right],
\end{gather*}
where $\delta_i$, $i=1,\dots,5$, are arbitrary constants, $\delta_1\delta_4\ne0$.
The corresponding transformations are applicable to any equation from the subclass~$\mathcal E_2$,
essentially depend on its arbitrary elements and, therefore,
form the generalized equivalence group~$G^{\sim}(\mathcal E_2)$ of the subclass~$\mathcal E_2$.
$G^{\sim}(\mathcal E_2)$ is a nontrivial conditional generalized equivalence group of the class~\eqref{class_vFH}
since the transformations from~$G^{\sim}(\mathcal E_2)$ with $\delta_5\ne0$ are not induced
by transformations from~$G^{\sim}_{FH}$.
The subclass~$\mathcal E_2$ is normalized in the generalized sense and closed
with respect to point transformations in the whole class~\eqref{class_vFH}.
Moreover, the transformations of arbitrary elements
\[
\tilde p=\frac p{\delta_1{}^{\!2}}, \quad
\tilde q=\frac q{\delta_1}-2\frac{\delta_3}{\delta_1}p, \quad
\tilde s_0=s_0+4\frac{\delta_3p}{\delta_1{}^{\!2}}\frac{q-\delta_3p}{(m-1)^2}, \quad
\tilde\delta=\frac{\delta_4^{1-m}}{\delta_1{}^{\!2}}\delta
\]
do not depend on the parameters~$\delta_2$ and~$\delta_5$, i.e.,
for any fixed values of the arbitrary elements the admissible transformations with
$\delta_1=1$ and~$\delta_3=0$ form the point symmetry group of the corresponding equation
(compare with case~\ref{TableLieSymHF_exp_px2_2op} of table~\ref{TableLieSymHF}).
This implies that the subclass~$\mathcal E_2$ is semi-normalized in the usual sense,
and its usual equivalence group is induced by~$G^{\sim}_{FH}$.
In fact, the generalized equivalence group~$G^{\sim}(\mathcal E_2)$ is generated by
the transformations from~$G^{\sim}_{FH}$ and point symmetry transformations of
equations from the subclass~$\mathcal E_2$.

Suppose that $K_0=0$. Then also $\tilde K_0=0$ due to equation~($K_0$).
Equation ($p$) can be rewritten in the form
\[
\left(\frac1{T_t}\right)_t=-p'\frac1{T_t}+\tilde p', \quad\mbox{where}\quad
p'=\frac{-8p}{m-1}, \quad \tilde p'=\frac{-8\tilde p}{m-1}.
\]
We integrate this equation and present the general solution in such form that
continuous dependence of it on the parameters~$p$ and~$\tilde p$ is obvious:
\begin{gather*}
p\tilde p\not=0\colon \quad
\frac{e^{\tilde p' T}-1}{\tilde p'}=\delta_1{}^{\!2}\frac{e^{p' t}-1}{p'}+\delta_2,
\qquad
p=0,\ \tilde p\not=0\colon \quad \frac{e^{\tilde p' T}-1}{\tilde p'}=\delta_1{}^{\!2}t+\delta_2,
\\[.5ex]
p\not=0,\ \tilde p=0\colon \quad T=\delta_1{}^{\!2}\frac{e^{p' t}-1}{p'}+\delta_2,
\qquad
p=\tilde p=0\colon \quad T=\delta_1{}^{\!2}t+\delta_2.
\end{gather*}
The derivative~$T_t$ can be presented in the uniform way as $T_t=\delta_1{}^{\!2}e^{p' t-\tilde p' T}$.
The expression for~$\zeta$ can be easily found from equation~($\delta$) if $T$ and $\sigma$ are known.
We will not adduce it since it is quite cumbersome.

In contrast to the case~$K_0\ne0$, here nontrivial admissible transformations exist
for both the zero and nonzero values of $(k,\kappa)$.

If $(k,\kappa)\ne(0,0)$ then $\sigma=\varepsilon\sqrt{T_t}\,\nu-\tilde\nu$ due to equation~($\nu$).
Imposing the condition $T_{tt}$ for nontrivial admissible transformations to exist,
we split equation~($q$) with respect to~$t$ and then obtain $q-2p\nu=\tilde q-2\tilde p\tilde\nu=0$.
There are no other constraints to be imposed on arbitrary elements.
The subclass singled out from class~$\mathcal E$ by the conditions
$K_2=K_1=K_0=0$, $(k,\kappa)\ne(0,0)$ and $q=2p\nu$ will be denoted by~$\mathcal E_3$.
The constructed transformations are applicable to any equation from the subclass~$\mathcal E_3$,
essentially depend on its arbitrary elements and, therefore,
form the five-parametric generalized equivalence group~$G^{\sim}(\mathcal E_3)$ of the subclass~$\mathcal E_3$.
As parameters, we can take $\delta_1$, $\delta_2$, $\tilde p$, $\tilde\nu$ and $\tilde\delta$.
The subclass~$\mathcal E_3$ is normalized in the generalized sense and closed
with respect to point transformations in the whole class~\eqref{class_vFH}.
Putting $\tilde p=p$, $\tilde\nu=\nu$ and $\tilde\delta=\delta$, we construct the point symmetry transformation of
the corresponding equation.
Therefore, any equation from~$\mathcal E_3$ has a two-parametric group of point symmetries and can
be reduced by point transformations to the equation from the same subclass with $p=\nu=q=0$, $\delta=\pm1$
and the same values of $k$ and $\kappa$.
Admissible transformations are not generated by transformations from~$G^{\sim}_{FH}$
and point symmetry transformations of single equations only if $p=0$ and $\tilde p\ne0$ or, conversely,
$p\ne0$ and $\tilde p=0$.
The subclasses of~$\mathcal E_3$ associated with the additional constraints $p\ne0$ or $p=0$
are semi-normalized in the usual sense, and their usual equivalence groups are induced by~$G^{\sim}_{FH}$.

Let now $k=\kappa=0$.
We denote the corresponding subclass of~$\mathcal E$ by $\mathcal E_4$.
The integration of equation~($q$) with respect to~$\sigma$ implies that
\[
\sigma=\delta_3\exp\frac{4\tilde pT}{m-1}
+\left\{\begin{array}{cl}
-\dfrac{\tilde q}{2\tilde p},&\tilde p\ne0 \\[3ex] \dfrac{2\tilde qT}{m-1},&\tilde p=0
\end{array}\right\}
-\delta_1\left\{\begin{array}{cl}
-\dfrac{q}{2p},&p\ne0 \\[3ex] \dfrac{2qt}{m-1},&p=0
\end{array}\right\}.
\]
Analogously to the above cases,
the constructed transformations are applicable to any equation from the subclass~$\mathcal E_4$,
essentially depend on its arbitrary elements and, therefore,
form the generalized equivalence group~$G^{\sim}(\mathcal E_4)$ of the subclass~$\mathcal E_4$,
parameterizable by six parameters (i.e., $\delta_1$, $\delta_2$, $\delta_3$, $\tilde p$, $\tilde q$ and $\tilde\delta$).
So, the subclass~$\mathcal E_4$ is normalized in the generalized sense and closed
with respect to point transformations in the whole class~\eqref{class_vFH}.
Putting $\tilde p=p$, $\tilde q=q$ and $\tilde\delta=\delta$, we construct the point symmetry transformation of
the corresponding equation.
Therefore, any equation from~$\mathcal E_4$ possesses a three-parametric group of point symmetries and, moreover,
can be reduced by point transformations to the constant-coefficient equation $u_t=u_{xx}\pm u^m$.
Admissible transformations which are not generated by transformations from~$G^{\sim}_{FH}$
and point symmetry transformations of single equations exist only if $p=0$ and $\tilde p\ne0$ or, conversely,
$p\ne0$ and $\tilde p=0$.
The subclasses of~$\mathcal E_4$ associated with the additional constraints $p\ne0$ or $p=0$
are semi-normalized in the usual sense, and their usual equivalence groups are induced by~$G^{\sim}_{FH}$.

Note that the subclass~$\mathcal E_{4,p=0}$ can be united with the class~$\mathcal E_1$ having
the same equivalence group which is formed by the transformations with $T_{tt}=\sigma_{tt}=0$.
The resulted class~$\mathcal E_1'$ is singled out by the constraints $k=\kappa=p=s_2=s_1=0$.
Only after the association with the class~$\mathcal E_1'$,
the above conditional equivalence group becomes maximal.

We unite the results of this section  with theorem~\ref{TheoremOnGsimFH} in the following statement.

\begin{theorem}\label{TheoremOnRDfghN0FormOfFHforAdmTrans}
The equivalence group~$G^{\sim}_{FH}$ of
class~\eqref{class_vFH}, where $m\ne0,1,2$, is formed by the transformations
\[
\tilde t={\delta_1}^2 t+\delta_2,\quad \tilde x=\delta_1 x+\delta_3,
\quad
\tilde v=\delta_4 v, \quad
\tilde F=\delta_1^{-2}F,\quad
\tilde H=\delta_1^{-2}\delta_4^{1-m}H, \quad \tilde m=m,
\]
where $\delta_j$, $j=1,\dots,4$, are arbitrary constants, $\delta_1\delta_4\not=0$.
The parameter~$m$ is an invariant of any point transformation in class~\eqref{class_vFH}.
Admissible transformations which are not induced by elements of~$G^{\sim}_{FH}$ exist
only between equations with arbitrary elements of the general form
\[%\label{EqRDfghN0FormOfFHforAdmTrans}
H=\delta|x+\nu|^k e^{px^2+qx}, \quad F=s_2x^2+s_1x+s_0+\frac\kappa{(x+\nu)^2},
\]
where $k$, $\kappa$, $\delta$, $\nu$, $p$, $q$, $s_2$, $s_1$ and $s_0$ are constants satisfying
the conditions
\begin{gather*}
s_2=-\frac{4p^2}{(m-1)^2},\quad
s_1=-\frac{4pq}{(m-1)^2}\quad\mbox{and}\quad
(k=\kappa=0\quad\mbox{or}\quad K_0=q-2p\nu=0),\\
K_0:=s_0+\frac{q^2+4p(k+2)}{(m-1)^2}-\frac{2p}{m-1}.
\end{gather*}
The set of equations possessing nontrivial admissible transformations are partitioned into
four subclasses normalized in the generalized sense and closed under point transformations
in the whole class~\eqref{class_vFH} (for each subclass we indicate the additional constraints for arbitrary elements):
$\mathcal E_1$: $K_0\ne0$, $k=\kappa=p=0$;
$\mathcal E_2$: $K_0\ne0$, $k=\kappa=0$, $p\ne0$;
$\mathcal E_3$: $K_0=0$,  $(k,\kappa)\ne(0,0)$, $q=2p\nu$; and
$\mathcal E_4$: $K_0=k=\kappa=0$.
The subclass~$\mathcal E_2$ is semi-normalized in the usual sense,
and its usual equivalence group is induced by~$G^{\sim}_{FH}$.

The set of admissible transformations of class~\eqref{class_vFH} is generated by the equivalence group~$G^{\sim}_{FH}$
and the nontrivial conditional generalized equivalence groups~$G^{\sim}(\mathcal E_j)$, $j=1,\dots,4$.
\end{theorem}

\section{On nonclassical symmetries}\label{Section_NonclassicalSymmetries_n0}

The notion of nonclassical symmetry (called also conditional or $Q$-conditional symmetry) was
introduced by Bluman and Cole~\cite{Bluman&Cole1969}. A precise and rigorous definition of this notion
was suggested noticeably later~\cite{Fushchych&Tsyfra1987} (see also~\cite{Zhdanov&Tsyfra&Popovych1999}).
Since then there is an explosion of research activity in the area of investigation of nonclassical
symmetries. In a number of papers the reduction method with respect to nonclassical symmetries
was successfully applied to obtain
new non-Lie exact solutions of PDEs arising as models in different fields of physics, biology and chemistry.
Some of these works concern with diffusion equations (with reaction or convection terms or
without them). See, for example,~\cite{ArrigoHill1995,ArrigoHillBroadbridge1993,Cherniha2007,
ChernihaPliukhin2007,Clarkson&Mansfield1993,FushchichSerov1990,Serov1990,Zhdanov&Lahno1998}.
Note that in fact it is more convenient to call nonclassical symmetries \emph{reduction operators}~\cite{Popovych&Vaneeva&Ivanova2007}.

We briefly review necessary definitions and statements on nonclassical symmetries 
\cite{Fushchych&Zhdanov1992,Popovych&Vaneeva&Ivanova2007,Zhdanov&Tsyfra&Popovych1999}, 
adopting them for the case of one second-order PDE with two independent variables, relevant for this paper. 
Consider a second order differential equation~$\mathcal L$ of the form~$L(t,x,u_{(2)})=0$
for the unknown function $u$ of the two independent variables~$t$ and~$x$, 
where $u_{(2)}=(u,u_t,u_x,u_{tt},u_{tx},u_{xx})$. 
Let $\mathcal Q$ denote the set of first-order differential operators of the general form
\[
Q=\tau(t,x,u)\p_t+\xi(t,x,u)\p_x+\eta(t,x,u)\p_u, \quad (\tau,\xi)\not=(0,0).
\]

\begin{definition}\label{DefinitionOfCondSym_n0}
The differential equation~$\mathcal L$ is called
\emph{conditionally invariant} with respect to an operator $Q$ if
the relation
$%\label{eq_cond_sym_criterion}
Q_{(2)}L(t,x,u_{(2)})\bigl|_{\mathcal L\cap\mathcal{Q}^{(2)}}=0
$
holds, which is called the \emph{conditional (or nonclassical) invariance criterion}.
Then $Q$ is called \emph{conditional symmetry}
(or nonclassical symmetry, $Q$-conditional symmetry or reduction operator)
of the equation~$\mathcal L$.
\end{definition}

The symbol $Q_{(2)}$ stands for the standard second prolongation of~$Q$.
$\mathcal Q^{(2)}$ is the manifold determined in the second order jet space by the differential
consequences of the characteristic equation~$\eta-\tau u_t-\xi u_x=0$, 
which have, as differential equations, orders not greater than two.

We denote the set of reduction operators of the equation~$\mathcal L$ by $\mathcal Q(\mathcal L)$.
Any Lie symmetry operator of~$\mathcal L$ belongs to~$\mathcal Q(\mathcal L)$. 
Sometimes $\mathcal Q(\mathcal L)$ is exhausted by the operators equivalent to Lie symmetry ones
in the sense of the following definition.

\begin{definition}\label{DefinitionEqvivOpCondSym_n0}
Operators $\widetilde Q$ and $Q$ are called \emph{equivalent} ($\widetilde Q\sim Q$)
if they differ by a multiplier being a non-vanishing function of~$t$, $x$ and~$u$:
$\widetilde Q=\lambda Q$, where $\lambda=\lambda(t,x,u)$, $\lambda\not=0$.
\end{definition}

\looseness=-1
We denote the result of factorization of~$\mathcal Q$ with respect to this equivalence relation by~$\mathcal Q_{\rm f}$.
Elements of~$\mathcal Q_{\rm f}$ will be identified with their representatives in~$\mathcal Q$.
If the equation~$\mathcal L$ is conditionally invariant with respect to the operator~$Q$,
then it is conditionally invariant with respect to any operator equivalent to~$Q$
\cite{Fushchych&Zhdanov1992,Zhdanov&Tsyfra&Popovych1999}.
Therefore, the equivalence relation on~$\mathcal Q$ induces a well-defined equivalence relation on~$\mathcal Q(\mathcal L)$;
and the factorization of $\mathcal Q$ with respect to this equivalence relation
can be naturally restricted on~$\mathcal Q(\mathcal L)$ that
results in the subset~$\mathcal Q_{\rm f}(\mathcal L)$ of $\mathcal Q_{\rm f}$.
As in the whole set~$\mathcal Q_{\rm f}$, we identify elements of~$\mathcal Q_{\rm f}(\mathcal L)$ with their
representatives in~$\mathcal Q(\mathcal L)$.
In fact, nonclassical symmetries should be studied up to the above equivalence relation. 
The elements of~$\mathcal Q(\mathcal L)$ which are not equivalent to Lie invariance operators of~$\mathcal L$ 
will be called \emph{pure nonclassical} symmetries of~$\mathcal L$. 

For any $(1+1)$-dimensional evolution equation, the case of reduction operators with $\tau=0$ is singular. 
Since in this case $\xi\not=0$, up to the equivalence of reduction operators
we can assume $\xi=1$ that implies $Q=\p_x+\eta\p_u$.
The conditional invariance criterion results in only a single determining equation on
the coefficient~$\eta$, which is reduced with a non-point transformation to the initial equation, 
where $\eta$ becomes a parameter~\cite{Fushchych&Shtelen&Serov&Popovych1992,Zhdanov&Lahno1998}.
This is why the case $\tau=0$ is called ``no-go''; and it has to be excluded from consideration under
the classification of nonclassical symmetries.
Nevertheless, it is possible to find some reduction operators of the form $Q=\p_x+\eta\p_u$ using
ans\"atze for the coefficient~$\eta$~\cite{Clarkson&Mansfield1993,Fushchych&Shtelen&Serov&Popovych1992,Gandarias2001}.
For example, $\eta$ can be assumed to be a polynomial in $u$.

We can essentially simplify and order classification of reduction operators, additionally taking into account
Lie symmetry transformations of an equation.

\begin{lemma}\label{LemmaOnMappingsOfOps}
Any point transformation~$g$: $\tilde t=T(t,x,u)$, $\tilde x=X(t,x,u)$, $\tilde u=U(t,x,u)$ 
induces a one-to-one mapping $g_*$ of~$\mathcal Q$ into itself.
Namely, any operator~$Q$ is mapped to the operator $g_*Q$ with the coefficients 
$\tilde\tau=QT$,
$\tilde\xi)=QX$,
$\tilde\eta=QU$.
If~$Q'\sim Q$ then  $g_* Q'\sim g_* Q$.
Therefore, the factorized mapping~$g_{\rm f} \colon\mathcal Q_{\rm f}\to\mathcal Q_{\rm f}$
is also well-defined and one-to-one.
\end{lemma}

Lemma~\ref{LemmaOnMappingsOfOps} results in appearing equivalence relation between operators, 
which differs from usual one described in  definition~\ref{DefinitionEqvivOpCondSym_n0}.

\begin{definition}\label{DefinitionOfEquivInvFamiliesWrtGroup}
Operators $\widetilde Q$ and $Q$ are \emph{equivalent with respect to a group $G$} 
of point transformations ($\widetilde Q\sim Q \bmod G$) if there exists an element $g\in G$
for which $\widetilde Q\sim g_*Q$.
\end{definition}

The problem of finding reduction operators is 
more complicated than the similar problem for Lie symmetries because the first problem is reduced
to the integration of an overdetermined system of nonlinear PDEs, whereas in the case of Lie symmetries
one deals with a more overdetermined system of linear PDEs.
The question occurs: could we use equivalence and gauging transformations in investigation
of  reduction operators as we do for finding Lie symmetries?
The following statements give the positive answer.

\begin{lemma}\label{LemmaOnInducedMapping}
Given any point transformation $g$ between equations~$\mathcal L$ and $\tilde{\mathcal L}$,
$g_*$ is a one-to-one map of~$\mathcal Q(\mathcal L)$ onto~$\mathcal Q(\tilde{\mathcal L})$.
The factorized mapping $g_{\rm f}$ between $\mathcal Q_{\rm f}(\mathcal L)$
and $\mathcal Q_{\rm f}(\tilde{\mathcal L})$ also is bijective.
\end{lemma}

\begin{corollary}\label{CorollaryOnEquivReductionOperatorWrtSymGroup}
Let $G$ be a Lie symmetry group of an equation~$\mathcal L$.
Then the equivalence of operators from $\mathcal Q$ with respect to~$G$
induces equivalence relations in~$\mathcal Q(\mathcal L)$ and in~$\mathcal Q_{\rm f}(\mathcal L)$.
\end{corollary}

Let $G^\sim=G^\sim(\mathcal L|_{\mathcal S})$ be the equivalence group
of a class~$\mathcal L|_{\mathcal S}$ of differential equations and
$P=P(\mathcal L|_{\mathcal S})$ denote the set of the pairs each of which consists of
an arbitrary element $\theta$ from the set~$\mathcal S$ and an operator~$Q$ from~$\mathcal Q(\mathcal L_\theta)$.
(Hereafter notations of section~\ref{SectionOnClassMappingsAndGroupClassification} are used.)
We will call $P(\mathcal L|_{\mathcal S})$ the \emph{set of nonclassical symmetries}
(or the set of reduction operators) of the class~$\mathcal L|_{\mathcal S}$.

\begin{corollary}\label{CorollaryOnEquivReductionOperatorWrtEquivGroup}
The action of transformations from~$G^\sim$ in $\mathcal S$ and
$\{\mathcal Q(\mathcal L_{\theta})\,|\,\theta\in{\cal S}\}$
together with the usual equivalence relation in $\mathcal Q$ 
naturally generates an equivalence relation in~$P$.
\end{corollary}

\begin{definition}\label{DefinitionOfEquivOfRedOperatorsWrtEquivGroup}
Let $\mathcal L_\theta,\mathcal L_{\theta'}\in\mathcal L|_{\mathcal S}$,
$Q\in\mathcal Q(\mathcal L_\theta)$, $Q'\in\mathcal Q(\mathcal L_{\theta'})$.
The pairs~$(\theta,Q)$ and~$(\theta',Q')$
are called {\em $G^{\sim}$-equivalent} if there exists $g\in G^{\sim}$
which transforms $\mathcal L_\theta$ to $\mathcal L_{\theta'}$, and $Q'\sim g_*Q$.
\end{definition}

The classification of reduction operators with respect to~$G^{\sim}$ will be understood as
the classification in~$P$ with respect to the above equivalence relation.
This problem can be investigated in the way that is similar to the usual group classification in classes
of differential equations.
Namely, we construct firstly the reduction operators that are defined for all values of the arbitrary elements.
Then we classify, with respect to the equivalence group, the values of arbitrary elements for each of that
the corresponding equation admits additional reduction operators.

In an analogues way, we can also introduce equivalence relations on~$P$, which are
generated by either generalizations of the usual equivalence group of~$\mathcal L|_{\mathcal S}$ or
all admissible point transformations in pairs of equations from~$\mathcal L|_{\mathcal S}$.

\begin{corollary}\label{CorollaryOnReductionOperatorsOfSimilarClasses}
Similar classes have similar sets of reduction operators.
Any similarity transformation $\Psi$ between classes $\mathcal L|_{\mathcal S}$
and $\mathcal L'|_{\mathcal S'}$ generates a bijection 
$\bar\Psi\colon P(\mathcal L|_{\mathcal S})\to P(\mathcal L'|_{\mathcal S'})$
via the rule $(\theta'\!,Q')=\bar\Psi(\theta,Q)$
if $\theta'=\Psi\theta$ and $Q'=(\Psi|_{(x,u)})_*Q$.
Here $(\theta,Q)\in P(\mathcal L|_{\mathcal S})$,
$(\theta'\!,Q')\in P(\mathcal L'|_{\mathcal S'})$.
\end{corollary}

\begin{corollary}
A point-transformation mapping between classes of differential equations induces a mapping between
the corresponding sets of reduction operators. 
Namely, if a class $\mathcal L'|_{\mathcal S'}$ is the image of
a class $\mathcal L|_{\mathcal S}$ under a family of point transformations
$\varphi_\theta\colon(x,u)\to(x',u')$, $\theta\in\mathcal{S}$, then
the image of $(\theta,Q)\in P(\mathcal L|_{\mathcal S})$ is
$(\theta'\!,Q')\in P(\mathcal L'|_{\mathcal S'})$, where
$\mathcal L'_{\theta'}=\mathrm{pr}_p\varphi_\theta\mathcal L_\theta$ and
$Q'=(\varphi_{\tilde\theta})_*Q$.
\end{corollary}

Moreover, the similar statement in the opposite direction is also true.

\begin{proposition}
The set of reduction operators of the initial class
$\mathcal L|_{\mathcal S}$ is reconstructed from
the one of its point-transformation image $\mathcal L'|_{\mathcal S'}$.
\end{proposition}

\begin{proof}
Suppose that the family of point transformations $\varphi_\theta\colon(x,u)\to(x',u')$, $\theta\in\mathcal{S}$,
maps the class $\mathcal L|_{\mathcal S}$ onto the class $\mathcal L'|_{\mathcal S'}$.
Let $(\theta'\!,Q')\in P(\mathcal L'|_{\mathcal S'})$ and let
$\mathcal L_\theta$ be some equation mapped to $\mathcal L'_{\theta'}$.
Then $(\theta,Q)\in P(\mathcal L|_{\mathcal S})$, where
$Q=((\varphi_\theta)^{-1})_*Q$.
Each nonclassical symmetry of an equation from the class~$\mathcal L|_{\mathcal S}$ is obtainable in the way described.
\end{proof}

Thus, equivalence and gauging
transformations can essentially simplify the problem of finding nonclassical symmetries.
Moreover, their application can appear to be a crucial point in solving the problem.
For example, the group classification problem for class~\eqref{eqRDfghPower} becomes tame only after
gauging of arbitrary elements by equivalence transformations and mappings to other classes.
Each Lie symmetry operator is a reduction operator.
Hence under the classification of nonclassical symmetries we need to surmount, at least, obstacles
similar to that under the group classification.
In fact, the obstacles will be much more essential since the system of determining equations for
nonclassical symmetries is nonlinear and less overdetermined.

Below we sketch the procedure of application of equivalence transformations, gauging of arbitrary elements
and mappings between classes of equations to classification of nonclassical symmetries.
The procedure will be described using class~\eqref{eqRDfghPower} as an example.

1. Similarly to the group classification, at first we gauge class~\eqref{eqRDfghPower}
to subclass~\eqref{class_f=g} constrained by the condition~$f=g$.
Further the cases $m=2$ and $m\neq2$ should be considered separately.
If $m=2$, class~\eqref{class_f=g} is mapped to the imaged class~\eqref{class_vFH}
by transformation~\eqref{gauge}. If $m\not=2$, class~\eqref{class_vFH_m2} is mapped to the double-imaged
class~\eqref{class_GH}
by the transformation
\[w=\sqrt{|f|}u-\frac {|f|}{2h}\bigl(\sqrt{|f|}\,\bigr)_{xx}\]
which is the composition
of transformations~\eqref{gauge} and~\eqref{gauge_m=2}.

2. Since nonclassical symmetries of constant coefficient equations from
the imaged classes are well investigated (see
below for more details), they should be excluded from the consideration.
In view of lemma 3, this also concerns variable coefficient equations from classes~\eqref{class_vFH}
and~\eqref{class_GH} which are equivalent with respect to point transformations to
constant coefficient ones, namely equations associated with cases $1|_{q\neq0}$, $2|_{q\neq0}$, $4$ and $6$ of
tables~\ref{TableLieSymHF} and~\ref{TableLieSymHG} and equations reduced to them
by  transformations from the corresponding equivalence groups. As a result, only equations
from classes~\eqref{class_vFH} and~\eqref{class_GH} which are inequivalent with
respect to all point transformations to constant coefficient ones should be studied.

3. Reduction operators should be classified up to the equivalence  relations
generated by the equivalence groups or even by the whole sets of admissible transformations.
Only the nonsingular case $\tau\ne0$ (reduced to the case $\tau=1$) should be considered.
All obtained operators equivalent to Lie symmetry ones should be neglected.

4. Preimages of the obtained nonclassical symmetries and of equations
admitting them should be found using backward gauging transformations 
and mappings induced by these transformations on the sets of operators.

We show how to derive some nonclassical symmetries for equations from class~\eqref{class_f=g}
using known results for constant coefficient equations from the imaged class~\eqref{class_vFH}.

Constant coefficient equations from the imaged and double-imaged classes belong to the wider class of a
quasilinear heat equations with a source of the general form 
$v_t=v_{xx}+q(v)$.
Lie and nonclassical symmetries  of these equations were investigated in~\cite{Dorodnitsyn1979;1982} 
and \cite{ArrigoHillBroadbridge1993,Clarkson&Mansfield1993,FushchichSerov1990,Serov1990}, respectively.
Their non-Lie exact solutions were constructed by the reduction method
in~\cite{ArrigoHillBroadbridge1993,Clarkson&Mansfield1993}, see appendix~\ref{Section_ExactSolutionsOfCubicHeatEq}.
The equation $v_t=v_{xx}+q(v)$ possesses
pure nonclassical symmetry operators with nonvanishing coefficients of $\partial_t$ 
if and only if $q$ is a cubic polynomial in $v$.
Thus, in the case $q=\delta v^3+\varepsilon v$, where $\delta\ne0$, such operators 
are exhausted, up to the equivalence with respect to the corresponding Lie symmetry groups, by the following:
\begin{gather}\label{EqReductionOpsOfNHEs}\arraycolsep=0ex
\begin{array}{l}
\delta<0\colon\quad\partial_t\pm\frac32\sqrt{-2\delta}\,v\partial_x+\frac32(\delta v^3+\varepsilon v)\partial_v,\\[1ex]
\varepsilon=0\colon\quad\partial_t-\frac3x\partial_x-\frac{3}{x^2}v\partial_v,\\[1ex]
\varepsilon<0\colon\quad\partial_t+3\mu\tan(\mu x)\partial_x-3\mu^2 \sec^2(\mu x)v\partial_v,\\[1ex]
\varepsilon>0\colon\quad\partial_t-3\mu\tanh(\mu x)\partial_x+3\mu^2 {\rm sech}^2(\mu x)v\partial_v,\\[1ex]
\phantom{\varepsilon>0\colon\quad}\partial_t-3\mu\coth(\mu x)\partial_x-3\mu^2 {\rm cosech}^2(\mu x)v\partial_v,
\end{array}
\end{gather}
where $\mu=\sqrt{|\varepsilon|/2}$. 
Note that the last operator was missed in~\cite{ArrigoHillBroadbridge1993,Clarkson&Mansfield1993}. 

Finding the preimages of equations with such values of~$q$ with respect to transformation~\eqref{gauge}
and the preimages of the corresponding reduction operators according to formula~\eqref{tr_op}, we obtain
the cases presented in table~\ref{TableNonclassicalSymhf_m3}.

\begin{center}\footnotesize
\renewcommand{\arraystretch}{1.7}
\refstepcounter{table}\label{TableNonclassicalSymhf_m3}\textbf{Table~\thetable.} Nonclassical
symmetries of equations of the form $f(x)u_t=(f(x)u_x)_x+\delta f(x)^2u^3$, $f(x)=\zeta(x)^2$.
\\[2ex]
\begin{tabular}{|c|c|l|}
\hline
N&$\zeta(x)$&\hfil Reduction operators \\
\hline
1&$c_1x+c_2$&$\partial_t\pm\frac32\sqrt{-2\delta}\zeta u\partial_x+
\frac32(\delta \zeta^2 u\mp c_1\sqrt{-2\delta})u^2\partial_u$,\\
&&$\partial_t-\dfrac3x\partial_x-\dfrac{3c_2}{x^2\zeta_{_{_{}}}}u\partial_u$\\
\hline
&&$\partial_t\pm\frac32\sqrt{-2\delta}\zeta u\partial_x+
\frac32(\delta {\zeta}^2u^2\mp\sqrt{-2\delta}\zeta_xu+\varepsilon)_{}u\partial_u$,
\\
2&$c_1\sin(\sqrt{\varepsilon}x)+c_2\cos(\sqrt{\varepsilon}x)$&$\partial_t-3\mu\tanh(\mu x)\partial_x+
3\mu \left(\dfrac{\zeta_x}{\zeta}\tanh(\mu x)+\mu\,{\rm sech}^2(\mu x)\right)_{_{}}u\partial_u$,
\\
&&$\partial_t-3\mu\coth(\mu x)\partial_x+3\mu \left(\dfrac{\zeta_x}{\zeta}\coth(\mu x)
-\mu\,{\rm cosech}^2(\mu x)\right)_{_{}}u\partial_u$\\
\hline
3&$c_1\sinh(\sqrt{|\varepsilon|}x)+c_2\cosh(\sqrt{|\varepsilon|}x)$&$\partial_t\pm\frac32\sqrt{-2\delta}\zeta u\partial_x+
\frac32(\delta {\zeta}^2u^2\mp\sqrt{-2\delta}\zeta_xu+\varepsilon)u\partial_u$,\\
&&$
\partial_t+3\mu\tan(\mu x)\partial_x-3\mu \left(\dfrac{\zeta_x}{\zeta}\tan(\mu x)+\mu\sec^2(\mu x)\right)_{_{}}u\partial_u$\\
\hline
\end{tabular}
\\[2ex]
\parbox{150mm}{$c_1^2+c_2^2\neq0$. In case 2 $\varepsilon>0$. In case 3 $\varepsilon<0$, $\mu=\sqrt{{|\varepsilon|}/2}$.}
\end{center}

\begin{note}
There exist two ways to use mappings between classes of equations in the investigation of nonclassical symmetries.
Suppose that nonclassical symmetries of equations from the imaged class are known.
The first way is to take the preimages of both the constructed operators and the equations possessing them.
Then we can reduce the preimaged equations with respect to the corresponding preimaged operators
to find non-Lie solutions of equations from the initial class.
The above way seems to be non-optimal since the ultimate goal of the investigation of nonclassical symmetries is
the construction of exact solutions.
This observation is confirmed by the fact that the equations from the imaged class and the associated nonclassical symmetry operators
have, as a rule, a simpler form and therefore, are more suitable than their preimages.
Reduced equations of the imaged class are also simpler to be integrated.
Moreover, it happens that preimages of uniformly parameterized similar equations do not have similar forms and belong to
different parameterized families.
As a result, making reductions in the initial class, we have to deal with a number of different ans\"atze and reduced equations
although this is equivalent to the consideration of a single ansatz and the corresponding reduced equation within the imaged classes.
This is why the second way based on the implementation of reductions in the imaged classes and preimaging of the
obtained exact solutions instead of preimaging the corresponding reduction operators is preferable.
\end{note}

\section{Exact solutions}\label{Section_OnRDfghExactSolutions_n0}

\looseness=-1
In this section new Lie and non-Lie solutions for the equations
from the initial class as well as for ones from the related classes are constructed 
to just illustrate possible applications of the classification results obtained. 
We apply two methods: the classical Lie reduction involving the maximal Lie invariance algebras of equations from the obtained 
classification lists (subsection~\ref{Subsection_OnRDfghSimilaritySolutions_n0}) 
and the generation of new solutions from known ones of equations which are equivalent to equations under consideration 
with respect to point transformations. 
Different kinds of equivalence transformations and general admissible transformations are used under the second approach.

\subsection{Similarity solutions}\label{Subsection_OnRDfghSimilaritySolutions_n0}

%The Lie symmetry operators enable us to construct exact solutions of equations from the investigated classes, using the
%reduction method. This technique is well known and quite algorithmic (see, e.g.~\cite{Ovsiannikov1982,Olver1986}).
Group analysis of a class of equations includes
finding similarity solutions by the reduction method for all equations possessing nontrivial
symmetry properties.
For this purpose, at first, the optimal sets of subalgebras should be constructed for each kind of
maximal Lie invariance algebras arising under group classification,
and then the reductions with respect to obtained subalgebras should be done.
Knowledge of the additional equivalence transformations allows us to simplify this problem by reducing
the number of cases to be considered.
For example, if we like to construct similarity solutions for equations from class~\eqref{class_vFH},
it is easier to do reductions only for cases inequivalent with respect to point transformations,
i.e., for cases $0$, $1|_{q=0}$, $2|_{q=0}$, $3$, and $5$ of table~\ref{TableLieSymHF}.
Then the obtained exact solutions and additional equivalence
transformations~\eqref{addit_tr_1} and \eqref{addit_tr_2} can be used to generate exact solutions for equations
from class~\eqref{class_vFH} with arbitrary elements presented by cases $1|_{q\neq0}$, $2|_{q\neq0}$, $4$, and $6$
of table~\ref{TableLieSymHF}.

In detail we consider the case~$\ref{TableLieSymHF_exp_qx_3op}|_{q=0}$ of table~\ref{TableLieSymHF}, i.e., 
the equation $v_t=v_{xx}+\delta v^m$ which possesses the three-dimensional Lie symmetry algebra~$\mathfrak g$
generated by the basis
operators \[X_1=\partial_t,\quad X_2=\partial_x,\quad X_3=2t\partial_t+x\partial_x+\frac2{1-m}v\partial_v.\]
These operators satisfy the commutations relations
$[X_1,X_2]=0$, $[X_1,X_3]=2X_1$, $[X_2,X_3]=X_2$.
An optimal set of subalgebras of the algebra~$\mathfrak g$ can be easily constructed with application of the standard
technique~\cite{Ovsiannikov1982,Olver1986}. Another way is to take the set from~\cite{Patera&Winternitz1977},
where optimal sets of subalgebras are listed for all three- and four-dimensional algebras.
A complete list of inequivalent one-dimensional subalgebras of the algebra~$\mathfrak g$ is exhausted by 
the subalgebras 
$\langle X_3\rangle$, $\langle X_2\rangle$, $\langle X_2-X_1\rangle$, $\langle X_2+X_1\rangle$, and $\langle X_1\rangle$.
This list can be reduced if we additionally use the discrete symmetry $(t,x,v)\rightarrow(t,-x,v)$, which
maps $\langle X_2+X_1\rangle$ to $\langle X_2-X_1\rangle$, thereby reducing the number of inequivalent subalgebras to four.

The optimal set of two-dimensional subalgebras is formed by the subalgebras
$\langle X_3,\,X_1\rangle$, $\langle X_3,\,X_2\rangle$, and $\langle X_1,\,X_2\rangle$.
Lie reduction to algebraic equations with the latter two-dimensional subalgebra leads only to the trivial zero solution.
Below we list all the other subalgebras from the optimal set as well as the corresponding ans\"atze and reduced equations.
Solutions of some reduced equations are presented.
\begin{gather*}
\langle X_3\rangle\colon\quad v=t^\frac1{1-m}\varphi(\omega),\quad\omega=\frac x{\sqrt t},\quad
\varphi_{\omega\omega}+\frac12\omega\varphi_{\omega}+\frac{1}{m-1}\,\varphi+\delta\varphi^m=0;\\
\langle X_2\rangle\colon\quad v=\varphi(\omega),\quad\omega=t,\quad
\varphi_{\omega}=\delta\varphi^m,\quad
\varphi=\left(\delta(1-m)\omega+C\right)^{\frac1{1-m}};\\[1ex]
\langle X_2-X_1\rangle\colon\quad v=\varphi(\omega),\quad\omega=x+t,\quad
\varphi_{\omega\omega}-\varphi_{\omega}+\delta\varphi^m=0;\\[1ex]
\langle X_1\rangle\colon\quad v=\varphi(\omega),\quad\omega=x,
\quad\varphi_{\omega\omega}+\delta \varphi^m=0;\\
\langle X_3,\,X_1\rangle\colon\quad v=\hat Cx^{\frac2{1-m}},\quad \tfrac{2(1+m)}{(1-m)^2}\hat C+\delta {\hat C}^m=0,\quad
\hat C=\left(-{\tfrac{\delta(1-m)^2}{2(1+m)}}\right)^{\frac 1{1-m}};\\[.5ex]
\langle X_3,\,X_2\rangle\colon\quad v=\hat Ct^{\frac1{1-m}},\quad \tfrac{1}{1-m}\hat C+\delta {\hat C}^m=0,\quad
\hat C=\left(\delta(1-m)\right)^{\frac 1{1-m}}.
\end{gather*}

Two kinds of exact solutions can be constructed for arbitrary values of $m$: 
the $x$-free solution  $v=\left(\delta(1-m)t\right)^\frac 1{1-m}$ and 
the stationary solution $v=\left(-{\frac{\delta(1-m)^2}{2(1+m)}}\right)^{\frac 1{1-m}}x^{\frac 2{1-m}}$. 
There are a number of other known exact solutions for particular values of $m$. 
Some of such exact solutions are presented in section~\ref{Subsection_GenerationOfSolutions_n0}. 

For the other cases from table~\ref{TableLieSymHF}, which are inequivalent with respect to point transformation,
we list only solutions without used ans\"atze and reduced equations.

\medskip
$1_{q=0}$:\quad
$v=\Bigl(Ce^{a_1(1-m)t}-\frac{\delta}{\tilde a_1}\Bigr)^\frac 1{1-m}$;

%\medskip $2_{q=0}$:\quad $v=\left(C+\delta(1-m)t\right)^\frac 1{1-m}$,\quad
%$v=\left(-\frac{\delta(1-m)^2}{2(1+m)}\right)^{\frac 1{1-m}}x^{\frac 2{1-m}}$;

\medskip
3:\quad
$v=\Bigl(-\tfrac{(k+2)(m+k+1)}{\delta(1-m)^2}-\tfrac{a_2}\delta\Bigr)^{\frac 1{m-1}}x^\frac{k+2}{1-m}$;

\medskip
5:\quad
$v=\left(\frac\delta{\beta(1-a_3)}+Ce^{2pt(1-a_3)}\right)^{\frac 1{1-m}}e^{-\frac{\beta}2x^2}$.

\medskip
Applying transformations~\eqref{addit_tr_1} and \eqref{addit_tr_2}, we generate exact solutions for the rest of equations
from class~\eqref{class_vFH} with the value of the arbitrary elements presented in table~\ref{TableLieSymHF}.

\medskip
1:\quad
$v=e^{\alpha x}\Bigl(Ce^{(\alpha^2+a_1)(1-m)t}-\frac{\delta}{\alpha^2+a_1}\Bigr)^\frac 1{1-m}$;

\medskip
2:\quad
$v=e^{\alpha x}\bigl(\delta(1-m)t+C\bigr)^\frac 1{1-m}$,\quad
$v=\left(-{\frac{\delta(1-m)^2}{2(1+m)}}\right)^{\frac 1{1-m}}(x+2\alpha t)^{\frac 2{1-m}}e^{\alpha x}$;

\medskip
4:\quad 
$v=\Bigl(-\tfrac{(k+2)(m+k+1)}{\delta(1-m)^2}-\tfrac{a_2}\delta\Bigr)^{\frac 1{m-1}}x^\frac{k+2}{1-m}e^{-\frac{\beta}{2}x^2}$;

\medskip
6:\quad 
$v=\left(\frac
{\delta (m-1)}{4\beta}+Ce^{4\beta t}\right)^{\frac 1{1-m}}e^{-\frac{\beta}2x^2}$,\quad
$v=\left(-{\frac
{\delta(1-m)^2}{2(1+m)}}\right)^{\frac 1{1-m}}x^{\frac 2{1-m}}e^{-\frac{\beta}2x^2}$.

\medskip
Similarity solutions of equations from the double-imaged class with arbitrary elements presented in table~\ref{TableLieSymHG} 
can be directly found by the reduction method with respect to subalgebras of the corresponding Lie invariance algebras 
or constructed from Lie solutions for equations from class~\eqref{class_vFH_m2} with $m=2$ via gauging transformation~\eqref{gauge_m=2}.
Using the reduction method, we construct the following exact solutions. 
(Below the numerations of cases coincides with numeration of table~\ref{TableLieSymHG}.
$\chi=\gamma\tan(\gamma t)$ if $\Delta<0$; 
$\chi\in\{t^{-1},0\}$ if $\Delta=0$; 
$\chi\in\{\gamma\tanh(\gamma t),\gamma\coth(\gamma t),\pm\gamma\}$ if $\Delta>0$; 
$\gamma=\sqrt{|\Delta|}$.)

\medskip
1 and 2:\quad $w=-\dfrac{q^2+2\chi(t)}{2\delta e^{qx}}$,\quad where \quad $\Delta=\dfrac{q^4}4-\delta b_1$;

\medskip
3:\quad $w=\dfrac{-(k+2)(k+3)\pm\sqrt{(k+2)^2(k+3)^2-4 b_2}}{2\delta x^{k+2}}$;

\medskip
4:\quad $w=-\dfrac{2px^2(2px^2+2k+3)+(k+2)(k+3)\pm\sqrt{(k+2)^2(k+3)^2-4 b_2}}{2\delta x^{k+2}e^{px^2}}$;

\medskip
5 and 6:\quad 
$w=-\dfrac{2p^2x^2-p+\chi(t)}{\delta e^{px^2}}$,\quad where \quad $\Delta=p^2(5-b_3)$ and $\Delta=16p^2$, respectively;

\medskip
6:\quad 
$w=-\dfrac{px^2(2px^2+3)+6}{\delta x^2e^{px^2}}$.

\medskip

In order to obtain similarity solutions of equations from the initial class~\eqref{class_f=g} we also have two possibilities,
namely, the direct reduction with respect to Lie symmetry operators adduced in table~\ref{TableLieSym_initial_class}
or the generation of solutions from the similarity solutions of equations from the imaged classes using the
gauging transformations~\eqref{gauge} or~\eqref{gauge_m=2}, respectively. It is easy to see that the imaged class~\eqref{class_vFH}
(and~\eqref{class_GH}) has simpler structure of equations and corresponding Lie symmetry operators.
So, the optimal way is applying of gauging transformations to similarity solutions of the imaged classes adduced above.
In particular, the following solutions are obtained.

%\medskip $u_t=u_{xx}+\delta e^{qx}u^m\colon$\quad $u=e^{\alpha x}\Bigl(Ce^{\alpha qt}-\frac\delta{\alpha^2}\Bigr)^\frac 1{1-m}$;

\medskip
$\cos^2\!x\,u_t=(\cos^2\!x\,u_x)_x+\delta e^{qx}\cos^{m+1}\!x\,u^m$:\quad
$u=\Bigl(Ce^{(\alpha^2+1)(1-m)t}
-\frac{\delta}{\alpha^2+1}\Bigr)^\frac 1{1-m}e^{\alpha x}\sec x$;

\medskip
$e^xu_t=(e^xu_x)_x+\delta e^{rx}u^m$:\quad 
$u=\Bigl(Ce^{\frac{(r-1)(r-m)}{1-m}t}-\frac{\delta(1-m)^2}{(r-1)(r-m)}\Bigr)^\frac1{1-m}e^{\frac{r-1}{1-m} x}$;

\medskip
$e^xu_t=(e^xu_x)_x+\delta e^{x}u^m$:\quad
$u=\left(C+\delta(1-m)t\right)^\frac 1{1-m}$,\quad
$u=\left(-{\frac{\delta(1-m)^2}{2(1+m)}}\right)^{\frac 1{1-m}}(x+t)^{\frac 2{1-m}}$;

\medskip
$x^{\lambda}u_t=(x^{\lambda}u_x)_x+\delta x^{\gamma}u^m$:\quad
$u=\left(\frac{\delta(1-m)^2}{(2-\lambda+\gamma)(m(\lambda-1)-\gamma-1)}\right)^\frac 1{1-m}x^\frac{2-\lambda+\gamma}{1-m}$;

\medskip
$x\cos^2(\ln|x|^\rho)\,u_t=(x\cos^2(\ln|x|^\rho)\,u_x)_x+\delta x^l\cos^{m+1}(\ln|x|^\rho)\,u^m$:

\smallskip\qquad
$u=\Bigl(\frac{-4\delta(1-m)^2}{(2l-m+3)(2l+m+1)+(1+4\rho^2)(1-m)^2}\Bigr)^{\frac 1{1-m}}x^\frac{l+1}{1-m}\sec(\ln|x|^\rho)$,

\medskip
$x^{-1}M^2_{\kappa,\mu}(\beta x^2)u_t=\left(x^{-1}M^2_{\kappa,\mu}(\beta x^2)u_x\right)_x+
\delta x^se^{px^2}M^{m+1}_{\kappa,\mu}(\beta x^2)u^m$, \quad $\beta=\frac {2p}{m-1}$:

\medskip\qquad
$\kappa=\frac{s+3}{2(1-m)}$:\quad
$u=\Bigl(\frac{-4\delta(1-m)^2}{(2s+m+5)(2s+3m+3)+(1-16\mu^2)(1-m)^2}\Bigr)^{\frac 1{1-m}}
\,\dfrac{x^{\frac{s+3}{1-m}}e^{-\frac{\beta}2x^2}}{M_{\kappa,\mu}(\beta x^2)}$;

\medskip\qquad
$s=-\frac{m+1}2$, $\mu=\frac14$:\quad
$u=\left(\frac\delta{\beta(1-4\kappa)}+Ce^{2pt(1-4\kappa)}\right)^{\frac 1{1-m}}
\dfrac{\sqrt{x}\,e^{-\frac{\beta}2x^2}}{M_{\kappa,\mu}(\beta x^2)}$;

\medskip\qquad
$s=-\frac{m+1}2$, $\mu=\frac14$, $\kappa=\frac{5-m}{4(1-m)}$:\quad

\qquad\qquad
$u=\left(\frac
{\delta (m-1)}{4\beta}+Ce^{4\beta t}\right)^{\frac 1{1-m}}\dfrac{\sqrt{x}\,e^{-\frac{\beta}2x^2}}{M_{\kappa,\mu}(\beta x^2)}$,\quad
$u=\left(-{\frac{\delta(1-m)^2}{2(1+m)}}\right)^{\frac 1{1-m}}\dfrac{x^{2\kappa}e^{-\frac{\beta}2x^2}}{M_{\kappa,\mu}(\beta x^2)}$.

\medskip

Due to possibility of finding more solutions for the double-imaged class, in the case
$m=2$ we can construct more exact solutions of equations from class~\eqref{class_f=g} 
using gauge transformations:

\medskip
$e^x\,u_t=(e^x\,u_x)_x+\delta e^{(q+1)x}\,u^2\colon$\quad
$u=-\dfrac{\gamma+\chi(t)}{\delta e^{qx}}$,\quad $\gamma:=\tfrac{q^2-q}2$;

\medskip
$\cos^2\!x\,u_t=(\cos^2\!x\,u_x)_x+\delta e^{qx}\cos^{3}\!x\,u^2\colon$\quad 
$u=-\dfrac{\gamma+\chi(t)}{\delta e^{qx}\cos x}$,\quad $\gamma:=\tfrac{q^2+1}{2}$,

\medskip\noindent
where $\chi\in\{\gamma\tanh(\gamma t),\gamma\coth(\gamma t),\gamma\}$.

All the solutions obtained can be transformed by equivalence transformations to
similarity solutions of equations from classes~\eqref{class_f=g}, \eqref{class_vFH} and \eqref{class_GH} 
with more complicated coefficients.

\subsection{Generation of exact solutions by point transformations}\label{Subsection_GenerationOfSolutions_n0}

Constant coefficient reaction--diffusion equations are better studied than variable coefficients ones.
Collections of their known exact solutions are presented, e.g., in~\cite{Ibragimov1994V1;2,Polyanin&Zaitsev2004}.
Using the mappings between classes via point transformations 
and known exact solutions of constant coefficient equations from imaged or double-imaged classes,
we construct new exact solutions for variable coefficient equations from class~\eqref{class_f=g}.

Constant coefficient nonlinear equations from class~\eqref{class_vFH} have the form
\begin{equation}\label{equation_vFH_constant coef}
v_t=v_{xx}+\delta v^m+\varepsilon v,
\end{equation}
where $\delta=\pm1\bmod G^{\sim}_{FH}$, $\varepsilon\in\{-1,0,1\}\bmod G^{\sim}_{FH}$,
and arise in heat and mass transfer, combustion theory, biology and ecology. 
For the values $\delta=-\varepsilon=-1$, equation~\eqref{equation_vFH_constant coef} 
generalizes the well-known Fisher equation ($m=2$), 
and belongs to the class of Kolmogorov--Petrovskii--Piskunov equations if $m>1$. 
It describes the mass transfer in a two-component medium at rest with a volume chemical reaction of quasi-first order. 
The kinetic function $f(v) = v(1-v)$ models also an autocatalytic chain reaction in combustion theory.
The following exact solutions of~\eqref{equation_vFH_constant coef} are known~\cite{NikitinBarannyk2004,Polyanin&Zaitsev2004}:
\begin{gather}\label{solution_KPP_eq}
v=\bigl(\beta+C\exp(\lambda t+\mu x) \bigr)^\frac2{1-m},
\end{gather}
where $C$ is an arbitrary constant and the parameters $\lambda$, $\mu$, and $\beta$ are given by
\[
\lambda=\frac{\varepsilon(1-m)(m+3)}{2(m+1)}, \quad
\mu=\pm\sqrt{\frac{\varepsilon(1-m)^2}{2(m+1)}}, \quad
\beta=\pm\sqrt{-\frac{\delta}{\varepsilon}}.
\]
If $\delta=-\varepsilon=-1$, solution~\eqref{solution_KPP_eq} 
is also presented in the form~\cite{Wang1988,HeremanTakaoka1990} 
\begin{equation}\label{solution_Fisher}
v=\left[\frac12-\frac12\tanh\left(\frac{m-1}{2\sqrt{2m+2}}
\left(x-\frac{(m+3)t}{\sqrt{2m+2}}\right)\right)\right]^\frac2{m-1}.
\end{equation}
In the case $m=2$, it is the classical traveling wave solution
of the Fisher equation found by Ablowitz and Zeppetella
as early as in 1979~\cite{AblowitzZeppetella1979}. 

Equations~\eqref{equation_vFH_constant coef} are the images,
with respect to transformation~\eqref{gauge}, of the equations from
class~\eqref{class_f=g} with the coefficients 
\begin{equation}\label{eq_preimage_of_KPP_eq}
f=\theta^2,\quad h=\delta\theta^{m+1}, 
\end{equation}
where
$\theta=c_1x+c_2$ if $\varepsilon=0$, 
$\theta=c_1\sin x+c_2\cos x$ if $\varepsilon=1$ and 
$\theta=c_1\sinh x+c_2\cosh x$ if $\varepsilon=-1$. 
Applying the transformation $u=v/|\theta|$ being the inverse of~\eqref{gauge} to known solution~\eqref{solution_KPP_eq}, 
we obtain new solutions of equation~\eqref{class_f=g} with the above values of $f$ and $h$:
\[
u=|\theta(x)|^{-1}\left[\beta+C\exp(\lambda t+\mu x)\right]^\frac2{1-m}.
\]

The class of equations~\eqref{equation_vFH_constant coef} depending on values of the constants
$\delta$, $\varepsilon$ and $m$ contains
interesting subclasses which were investigated by many researchers from different points of view.
As a result, exact solutions differing from~\eqref{solution_KPP_eq} were constructed for some of these equations.
Below we consider several important equations of form~\eqref{equation_vFH_constant coef}
and use their exact solutions for the generation of new exact solutions for equations from class~\eqref{class_f=g}
with the coefficients given by~\eqref{eq_preimage_of_KPP_eq}.

Due to the connection established by the gauging transformation~\eqref{gauge},
we can easily construct also real non-Lie exact solutions for any equation
from class~\eqref{class_f=g} with $m=3$ and coefficients given by~\eqref{eq_preimage_of_KPP_eq},
as well as for all equations equivalent to them with respect to transformations from the group~${\hat G_{f=g}}^{\sim}$.
For example, if $c_1=0$, $c_2=1$, $\delta=1$ and $\varepsilon=-1$ then $\theta=\cosh x$ and 
$(f,h)=(\cosh^2\!x,\,\cosh^4\!x)$. 
Therefore, applying the transformation $u=v/\cosh x$ to the solution presented for the chosen values of $\delta$ and $\varepsilon$ 
in appendix~\ref{Section_ExactSolutionsOfCubicHeatEq}, we obtain the non-Lie exact solution
\[
u=\frac12 C_1\,{\rm sech} (x)\exp\left(-\tfrac32t\right)\sin\left(\tfrac{\sqrt2}2x\right)
{\rm sd}\left(C_1\exp\left(-\tfrac32t\right)\cos\left(\tfrac{\sqrt2}2x\right)+C_2,\tfrac{\sqrt2}2\right),\quad C_1\neq0.
\]
of the equation $\cosh^2\!x\,u_t=\left(\cosh^2\!x\,u_x\right)_x+\cosh^4\!x\,u^3$.
Hereafter ${\rm cn}(z,k)$, ${\rm sn}(z,k)$, ${\rm ds}(z,k)$, and ${\rm sd}(z,k)$
are Jacobian elliptic functions \cite{WhittakerWatson}.

\looseness=-1
In section~\ref{SectionClassificationOfAdmTrans_n0} the sets of all additional equivalence
transformations for the classes~\eqref{class_vFH}, \eqref{class_GH} and~\eqref{class_f=g}
are exhaustively described.
Some of these transformations also map variable coefficient equations to constant coefficient
ones whose exact solutions are often known.
Thus, cases $2|_{q\neq0}$ and $6$ of table~\ref{TableLieSymHF}
(resp. cases $2.2$ and $6$ of table~\ref{TableLieSym_initial_class})
are equivalent to case $2|_{q=0}$ (resp. case $2.1$) of the same table
with respect to point transformations~\eqref{addit_tr_1} and \eqref{addit_tr_2} with $k=0$ 
(resp.~\eqref{addit_tr_1fh} and~\eqref{addit_tr_3fh}).

The application of the inverse of transformation~\eqref{addit_tr_1} to the non-Lie solutions~\eqref{solutions_m3_CM1994}
of equation~\eqref{equation_vFH_constant coef} with $m=3$ and $\varepsilon=0$, presented in appendix~\ref{Section_ExactSolutionsOfCubicHeatEq}, 
results in the non-Lie solutions
\begin{gather*}
\delta=-1\colon\quad
v=2\sqrt2e^{-\frac{qx}2}(x-qt)\,{\rm ds}\!\left(\omega,\tfrac{\sqrt2}2\right),\quad
v=\sqrt2e^{-\frac{qx}2}(x-qt)\dfrac{1+{\rm cn}\left(\omega,\tfrac{\sqrt2}2\right)}{{\rm sn}\left(\omega,\tfrac{\sqrt2}2\right)},\\
\delta=1\colon\phantom{-}\quad v={\sqrt2}e^{-\frac{qx}2}(x-qt)\,{\rm sd}\!\left(\omega,\tfrac{\sqrt2}2\right),
\end{gather*}
where $\omega=(x-qt)^2+6t$, of the equation
$%\label{eq_for_transformed_solutions}
v_t=v_{xx}+\delta e^{qx}v^3-\frac14q^2v.
$
In the same way 
stationary solutions~\eqref{EqStationarySolutionsNHEsM3} leads to the nonstationary solutions
\begin{gather*}\begin{array}{l}
\delta=-1\colon\quad
v=\dfrac{\sqrt2}{\zeta}\,e^{-\frac{qx}2}, \quad
v=\sqrt2\,e^{-\frac{qx}2}\,{\rm ds}\left(\zeta,\tfrac{\sqrt2}2\right), \quad
v=\dfrac{\sqrt2}2\,e^{-\frac{qx}2}\,\dfrac{1+{\rm cn}\left(\zeta,\tfrac{\sqrt2}2\right)}{{\rm sn}\left(\zeta,\tfrac{\sqrt2}2\right)},
\\[1ex]
\delta=1\colon\phantom{-}\quad 
v=\dfrac{\sqrt2}2\,e^{-\frac{qx}2}\,{\rm sd}\!\left(\zeta,\tfrac{\sqrt2}2\right),
\end{array}\end{gather*}
where $\zeta=x-qt$, 
and the scale-invariant solution~\eqref{solution_m3_scale} is mapped to the solution
\[v=\frac{2\sqrt2(x-qt)}{(x-qt)^2+6t}\,e^{-\frac{qx}2}.\]

Transforming solutions~\eqref{solutions_m3_CM1994}--\eqref{EqStationarySolutionsNHEsM3} 
via the inverse of~\eqref{addit_tr_2} with $k=0$,
we obtain the following solutions of the equation
$v_t=v_{xx}+\delta e^{px^2}v^3-p(px^2+1)v$:
\begin{gather*}
\delta=-1\colon\quad 
v=2\sqrt2x\,e^{-\frac{p}2x^2-4pt}{\rm ds}\!\left(\hat\omega,\tfrac {\sqrt2}2\right),\quad
v=\sqrt2\,x\,e^{-\frac{p}2x^2-4pt}\dfrac{1+{\rm cn}\left(\hat\omega,\tfrac{\sqrt2}2\right)}{{\rm sn}\left(\hat\omega,\tfrac{\sqrt2}2\right)},
\\
\phantom{\delta=-1\colon\quad}
v=\frac{4\sqrt2\,p\,x}{2px^2-3}\,e^{-\frac{p}2x^2},\quad 
v=\frac{\sqrt2}xe^{-\frac p2x^2},\quad 
v=\sqrt2e^{-\frac{p}2x^2-2pt}{\rm ds}\left(\check\omega,\tfrac {\sqrt2}2\right),
\\
\phantom{\delta=-1\colon\quad}
v=\dfrac{\sqrt2}2\,e^{-\frac{p}2x^2-2pt}\,\dfrac{1+{\rm cn}\left(\check\omega,\tfrac{\sqrt2}2\right)}
{{\rm sn}\left(\check\omega,\tfrac{\sqrt2}2\right)};
\\
\delta=1\colon\phantom{-}\quad 
v={\sqrt2}x\,e^{-\frac{p}2x^2-4pt}{\rm sd}\left(\hat\omega,\tfrac {\sqrt2}2\right),\quad 
v=\frac{\sqrt2}2e^{-\frac{p}2x^2-2pt}{\rm sd}\left(\check\omega,\tfrac {\sqrt2}2\right).
\end{gather*}
Here $\hat\omega=e^{-4pt}\left(x^2-\tfrac 3{2p}\right)$ and $\check\omega=e^{-2pt}x$.

In analogous way, acting on solutions~\eqref{solutions_m3_CM1994}--\eqref{EqStationarySolutionsNHEsM3} 
by the inverses of transformations~\eqref{addit_tr_1fh} and~\eqref{addit_tr_3fh},
we can construct exact solutions of the equations $e^xu_t=(e^xu_x)_x+\delta e^{x}u^3 $ and
\[
x^{-1}M^2_{-\frac14,\frac14}(p x^2)u_t=\left(x^{-1}M^2_{-\frac14,\frac14}(p x^2)u_x\right)_x+
\delta x^{-2}e^{px^2}M^{4}_{-\frac14,\frac14}(p x^2)u^3
\]
from class~\eqref{class_f=g}, respectively.
Note that in the case $p>0$ the above Whittaker function is expressed via the error function%${\rm erf}(z)=\frac{2}{\sqrt{\pi}}\int_0^ze^{-t^2}dt$
: $M_{-\frac14,\frac14}(p x^2)=\frac12\,\sqrt\pi\,\sqrt[4]{px^2}\,e^{\frac p2\,x^2}{\rm erf}( \sqrt{px^2})$. %\cite{AbramowitzStegun}.

\section{Conclusion}\label{Conclusion_n0}

\looseness=-1
The group classification problem for class~\eqref{eqRDfghPower} has demanded a development of new tools.
A brief scheme of the used technique is as follows.
We construct the generalized extended equivalence group of class~\eqref{eqRDfghPower} and then apply it
to gauge arbitrary elements, reducing their number. The~used gauge is multi-step.
At first, we put $g=f$ with usual equivalence transformations.
Even this step is not obvious since the gauge $g=1$ seems simpler than $g=f$ but this is not the case.
Further gauging is not possible within class~\eqref{eqRDfghPower}.
This is why the next step is to gauge arbitrary elements via a mapping of the gauged class $g=f$ to another class
having an equivalence group of a simpler structure.
Since the case $m=2$ possesses an extension of the (conditional) equivalence group in comparison
with the general values of $m$, an additional mapping to one more class is necessary in this case.
As a result, the final imaged classes admit only the usual equivalence groups parameterized by four constant
parameters. Therefore, these classes are much more appropriate for group classification than class~\eqref{eqRDfghPower}.
The used mappings are not one-to-one but the preimages of the same equations are equivalent within the initial class,
i.e., the mappings are agreed with the structure of the equivalence groups.
This allows us to perform at first the group classification of the imaged classes
and then to derive the group classification of the initial class.

In conclusion we would like to emphasize that
the complete solution of the group classification problem
for class~\eqref{eqRDfghPower} became possible only due to usage of
the method base on simultaneous application of gauging
and generalized extended equivalence transformations.
This method can be applied for solving of similar problems for other classes of differential equations and
extended, e.g., to investigation of reduction operators, conservation laws and potential symmetries.

The natural continuation of the presented study is to perform extended group analysis
of the wider class of variable coefficient reaction--diffusion equations of the general form
\begin{equation}\label{equation_RD_general}
f(x)u_t=\left(g(x)A(u)u_x\right)_x+h(x)B(u),
\end{equation}
where $fgA\neq0$. \looseness=-1
The case of $A$ and $B$ being power functions is investigated successfully in~\cite{VJPS2007}
and in the present paper.
Our practice shows that the case of power nonlinearities similar to $A$ and $B$ usually is the most complicated
(see, e.g.,~\cite{Ivanova&Sophocleous2006,Ivanova&Popovych&Sophocleous2006I,Popovych&Ivanova2004NVCDCEs}).
The group classification for the case of exponential nonlinearities can be derived from
the corresponding results for power nonlinearities by  means of using
the singular limiting processes called equation contractions~\cite{Ivanova&Popovych&Sophocleous2006II}.
Thus, one going to classify Lie symmetries of equations from class~\eqref{equation_RD_general}
has every prospect of success.

\subsection*{Acknowledgements}
%\bigskip\noindent{\bf Acknowledgements.}
The research of R.\,P. was supported by Austrian Science Fund (FWF), Lise Meitner project M923-N13 and START-project Y237. 
O.\,V. and R.\,P. are grateful for the hospitality and financial support by the University of Cyprus.
The authors also wish to thank the referees for their suggestions for the improvement of this paper.

\appendix

\section{Exact solutions of cubic heat equation}\label{Section_ExactSolutionsOfCubicHeatEq}

We consider equation~\eqref{equation_vFH_constant coef} with $m=3$. 
For the values $\delta=-\varepsilon=-1$ 
it is the real Newell--Whitehead equation whose exact solution was found in~\cite{Cariello&Tabor1989}. 
This equation is a partial case of the Fitzhugh--Nagumo equation
$v_t=v_{xx}-v(1-v)(a-v)$ with $a=-1$, which arises in population genetics and models the transmission
of nerve impulses. 
A catalogue of exact solutions of the Fitzhugh--Nagumo equation is presented, e.g., in~\cite{Polyanin&Zaitsev2004}. 

Equations~\eqref{equation_vFH_constant coef} with $m=3$ are interesting also with the nonclassical
symmetry point of view (see section~\ref{Section_NonclassicalSymmetries_n0} for more details).
A number of exact solutions of equation~\eqref{equation_vFH_constant coef} with $m=3$ and of other
constant coefficient quasilinear diffusion equations with cubic source terms
were constructed in~\cite{ArrigoHillBroadbridge1993}
and~\cite{Clarkson&Mansfield1993} (and in implicit form in~\cite{FushchichSerov1990}) by the reduction method with nonclassical symmetry operators.
Since these solutions are used in section~\ref{Subsection_GenerationOfSolutions_n0} to derive new exact solutions 
of variable coefficient equations from class~\eqref{eqRDfghPower}, 
here we arrange and complete lists of exact solutions presented in~\cite{ArrigoHillBroadbridge1993,Clarkson&Mansfield1993} by missed ones.
Thus, the non-Lie real exact solutions of equation~\eqref{equation_vFH_constant coef} with $m=3$ and $\varepsilon\neq0$ are the following:
\begin{gather*}
\delta=-1,\, \varepsilon=1\colon\phantom{-}\,
v=\frac{C_1\exp\left(
\frac{\sqrt2}2x\right)-C_1'\exp\left(-\frac{\sqrt2}2x\right)}
{C_2\exp\left(-\frac32t\right)+C_1\exp\left(\frac{\sqrt2}2x\right)+C_1'\exp\left(-\frac{\sqrt2}2x\right)},
\\[.5ex]
\phantom{\delta=-1,\, \varepsilon=-1\colon\,}
v=C_1\exp\left(\tfrac32t\right)\sinh\left(\tfrac{\sqrt2}2x\right)
{\rm ds}\left(C_1\exp\left(\tfrac32t\right)\cosh\left(\tfrac{\sqrt2}2x\right)+C_2,\tfrac{\sqrt2}2\right),
\\[.5ex]
\phantom{\delta=-1,\, \varepsilon=-1\colon\,}
v=C_1\exp\left(\tfrac32t\right)\cosh\left(\tfrac{\sqrt2}2x\right)
{\rm ds}\left(C_1\exp\left(\tfrac32t\right)\sinh\left(\tfrac{\sqrt2}2x\right)+C_2,\tfrac{\sqrt2}2\right),
\\[.5ex]
\phantom{\delta=-1,\, \varepsilon=-1\colon\,}
v=\frac{C_1}2\exp\left(\tfrac32t\right)\sinh\left(\tfrac{\sqrt2}2x\right)\!\!
\frac{1+{\rm cn}\left(C_1\exp\left(\tfrac32t\right)\cosh\left(\tfrac{\sqrt2}2x\right)+C_2,\tfrac{\sqrt2}2\right)}{{\rm sn}
\left(C_1\exp\left(\tfrac32t\right)\cosh\left(\tfrac{\sqrt2}2x\right)+C_2,\tfrac{\sqrt2}2\right)},
\\[.5ex]
\phantom{\delta=-1,\, \varepsilon=-1\colon\,}
v=\frac{C_1}2\exp\left(\tfrac32t\right)\cosh\left(\tfrac{\sqrt2}2x\right)\!\!
\frac{1+{\rm cn}\left(C_1\exp\left(\tfrac32t\right)\sinh\left(\tfrac{\sqrt2}2x\right)+C_2,\tfrac{\sqrt2}2\right)}{{\rm sn}
\left(C_1\exp\left(\tfrac32t\right)\sinh\left(\tfrac{\sqrt2}2x\right)+C_2,\tfrac{\sqrt2}2\right)};
\\[1ex]
\delta=-1,\, \varepsilon=-1\colon\,
v=\frac{\sin\left(\frac{\sqrt2}2x\right)}{C_2\exp\left(\frac32t\right)+\cos\left(\frac{\sqrt2}2x\right)},
\\
\phantom{\delta=-1,\, \varepsilon=-1\colon\,}
v=C_1\exp\left(-\tfrac32t\right)\sin\left(\tfrac{\sqrt2}2x\right){\rm ds}
\left(C_1\exp\left(-\tfrac32t\right)\cos\left(\tfrac{\sqrt2}2x\right)+C_2,\tfrac{\sqrt2}2\right),
\\
\phantom{\delta=-1,\, \varepsilon=-1\colon\,}
v=\frac{C_1}2\exp\left(-\tfrac32t\right)\cos\left(\tfrac{\sqrt2}2x\right)\!\!
\frac{1+{\rm cn}\left(C_1\exp\left(-\tfrac32t\right)\sin\left(\tfrac{\sqrt2}2x\right)+C_2,\tfrac{\sqrt2}2\right)}{{\rm sn}
\left(C_1\exp\left(-\tfrac32t\right)\sin\left(\tfrac{\sqrt2}2x\right)+C_2,\tfrac{\sqrt2}2\right)};
\\
\delta=1,\, \varepsilon=1\colon\phantom{--}\,
v=\frac{C_1}2\exp\left(\tfrac32t\right)\sinh\left(\tfrac{\sqrt2}2x\right)
{\rm sd}\left(C_1\exp\left(\tfrac32t\right)\cosh\left(\tfrac{\sqrt2}2x\right)+C_2,\tfrac{\sqrt2}2\right),
\\
\phantom{\delta=1,\, \varepsilon=1\colon\phantom{--}}\,
v=\frac{C_1}2\exp\left(\tfrac32t\right)\cosh\left(\tfrac{\sqrt2}2x\right)
{\rm sd}\left(C_1\exp\left(\tfrac32t\right)\sinh\left(\tfrac{\sqrt2}2x\right)+C_2,\tfrac{\sqrt2}2\right);
\\
\delta=1,\, \varepsilon=-1\colon\phantom{-}\,
v=\frac{C_1}2\exp\left(-\tfrac32t\right)\sin\left(\tfrac{\sqrt2}2x\right)
{\rm sd}\left(C_1\exp\left(-\tfrac32t\right)\cos\left(\tfrac{\sqrt2}2x\right)+C_2,\tfrac{\sqrt2}2\right).
\end{gather*}
Here ${\rm cn}(z,k)$, ${\rm sn}(z,k)$, ${\rm ds}(z,k)$, and ${\rm sd}(z,k)=1/{\rm ds}(z,k)$
are Jacobian elliptic functions \cite{WhittakerWatson}.

\begin{note}
The constant~$C_2$ arising in the above solutions with elliptic functions is associated with the invariance of the corresponding reduced equations
with respect to the translations of the invariant independent variables.
These translations are pure hidden symmetries of the initial equations since they are not induced by Lie symmetries of the initial equations.
This is why the constant~$C_2$ cannot be omitted without loss of generality of constructed solutions. 
Therefore, the above families of solutions are essentially wider than presented in~\cite{ArrigoHillBroadbridge1993,Clarkson&Mansfield1993}.
\end{note}

\begin{note}
The above solutions involving hyperbolic functions can be partitioned into pairs invariant under the change $\sinh\leftrightarrow\cosh$.
The same statement is true for the operators of nonclassical symmetries, adduced in formula~\eqref{EqReductionOpsOfNHEs}.
The exact solutions corresponding to the last operator in~\eqref{EqReductionOpsOfNHEs} are also missed 
in~\cite{ArrigoHillBroadbridge1993,Clarkson&Mansfield1993}.
\end{note}

Up to the equivalence generated by translational and scaling symmetries
and two discrete symmetry transformations of alternating sign of the variables $x$ and $v$,
the following non-Lie real exact solutions are known 
for the equation~\eqref{equation_vFH_constant coef} with $m=3$ and $\varepsilon=0$ 
\cite{ArrigoHillBroadbridge1993,Clarkson&Mansfield1993,FushchichSerov1990}:
\begin{gather}\label{solutions_m3_CM1994}
\begin{array}{l}
\delta=-1\colon\quad
v=2\sqrt2\,x\,{\rm ds}\left(x^2+6t,\tfrac{\sqrt2}2\right), \quad
v=\sqrt2\,x\,
\dfrac{1+{\rm cn}\left(x^2+6t,\tfrac{\sqrt2}2\right)}
{{\rm sn}\left(x^2+6t,\tfrac{\sqrt2}2\right)},
\\
\delta=1\colon\phantom{-}\quad v=\sqrt2\,x\,
{\rm sd}\!\left(x^2+6t,\tfrac{\sqrt2}2\right).
\end{array}\end{gather}
In the case $\delta=-1$ this equation has the one more solution 
\begin{equation}\label{solution_m3_scale}
v=\dfrac{2\sqrt2\, x}{{x}^2+6t}
\end{equation}
which is invariant with respect to the same reduction operator (see~\eqref{EqReductionOpsOfNHEs}, the case $\varepsilon=0$) 
as solutions~\eqref{solutions_m3_CM1994} but is a Lie one since it is also invariant with respect to scale Lie symmetries. 
Solutions~\eqref{solutions_m3_CM1994} and~\eqref{solution_m3_scale} exhaust, up to the above equivalence, 
the set of nonzero solutions of this equation, having the form $v=x\varphi(\omega)$, where $\omega=x^2+6t$. 
Therefore, each solution constructed in section~3 of~\cite{NikitinBarannyk2004} is equivalent to a solution from~\eqref{solutions_m3_CM1994}.
The equation~\eqref{equation_vFH_constant coef} with $m=3$ and $\varepsilon=0$ also has similar stationary Lie solutions
\begin{gather}\label{EqStationarySolutionsNHEsM3}\begin{array}{l}
\delta=-1\colon\quad
v=\dfrac{\sqrt2}{x}, \quad
v=\sqrt2\,{\rm ds}\left(x,\tfrac{\sqrt2}2\right), \quad
v=\dfrac{\sqrt2}2\,\dfrac{1+{\rm cn}\left(x,\tfrac{\sqrt2}2\right)}
{{\rm sn}\left(x,\tfrac{\sqrt2}2\right)},
\\
\delta=1\colon\phantom{-}\quad v=\dfrac{\sqrt2}2\,{\rm sd}\!\left(x,\tfrac{\sqrt2}2\right).
\end{array}\end{gather}


\begin{thebibliography}{99}\itemsep=-0.3ex
\footnotesize

\bibitem{AblowitzZeppetella1979}
M.J. Ablowitz, A. Zeppetella,
Explicit solutions of Fisher's equation for a special wave speed,
Bull.Math. Biol. {\bf 41} (1979), 835--840.

%\bibitem{AbramowitzStegun}
%M. Abramowitz, I.A. Stegun (Editors),
%Handbook of mathematical functions with formulas, graphs, and mathematical tables,
%Dover Publications, Inc., New York, 1992.

\bibitem{Abramenko&Lagno&Samojlenko2002}
A.A. Abramenko, V.I. Lagno, A.M. Samoilenko,
Group classification of nonlinear evolution equations. II.
Invariance under solvable local transformation groups, Differ. Equ. {\bf 38} (2002), 502--509.

\bibitem{Akhatov&Gazizov&Ibragimov1989}
I.Sh.~Akhatov, R.K.~Gazizov and N.Kh.~Ibragimov, 
Nonlocal symmetries. A heuristic approach, 
%Itogi Nauki i Tekhniki, Current problems in mathematics. Newest results {\bf 34} (1989), 3--83 (in Russian); Translated in 
J. Soviet Math. {\bf 55} (1991), 1401--1450.

\bibitem{ArrigoHill1995}
D.J. Arrigo, J.M. Hill,
Nonclassical symmetries for nonlinear diffusion and absorption,
Stud. Appl. Math. {\bf 94} (1995), 21--39.

\bibitem{ArrigoHillBroadbridge1993}
D.J. Arrigo, J.M. Hill, P. Broadbridge,
Nonclassical symmetry reductions of the linear diffusion equation with a nonlinear source,
IMA J. Appl. Math. {\bf 52} (1994), 1--24.

\bibitem{Basarab-Horwath&Lahno&Zhdanov2001}
P. Basarab-Horwath, V. Lahno, R. Zhdanov, The structure of Lie
algebras and the classification problem for partial differential
equations, Acta Appl. Math. {\bf 69} (2001), 43--94; arXiv:math-ph/0005013.

\bibitem{Bluman&Cole1969}
G.W. Bluman, J.D. Cole,
The general similarity solution of the heat equation,
J. Math. Mech. {\bf 18} (1969), 1025--1042.

\bibitem{BlumanKumei1989}
Bluman G.W. and Kumei S., Symmetries and differential equations, Springer-Verlag, New York, 1989.

\bibitem{Cariello&Tabor1989}
F. Cariello, M. Tabor,
Painleve expansions for nonintegrable evolution equations,
Physica D {\bf 39} (1989), 77--94.

\bibitem{Cherniha2007}
R. Cherniha,  New $Q$-conditional symmetries and
 exact solutions of some reaction-diffusion-convection
equations arising in mathematical biology, J. Math. Anal. Appl. {\bf 326} (2007), 783--799.

\bibitem{ChernihaPliukhin2007}
R. Cherniha, O. Pliukhin, New conditional symmetries and exact solutions of nonlinear
reaction--diffusion--convection equations, J. Phys. A: Math. Theor. {\bf 40} (2007), 10049--10070;
arXiv:math-ph/0612078.

\bibitem{Clarkson&Mansfield1993}
P.A. Clarkson, E.L. Mansfield, Symmetry reductions and exact solutions of
a class of nonlinear heat equations,
Physica D {\bf 70} (1994), 250--288.

\bibitem{crank1979}
J. Crank, The Mathematics of Diffusion, 2nd ed., Oxford, London,
1979.

\bibitem{Dorodnitsyn1979;1982}
V.A. Dorodnitsyn, Group properties and invariant solutions of a
nonlinear heat equation with a source or a sink, Preprint N~57,
Moscow, Keldysh Institute of Applied Mathematics of Academy of
Sciences USSR, 1979;
%\bibitem{Dorodnitsyn1982}V.A. Dorodnitsyn, 
On invariant solutions of non-linear heat equation with a sourse, 
Zhurn. Vych. Matemat. Matemat. Fiziki {\bf 22} (1982), 1393--1400 (in Russian).

\bibitem{FushchichSerov1990}
W.I. Fushchich, N.I. Serov, Conditional invariance and reduction of nonlinear heat equation,
Dokl. Akad. Nauk Ukrain. SSR Ser. A (1990), no. 7, 24--27 (in Russian).
%Availiable at http://www.imath.kiev.ua/$\sim$fushchych.

\bibitem{Fushchych&Shtelen&Serov&Popovych1992}
W.I. Fushchych, W.M. Shtelen, M.I. Serov, R.O. Popovych,
Q-conditional symmetry of the linear heat equation,
Proc. Acad. of Sci. Ukraine (1992), no 12, 28--33.

\bibitem{Fushchych&Tsyfra1987}
W.I. Fushchych, I.M. Tsyfra,
On a reduction and solutions of the nonlinear wave equations with broken symmetry,
J. Phys. A: Math. Gen. {\bf 20} (1987), L45--L48.

\bibitem{Fushchych&Zhdanov1992}
W.I. Fushchych, R.Z. Zhdanov,
Conditional symmetry and reduction of partial differential equations,
Ukrainian Math. J. {\bf 44} (1992), 970--982.

\bibitem{Gagnon&Winternitz1993}
Gagnon L. and Winternitz P., 
Symmetry classes of variable coefficient nonlinear Schr\"odinger equations, 
J.~Phys. A {\bf 26} (1993), 7061--7076.

\bibitem{Gandarias2001}
M.L. Gandarias,
New symmetries for a model of fast diffusion,
Phys. Lett. A {\bf 286} (2001), 153--160.

\bibitem{HeremanTakaoka1990}
W. Hereman, M. Takaoka,
Solitary wave solutions of nonlinear evolution and wave equations using a direct method and
MACSYMA,
J. Phys. A {\bf 23} (1990), 4805--4822.

\bibitem{Ibragimov1994V1;2}
N.H. Ibragimov (Editor), Lie group analysis of differential
equations --- symmetries, exact solutions and conservation laws,
V.1,2, CRC Press, Boca Raton, FL, 1994.

\bibitem{Ivanova&Sophocleous2006}
N.M. Ivanova, C. Sophocleous, On the group classification of
variable coefficient nonlinear diffusion--convection equations,
J. Comp. Appl. Math. {\bf 197} (2006), 322--344.

\bibitem{Ivanova&Popovych&Sophocleous2006I}
N.M. Ivanova, R.O. Popovych, C. Sophocleous, 
Group analysis of variable coefficient diffusion--convection equations. I. Enhanced group classification, 
arXiv:0710.2731.

\bibitem{Ivanova&Popovych&Sophocleous2006II}
N.M. Ivanova, R.O. Popovych, C. Sophocleous, 
Group analysis of variable coefficient diffusion--convection equations. II. Contractions and exact solutions, 
arXiv:0710.3049.

\bibitem{kamin&rosenau1982}
S. Kamin, P. Rosenau, Nonlinear thermal evolution in an
inhomogeneous medium, J. Math. Phys. {\bf 23} (1982), 1385--1390.

\bibitem{Kingston&Sophocleous1991}
J.G. Kingston, C. Sophocleous, On point transformations of a
generalised Burgers equation, Phys. Lett.~A {\bf 155} (1991), 15--19.

\bibitem{Kingston&Sophocleous1998}
J.G. Kingston, C. Sophocleous, On form-preserving point
transformations of partial differential equations, J. Phys. A:
Math. Gen. {\bf 31} (1998), 1597--1619.

\bibitem{Kingston&Sophocleous2001}
J.G. Kingston, C. Sophocleous, Symmetries and form-preserving
transformations of one-dimensional wave equations with
dissipation, Int. J. Non-Lin. Mech. {\bf 36} (2001), 987--997.

\bibitem{Lahno&Zhdanov&Magda2006}
V. Lahno, R. Zhdanov and O. Magda,
Group classification and exact solutions of nonlinear wave equations,
Acta Appl. Math. {\bf 91} (2006), 253--313.

\bibitem{Lagno&Samojlenko2002}
V.I. Lagno, A.M. Samoilenko,
Group classification of nonlinear evolution equations. I.
Invariance under semisimple local transformation groups, Differ. Equ. {\bf 38} (2002), 384--391.

\bibitem{Lahno&Spichak&Stognii2002}
V.I. Lahno, S.V. Spichak, V.I. Stognii, Symmetry analysis of
evolution type equations, Kyiv: Institute of Mathematics of NAS of Ukraine, 2002.

\bibitem{Lie1881}
S. Lie, \"Uber die Integration durch bestimmte Integrale von einer Klasse linear partieller
Differentialgleichung, Arch. for Math. {\bf 6}, no.~3 (1881), 328--368.
%(Translation by N.H. Ibragimov:
%S. Lie, On integration of a class of linear partial differential equations by means of
%definite integrals, {\it CRC Handbook of Lie Group Analysis of Differential Equations},
%Vol. 2, 1994, 473--508).

\bibitem{Lie1891}
S. Lie, Vorlesungen \"uber Differentialgleichungen mit bekannten infinitesimalen Transformationen, 
B.G. Teubner, Leipzig, 1891.

\bibitem{Meleshko1994}
S.V. Meleshko, Group classification of equations of two-dimensional
gas motions, Prikl. Mat. Mekh. {\bf 58} (1994), 56--62 (in
Russian); translation in J.~Appl. Math. Mech. {\bf 58} (1994),
629--635.

\bibitem{murray2002}
J.D. Murray, Mathematical Biology I: An Introduction, 3rd Ed.,
Springer, New York, 2002.

\bibitem{NikitinBarannyk2004}
A.G. Nikitin, T.A. Barannyk,
Solitary wave and other solutions for nonlinear heat equations,
Cent. Eur. J. Math. {\bf 2} (2004), 840--858; arXiv:math-ph/0303004.

\bibitem{Olver1986}
P. Olver, Applications of Lie groups to differential equations,
New-York, Springer-Verlag, 1986.

\bibitem{Ovsiannikov1959}
L.V. Ovsiannikov, Group properties of nonlinear heat equation,
Dokl. AN SSSR {\bf 125} (1959), 492--495.

\bibitem{Ovsiannikov1982}
L.V. Ovsiannikov, Group analysis of differential equations, 
Academic Press, New York, 1982.

\bibitem{Patera&Winternitz1977}
J. Patera, P. Winternitz, 
Subalgebras of real three and four-dimensional Lie algebras, 
J.~Math. Phys. {\bf 18} (1977), 1449--1455.

\bibitem{peletier1981}
L.A. Peletier, Applications of Nonlinear Analysis in the Physical Sciences, Pitman, London, 1981.

\bibitem{Polyanin&Zaitsev2004}
A.D. Polyanin and V.F. Zaitsev, Handbook of nonlinear partial differential equations,
Chapman \& Hall / CRC, Boca Raton, 2004.

\bibitem{Popovych2006c}
R.O. Popovych, Classification of admissible transformations of differential equations,
Collection of Works of Institute of Mathematics (Kyiv, Ukraine) {\bf 3} (2006), no. 2, 239--254.
%Availible at http://www.imath.kiev.ua/$\sim$appmath/Collections/collection2006.pdf.

\bibitem{Popovych&Ivanova2004NVCDCEs}
R.O. Popovych, N.M. Ivanova, New results on group classification of
nonlinear diffusion--convection equations, J. Phys. A: Math. Gen. {\bf 37}
(2004), 7547--7565; arXiv:math-ph/0306035.

\bibitem{Popovych&Kunzinger&Eshraghi2006}
R.O. Popovych, M. Kunzinger, H. Eshraghi, Admissible
transformations and normalized classes of nonlinear Schr\"odinger equations, 35 p;
arXiv:math-ph/0611061.

\bibitem{Popovych&Vaneeva&Ivanova2007}
R.O. Popovych, O.O. Vaneeva, N.M. Ivanova, Potential nonclassical
symmetries and solutions of fast diffusion equation. Phys. Lett. A
{\bf 362} (2007), 166--173; arXiv:math-ph/0506067.

\bibitem{Prokhorova2005}
M. Prokhorova, The structure of the category of parabolic equations, 24 p;
arXiv:math.AP/0512094.

\bibitem{Serov1990}
N. I. Serov, Conditional invariance and exact solutions of a nonlinear heat equation,
Ukrainian Math. J. {\bf 42} (1990), no. 10, 1216--1222.

\bibitem{touloukian1970}
Y.S. Touloukian, P.W. Powell, C.Y. Ho, P.G. Klemens, Thermophysical
properties of matter, vol. 1, Plenum, New York, 1970.

\bibitem{VJPS2007}
O.O. Vaneeva, A.G. Johnpillai, R.O. Popovych and C. Sophocleous,
Enhanced group analysis and conservation laws of variable
coefficient reaction--diffusion equations with power nonlinearities,
J. Math. Anal. Appl. {\bf 330} (2007), 1363--1386; arXiv:math-ph/0605081.

\bibitem{Wang1988}
X.Y. Wang,
Exact and explicit solitary wave solutions for the generalised Fisher equation,
Phys. Lett. A {\bf 131} (1988), 277--279.

\bibitem{WhittakerWatson}
E.T. Whittaker, G.N. Watson, A course of modern analysis%.
%An introduction to the general theory of infinite processes and of analytic functions;
%with an account of the principal transcendental functions
, Cambridge University Press, Cambridge, 1996.

\bibitem{Zhdanov&Lahno1998}
R.Z. Zhdanov, V.I. Lahno,
Conditional symmetry of a porous medium equation,  Physica D {\bf 122} (1998), 178--186.

\bibitem{Zhdanov&Lahno1999;2007}
R.Z. Zhdanov, V.I. Lahno,
Group classification of heat conductivity equations with a nonlinear source,  J. Phys. A {\bf 32} (1999), 7405--7418;
Group classification of the general second-order evolution equation: semi-simple invariance groups,
J. Phys. A: Math. Theor. {\bf 40} (2007), 5083--5103.

\bibitem{Zhdanov&Lahno2005}
R. Zhdanov, V. Lahno,
Group classification of the general evolution equation: local and quasilocal symmetries,
SIGMA {\bf 1} (2005), Paper 009, 7 pages; arXiv:nlin/0510003.

\bibitem{Zhdanov&Tsyfra&Popovych1999}
R.Z. Zhdanov, I.M. Tsyfra, R.O. Popovych,
A precise definition of reduction of partial differential equations,
J. Math. Anal. Appl. {\bf 238} (1999), 101--123; arXiv:math-ph/0207023.


\end{thebibliography}
\end{document}